\documentclass[twocolumn,showpacs,preprintnumbers,amsmath,amssymb,floatfix]{revtex4}

\usepackage{graphicx}
\usepackage{dcolumn} 
\usepackage{bm}

\newcommand{\met}{\mbox{$E_{T}\!\!\!\!\!\!\!/\,\,\,\,\,\,$}}

\newcommand{\rr}[2]{\raisebox{#2ex}[0pt]{#1}}
\begin{document}

\onecolumngrid
\begin{center}{\large \bf \boldmath Measurement of the azimuthal angle distribution
of leptons from $W$ boson decays as a function of the $W$ transverse momentum in $p\bar{p}$ collisions at $\sqrt s$=1.8 TeV}\end{center}

\date{March 30, 2005}
\thispagestyle{empty}
\font\eightit=cmti8
\def\r#1{\ignorespaces $^{#1}$}
\hfilneg
\begin{sloppypar}
\noindent
D.~Acosta,\r {14} T.~Affolder,\r 7 M.G.~Albrow,\r {13} D.~Ambrose,\r {36}   
D.~Amidei,\r {27} K.~Anikeev,\r {26} J.~Antos,\r 1 
G.~Apollinari,\r {13} T.~Arisawa,\r {50} A.~Artikov,\r {11}
% ...
\end{sloppypar}
\vskip .026in
\begin{center}
(CDF Collaboration)
\end{center}

\vskip .026in
\begin{center}
\r 1  {\eightit Institute of Physics, Academia Sinica, Taipei, Taiwan 11529, 
Republic of China} \\
\r 2  {\eightit Argonne National Laboratory, Argonne, Illinois 60439} \\
\r 3  {\eightit Istituto Nazionale di Fisica Nucleare, University of Bologna,
I-40127 Bologna, Italy} \\
\r 4  {\eightit Brandeis University, Waltham, Massachusetts 02254} \\
\r 5  {\eightit University of California at Davis, Davis, California  95616} \\
\r 6  {\eightit University of California at Los Angeles, Los 
Angeles, California  90024} \\ 
\r 7  {\eightit University of California at Santa Barbara, Santa Barbara, California 
93106} \\ 
\r 8 {\eightit Instituto de Fisica de Cantabria, CSIC-University of Cantabria, 
39005 Santander, Spain} \\
\r 9  {\eightit Carnegie Mellon University, Pittsburgh, Pennsylvania  15213} \\
\r {10} {\eightit Enrico Fermi Institute, University of Chicago, Chicago, 
Illinois 60637} \\
\r {11}  {\eightit Joint Institute for Nuclear Research, RU-141980 Dubna, Russia}
\\
\r {12} {\eightit Duke University, Durham, North Carolina  27708} \\
\r {13} {\eightit Fermi National Accelerator Laboratory, Batavia, Illinois 
60510} \\
\r {14} {\eightit University of Florida, Gainesville, Florida  32611} \\
\r {15} {\eightit Laboratori Nazionali di Frascati, Istituto Nazionale di Fisica
               Nucleare, I-00044 Frascati, Italy} \\
\r {16} {\eightit University of Geneva, CH-1211 Geneva 4, Switzerland} \\
\r {17} {\eightit Glasgow University, Glasgow G12 8QQ, United Kingdom}\\
\r {18} {\eightit Harvard University, Cambridge, Massachusetts 02138} \\
\r {19} {\eightit Hiroshima University, Higashi-Hiroshima 724, Japan} \\
\r {20} {\eightit University of Illinois, Urbana, Illinois 61801} \\
\r {21} {\eightit The Johns Hopkins University, Baltimore, Maryland 21218} \\
\r {22} {\eightit Institut f\"{u}r Experimentelle Kernphysik, 
Universit\"{a}t Karlsruhe, 76128 Karlsruhe, Germany} \\
\r {23} {\eightit Center for High Energy Physics: Kyungpook National
University, Taegu 702-701; Seoul National University, Seoul 151-742; and
SungKyunKwan University, Suwon 440-746; Korea} \\
\r {24} {\eightit Ernest Orlando Lawrence Berkeley National Laboratory, 
Berkeley, California 94720} \\
\r {25} {\eightit University College London, London WC1E 6BT, United Kingdom} \\
\r {26} {\eightit Massachusetts Institute of Technology, Cambridge,
Massachusetts  02139} \\   
\r {27} {\eightit University of Michigan, Ann Arbor, Michigan 48109} \\
\r {28} {\eightit Michigan State University, East Lansing, Michigan  48824} \\
\r {29} {\eightit Institution for Theoretical and Experimental Physics, ITEP,
Moscow 117259, Russia} \\
\r {30} {\eightit University of New Mexico, Albuquerque, New Mexico 87131} \\
\r {31} {\eightit Northwestern University, Evanston, Illinois  60208} \\
\r {32} {\eightit The Ohio State University, Columbus, Ohio  43210} \\
\r {33} {\eightit Osaka City University, Osaka 588, Japan} \\
\r {34} {\eightit University of Oxford, Oxford OX1 3RH, United Kingdom} \\
\r {35} {\eightit Universita di Padova, Istituto Nazionale di Fisica 
          Nucleare, Sezione di Padova, I-35131 Padova, Italy} \\
\r {36} {\eightit University of Pennsylvania, Philadelphia, 
        Pennsylvania 19104} \\   
\r {37} {\eightit Istituto Nazionale di Fisica Nucleare, University and Scuola
               Normale Superiore of Pisa, I-56100 Pisa, Italy} \\
\r {38} {\eightit University of Pittsburgh, Pittsburgh, Pennsylvania 15260} \\
\r {39} {\eightit Purdue University, West Lafayette, Indiana 47907} \\
\r {40} {\eightit University of Rochester, Rochester, New York 14627} \\
\r {41} {\eightit Rockefeller University, New York, New York 10021} \\
\r {42} {\eightit Instituto Nazionale de Fisica Nucleare, Sezione di Roma,
University di Roma I, ``La Sapienza," I-00185 Roma, Italy}\\
\r {43} {\eightit Rutgers University, Piscataway, New Jersey 08855} \\
\r {44} {\eightit Texas A\&M University, College Station, Texas 77843} \\
\r {45} {\eightit Texas Tech University, Lubbock, Texas 79409} \\
\r {46} {\eightit Institute of Particle Physics, University of Toronto, Toronto
M5S 1A7, Canada} \\
\r {47} {\eightit Istituto Nazionale di Fisica Nucleare, University of Trieste/\
Udine, Italy} \\
\r {48} {\eightit University of Tsukuba, Tsukuba, Ibaraki 305, Japan} \\
\r {49} {\eightit Tufts University, Medford, Massachusetts 02155} \\
\r {50} {\eightit Waseda University, Tokyo 169, Japan} \\
\r {51} {\eightit University of Wisconsin, Madison, Wisconsin 53706} \\
\r {52} {\eightit Yale University, New Haven, Connecticut 06520} \\
\end{center}

\begin{abstract}
We present the first measurement of
the $A_2$ and $A_3$ angular coefficients of the $W$ boson
produced in proton-antiproton collisions.
We study $W\rightarrow e \nu_e$ and $W\rightarrow \mu \nu_{\mu}$
candidate events 
produced in association with at least one jet at CDF, during Run Ia 
and Run Ib of the Tevatron at $\sqrt s$=1.8 TeV.  The corresponding 
integrated luminosity was 110 pb$^{-1}$.  
The jet balances 
the transverse momentum of the $W$ and introduces QCD 
effects in $W$ boson production.  The extraction of the angular
coefficients is achieved through the direct measurement of 
the azimuthal angle of the charged lepton in the Collins-Soper 
rest-frame of the $W$ boson.  The angular coefficients are measured
as a function of the transverse momentum of the $W$ boson.  The electron,
muon, and combined results are in good agreement with the Standard
Model prediction, up to order $\alpha_s^2$ in QCD.
\end{abstract}


\maketitle

\twocolumngrid
\section{\label{sec:1} Introduction}

Measurements of the $W$ boson differential cross section, as a function 
of energy and direction, provide information about the nature of both 
the underlying electroweak interaction, and the effects of chromodynamics
(QCD).
This differential cross section can be expressed us a function of 
the helicity cross sections of the $W$, allowing us to study the
$W$ polarization and associated asymmetries.  Because of the
difficulties in fully reconstructing a $W$ boson in three-dimensions 
at a hadron collider, the complete angular distribution of the $W$ 
has not been determined yet.  In this paper we present the first 
measurement of two of the four significant leading angular
coefficients of the $W$ boson produced at a hadron collider.

\begin{figure}
\includegraphics[scale=.27]{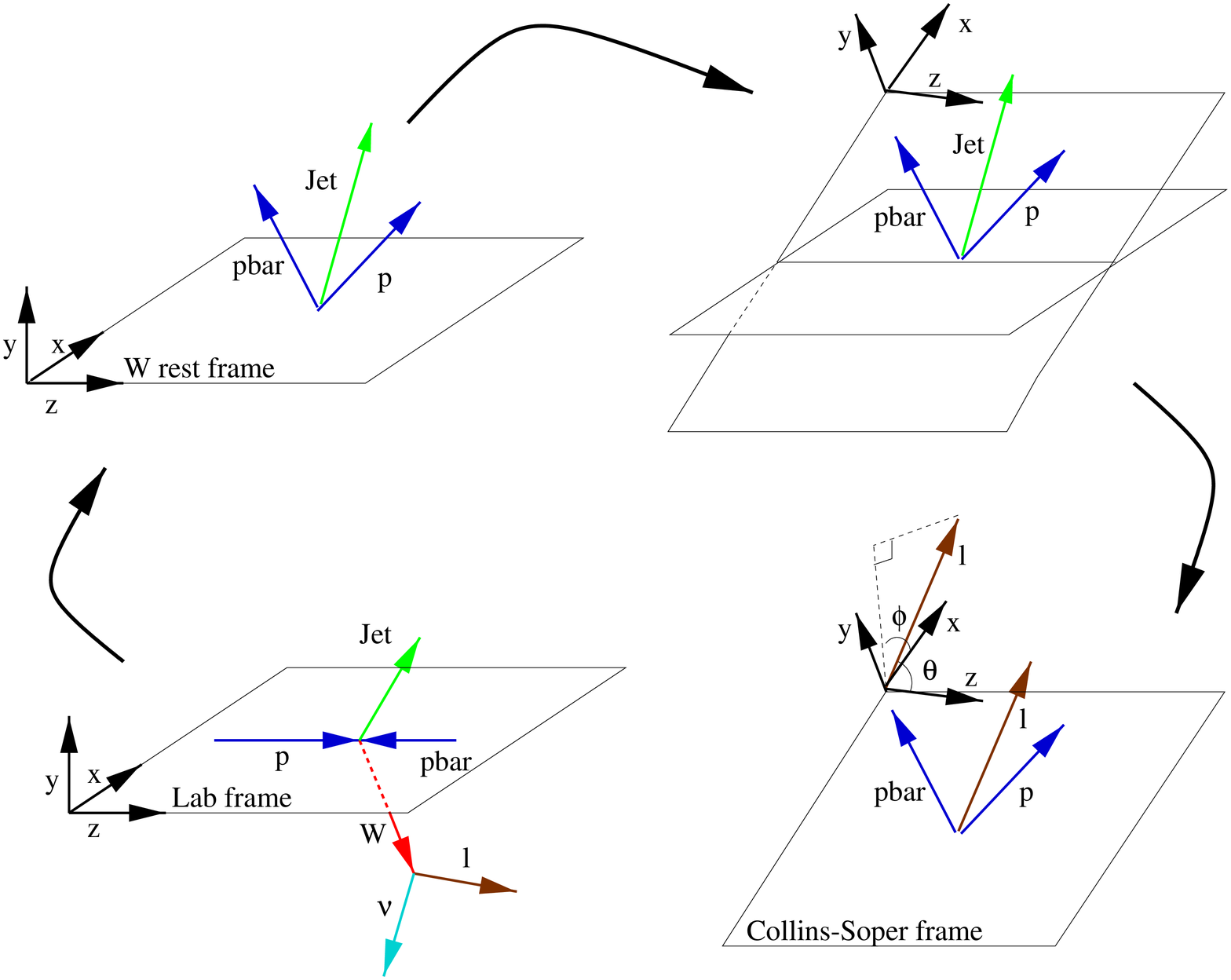} 
\caption{Transforming from the laboratory frame to the Collins-Soper frame.  We 
first boost to the $W$ rest-frame, then rotate the $x-z$ plane so that it coincides
with the $p$-$\bar{p}$ plane.  Finally we rotate the frame around the $y$-axis
so that the $z$-axis bisects the angle between $\vec{p}$ and $-\vec{\bar{p}}$.
The positive $y$ axis is selected to have the same direction as $\vec{p}_{\rm{CS}} \times \vec{\bar{p}}_{\rm{CS}}$.}
\label{csframe}
\end{figure}

The total differential cross section for $W$ boson production 
in a hadron collider is given by
\begin{eqnarray}
\frac{d{\sigma}}{d(p_T^W)^2dyd\cos{\theta}d\phi} &=& 
\frac{3}{16\pi}\frac{d\sigma^u}{d(p_T^W)^2dy}[(1+\cos^2{\theta})  \nonumber \\
&+& \frac{1}{2}A_0(1-3\cos^2{\theta})
+ A_1\sin{2\theta}\cos{\phi} \nonumber \\
&+&\frac{1}{2}A_2\sin^2{\theta}\cos{2\phi}
+ A_3\sin{\theta}\cos{\phi} \nonumber \\
&+& A_4\cos{\theta} +A_5 \sin^2 \theta \sin 2\phi  \nonumber \\
&+& A_6\sin{2\theta}\sin{\phi} 
+ A_7\sin{\theta}\sin{\phi}] 
\label{eq1}
\end{eqnarray}
where $p_T^W$ and $y$ are the transverse momentum and the rapidity of the $W$
in the laboratory frame, and $\theta$ and $\phi$ are the polar and azimuthal 
angles of the charged lepton from $W$ boson decay in the Collins-Soper (CS)
frame \cite{mirkesnpb}.  The factors $A_i(p_T^W,y)$ are the angular
coefficients of the $W$ boson, which are ratios of the helicity
cross sections of the $W$ and its total unpolarized cross section 
$d\sigma^u / d(p_T^W)^2dy$.
The CS frame \cite{cs} is the rest-frame of the $W$ with a $z$-axis that
bisects the angle between the proton direction and the direction opposite that of the antiproton 
(Figure \ref{csframe}), and it is used because in this frame we can in principle 
exactly reconstruct the azimuthal angle $\phi$ and the polar quantity
$|\cos{\theta}|$.
Our ignorance of the $W$ boson longitudinal momentum, which
is due to our inability to measure the longitudinal momentum
of the neutrino, only introduces a two-fold ambiguity on the sign
of $\cos\theta$.
It is common to integrate Equation (\ref{eq1}) over $y$ and
study the variation of the angular coefficients as a function 
of $p_T^W$.

To study the angular distribution of the $W$ we must choose
a particular charge for the boson.  In this paper we consider the $W^-$ bosons;
the $W^{+}$ bosons in our samples are $CP$ transformed to be treated 
as $W^{-}$ bosons.  The angular coefficients for the $W^+$ are 
obtained by $CP$ transforming Equation (\ref{eq1}) \cite{footnote1}.

If the $W$ is produced with no transverse momentum, it is
polarized along the beam axis, due to the V-A nature
of the weak interactions and helicity conservation.
In that case $A_4$ is the only non-zero coefficient.  
If only valence quarks contributed to $W$ production,
$A_4$ would equal 2, and the angular distribution given
by Equation (\ref{eq1}) 
would be $\sim (1+\cos\theta)^2$, a result that was 
first verified by the UA1 experiment \cite{ua1}.

If the $W$ is produced with non-negligible transverse momentum,
balanced by the associated production of jets, the rest of the
angular coefficients are present, and the cross section depends
on the azimuthal angle $\phi$ as well.  The last three angular
coefficients $A_5$, $A_6$, and $A_7$ are non-zero only if gluon 
loops are present
in the production of the $W$ boson.  Hence, in order to study all the
angular coefficients and associated helicity cross sections of 
the $W$ boson in a hadron collider, we must consider the production
of the $W$ with QCD effects up to order $\alpha_s^2$.

The importance of the determination of the $W$ angular coefficients 
is discussed in \cite{mirkesprd}, and summarized here.  It allows us to measure for the first time
the full differential cross section of the $W$ and study its polarization,
since the angular coefficients are directly 
related to the helicity cross sections.
It also helps us verify the QCD effects in the production of the
$W$ up to order $\alpha_s^2$.  For example, according to the Standard Model (SM),
$A_2$ is not equal to $A_0$ only if the effects of gluon loops are taken into account. 
In addition, $A_3$ is only affected by the gluon-quark interaction
and its measurement can be used to constrain the gluon parton distribution functions.
Moreover, the next-to-leading order angular 
coefficients $A_5$, $A_6$, and $A_7$ 
are $P$-odd and $T$-odd
and may play an important role in direct $CP$ violation effects in
$W$ production and decay \cite{hagiwara}.  Finally, quantitative 
understanding of the $W$ angular distribution could be used to test 
new theoretical models and to facilitate new discoveries.

In this paper we present the first measurement of the
$A_2$ and $A_3$ angular coefficients of the $W$ boson.
These coefficients fully describe the azimuthal differential
cross section of the $W$ boson, and they are two of the four significant
coefficients that describe the total differential cross section 
of the $W$, given that $A_1$ and the next-to-leading
order angular coefficients have considerably lower values \cite{mirkesprd,js}.
This measurement is accomplished using the azimuthal
angle of the charged lepton in the CS $W$ rest-frame
\cite{mythesis}, and is presented as a function of the
transverse momentum of the $W$ boson.  The CS polar angle
analysis is more sensitive to the $A_0$ and $A_4$ angular 
coefficients (see \cite{d0, lucio} for a measurement of $A_0$).
Because Equation (\ref{eq1}) arises solely from quantum field theory, 
without input from any
specific theoretical model of $W$ boson production, our experimental
results are thus model-independent.

\section{\label{sec:2} The CDF detector and event selection}

\subsection{The CDF detector}

The CDF detector
is described in detail in \cite{cdf}.
It is a general purpose detector of charged leptons,
hadrons, jets, and photons, produced from proton-antiproton
collisions at the Tevatron accelerator at Fermilab.  
The $W$ and $Z$ bosons are
detected through their decay leptons, while the transverse
momentum of the neutrinos is estimated from the missing
transverse energy of the events ($\met$).

The $z$-axis of the detector
coincides with the direction of the proton beam
and defines the polar angle $\theta_{\rm lab}$ in the 
laboratory frame.  
The $y$-axis points vertically upward and 
the $x$-axis is in the horizontal plane, so as to
form a right-handed coordinate system. 
The pseudorapidity, $\eta_{\rm lab} = -\ln[\tan(\theta_{\rm lab}/2)]$,
and the azimuthal angle $\phi_{\rm lab}$ are used to specify
detector physical areas.
 
The tracking system of CDF consists of the 
silicon vertex detector (SVX), the vertex time projection chamber (VTX)
and the central tracking chamber (CTC), all immersed in a 1.4~T
magnetic field produced by a superconducting
solenoid of length 4.8 m and radius 1.5 m.
The SVX, a four layer silicon micro-strip vertex detector,
is located immediately outside the beampipe.  It is used to 
find secondary vertices and provides
the impact parameter of tracks in the transverse $r-\phi_{\rm lab}$ plane.
The VTX, located outside the SVX, is a 
vertex time projection chamber that provides $r-z$ tracking information
up to a radius of 22 cm and pseudorapidity $|\eta_{\rm lab}|\leq 3.5$.
It measures the $z$-position of the primary vertex.
Finally, surrounding the SVX and the VTX is the CTC,
a 3.2 m long cylindrical drift chamber containing 
84 layers of sense wires arranged in five superlayers of axial
wires and four superlayers of stereo wires.  The axial superlayers
have 12 radially separated layers of sense wires, parallel to
the $z$-axis, that measure the $r-\phi_{\rm lab}$ position of the tracks.
The stereo superlayers have six layers of sense wires with 
alternate $\sim \pm 3^{\circ}$ stereo angles with respect to the beamline,
and measure a combination 
of $r-\phi_{\rm lab}$ and $z$ information.  The stereo and
axial data are combined to reconstruct the 3-dimensional track.
The CTC covers the pseudorapidity interval $|\eta_{\rm lab}|<1.0$ and
transverse momentum $p_T \geq 0.4$ GeV \cite{units}.
The combined momentum resolution of the tracking system
is $\delta p_T/p_T = \sqrt{(0.0009 p_T)^2+(0.0066)^2}$, where
$p_T$ is the transverse momentum in GeV.

The solenoid is surrounded by sampling calorimeters used
to measure the electromagnetic and hadronic energy of 
electrons, photons, and jets.  The calorimeters cover the
pseudorapidity range $|\eta_{\rm lab}|\leq 4.2$ and the azimuthal
angle range $0 \leq \phi_{\rm lab} \leq 2\pi$.  They are segmented
in $\eta_{\rm lab}-\phi_{\rm lab}$ towers pointing to the nominal interaction
point at the center of the detector.  The tower granularity
is ($\Delta \eta_{\rm lab} \times \Delta \phi_{\rm lab})=(0.1 \times 15^{\circ}$) in the 
central region ($0 \leq |\eta_{\rm lab}| \leq 1.1$)
and ($0.1 \times 5^{\circ}$) in the 
plug ($1.1 < |\eta_{\rm lab}| \leq 2.4$)
and forward ($2.4 < |\eta_{\rm lab}| \leq 4.2$)
regions.  Each region has an electromagnetic
calorimeter (CEM in the central region, PEM in the plug region,
and FEM in the forward region) followed by a hadron calorimeter
at larger radius from the beam (CHA, PHA, and FHA respectively).
The central calorimeters are segmented in 24 wedges per each 
half of the detector ($-1.1\leq \eta_{\rm lab} \leq 0$ and $0\leq \eta_{\rm lab} \leq 1.1$).
The CEM is an 18 radiation length lead-scintillator stack
with a position resolution of 2 mm and an energy resolution
of $\delta E_T / E_T = \sqrt{(13.5\% / \sqrt{E_T})^2+(2\%)^2}$,
where $E_T$ is the transverse energy in GeV.  Located 
six radiation lengths deep inside the CEM calorimeter
(184 cm from the beamline), proportional wire chambers
(CES) with additional cathode strip read-out provide shower
position measurements in the $z$ and $r-\phi_{\rm lab}$  directions.
The central hadron calorimeter (CHA) is an iron-scintillator
stack which is 4.5 interaction lengths thick and provides energy 
measurement with a resolution of 
$\delta E_T/E_T = \sqrt{(50\%/\sqrt{E_T})^2+(3\%)^2}$,
where $E_T$ is the transverse energy in GeV.

The central muon system consists of three components and is capable
of detecting muons with transverse momentum $p_T \geq 1.4$ GeV
and pseudorapidity $|\eta_{\rm lab}|<1.0$.  The Central Muon Chambers (CMU)
cover the region $|\eta_{\rm lab}|<0.6$ and consist of four layers of planar 
drift chambers outside the hadron calorimeter, allowing the reconstruction
of the muons which typically pass the five absorption lengths of material.
Outside the CMU there are three additional absorption lengths of material
(0.6 m of steel) followed by four layers of drift chambers, the Central
Muon Upgrade (CMP).  The CMP chambers cover the same pseudorapidity
region as the CMU, and they were introduced to limit the background
caused from punch-through pions.  Finally, the Central Muon Extension
chambers (CMX) cover the region $0.6 \leq \eta_{\rm lab} \leq 1.0$. These drift
chambers are sandwiched between scintillators (CSX).  Depending
on the incident angle, particles have to penetrate six to nine absorption
lengths of material to be detected in the CMX.  The particle candidate stub
provided by the muon system is matched with a track from the CTC 
in order to successfully reconstruct a muon.

\subsection{The CDF triggers}

CDF has a three-level trigger system designed to 
select events that can contain electrons, muons, 
jets, and $\met$.  The first
two levels are implemented in hardware, while the
third is a software trigger which uses a version of 
the offline reconstruction software optimized for
speed and implemented by a CPU farm.

At level-1, electrons were selected by the presence
of an electromagnetic trigger tower with energy above 
6 GeV (Run Ia) or 8 GeV (Run Ib), where 
one trigger tower consisted of two adjacent
physical towers (in pseudorapidity).  Muons were selected by the presence
of a track stub in the CMU or CMX, where there was also
signal in the CMP.

At level-2, electrons satisfied one of several 
triggers.  In Run Ia, the event passed the trigger if the energy 
cluster in the CEM was at least 9 GeV with a seed tower of 
at least 7 GeV, and a matching track with $p_T>9.2$ GeV was found
by the Central Fast Tracker (CFT), the fast hardware processor that 
matched CTC tracks in the $r-\phi_{\rm lab}$ plane with signals in the 
calorimeters and muon chambers.  It also passed the trigger if there
was an isolated cluster in the CEM calorimeter of at least
16 GeV.  The most common Run Ib level-2 electron trigger
requires the existence of a cluster in the CEM with at least 16 GeV 
and the existence of a matching track in the CFT
with $p_T>12$ GeV. 
The muon trigger at level-2 required a track of at least 9 GeV (Run Ia)
or 12 GeV (Run Ib) that 
matched a CMX stub (CMX triggers), both CMU and CMP stubs (CMUP triggers),
or a CMU stub but no CMP stub (CMNP triggers).

At level-3, reconstruction programs performed 3-dimensional 
track reconstruction.  
In the Run Ia level-3 electron trigger, most of the accepted events passed
the requirement that the CEM cluster
had $E_T > 18$ GeV, and was associated with 
a track of $p_T > 13$ GeV.  The transverse energy
of the cluster is defined as $E_T=E\sin\theta$,
where $E$ is the total energy deposited in the
CEM, and $\theta$ is the polar angle
measured from the event vertex to the centroid of the cluster.
Cuts were applied on the shape of the electron shower profile and 
the energy deposition patterns.
In the Run Ib level-3 electron trigger, 
CEM $E_T > 18$ GeV and CFT $p_T > 13$ GeV
requirements were applied.
The muon trigger at level-3 required that the CFT transverse momentum
was greater than 18 GeV,
the energy deposited in the hadron calorimeter was less than 6 GeV, 
the energy deposited in the electromagnetic calorimeter was less than 2 GeV,
and the extrapolated CTC track was no more than 2 centimeters away from the
muon stub in the CMU chambers and 5 centimeters in the CMP or CMX chambers 
in the $x$ direction.  Events that pass the level-3 trigger were recorded 
to tape for offline analysis.

\subsection{The datasets}

The events passing the three levels of our trigger system constitute
the inclusive high-$p_T$ electron and muon data samples.  We apply kinematic
and lepton identification cuts, described in Sections \ref{ds1} and \ref{ds2} 
to obtain the inclusive $W$ electron and muon datasets respectively.  Using these 
datasets we arrive at the $W$+jet datasets by applying the jet selection 
cuts described in Section \ref{ds3}.

\subsubsection{Inclusive $W$ Electron Selection\label{ds1}}

After passing the three levels of trigger requirements, the following event
selection cuts are applied to the inclusive electron data sample:

$\bullet$ The event must belong to a good run.

$\bullet$ $E_T^{e}\geq$ 20 GeV, \\
where $E_T^{e}$ is the transverse energy of the CEM cluster, corrected for
differences in response, non-linearities, and time-dependent changes.

$\bullet$ $|\eta_{\rm lab}^e|\leq 1$, \\
where $\eta_{\rm lab}^e$ is the pseudorapidity of the electron. 

$\bullet$ The electron must fall in a fiducial
part of the CEM calorimeter.

$\bullet$ ISO(0.4)$\equiv E_{\Delta R=0.4}^{\rm Excess}/E_T^{\rm cluster} <0.1$,\\ 
where $E_{\Delta R=0.4}^{\rm Excess}$ is the excess transverse energy
in a cone of size $\Delta R=\sqrt{(\Delta \phi_{\rm lab})^2+(\Delta \eta_{\rm lab})^2}$ 
centered on the direction of the electromagnetic cluster, and 
$E_T^{\rm cluster}$ is the transverse energy of that cluster.

$\bullet$ $E^{\rm HAD}/E^{\rm EM} < 0.055+0.00045 E^{e}$, \\
where $E^{\rm HAD}$ 
is the energy deposited in the hadron calorimeter, and $E^{\rm EM}$ is 
the energy deposited in the electromagnetic calorimeter.

$\bullet$ $L_{\rm SHR}\equiv 0.14\sum_i\frac{E_i^{\rm meas}-E_i^{\rm exp}}
{\sqrt{(0.14)^2E^{\rm meas}+(\Delta{E_i^{\rm exp}})^2}}<0.2$,\\
where $L_{\rm SHR}$ is the {\it lateral shower profile},
$E_i^{\rm meas}$ is the energy measured in the 
$i^{\rm{th}}$-tower adjacent to the seed tower, $E_i^{\rm exp}$
is the expectation for the energy in that tower,
$\Delta{E_i^{\rm exp}}$ is the uncertainty on the expected 
energy, and $0.14\sqrt{E^{\rm meas}}$
is the uncertainty in the measurement of the cluster energy.

$\bullet$ $\chi^2_{\rm CES}<10$. \\
We measure the shower profile along the z direction
using the CES strips and the shower profile along the
x direction using the CES wires.  By comparing the measured
x-shape and z-shape to the ones determined from test-beam studies
we extract the chi-squared quantities for the two directions.
The chi-squared we use is the average of the two.

$\bullet$ $0.5 \leq E^{\rm e}/p^{\rm e} \leq 2.0$,\\
where $E^{\rm e}$ is the corrected energy of the electron, and $p^{\rm e}$ is
the beam-constrained momentum of the electron, i.e., 
the momentum determined when the fit trajectory of the CTC hits is 
constrained to pass through the beam line.

$\bullet$ $|\Delta X|<1.5$ cm and $|\Delta Z|<3.0$ cm,\\
where $\Delta X$ and $\Delta Z$ are the difference in the $x$ and $z$ directions
respectively, between the extrapolated CTC track and the CES position of the shower.

$\bullet$ $|Z_{\rm VTX}| \leq 60$ cm,\\
where $Z_{\rm VTX}$ is the $z$ position of the primary vertex.

$\bullet$ Photon conversions are removed.

We next apply the following cuts:

$\bullet$ $\met > 20$ GeV,\\
where $\met$ is the missing transverse energy in the event, calculated
from the energy imbalance in the calorimeters, with a correction for the unclustered
energy -- calorimeter energy not taken into account by the jet clustering algorithm 
-- and possible presence of muons.

$\bullet$ $M_T^W > 40$ GeV,\\
where $M_T^W$ is the $W$ transverse mass.  This cut removes the background
from $W$ bosons decaying into tau leptons which subsequently decay into electrons.

$\bullet$ The event must not be consistent with a $Z$ decaying into
two observed leptons, or a $Z$ in which one of the decay tracks has not been identified.

The 73363 events passing these cuts constitute our inclusive $W$ electron
data sample (Run Ia: 13290 events and Run Ib: 60073 events), corresponding 
to an integrated luminosity of 110 pb$^{-1}$ (Run Ia: $19.65 \pm 0.71$ pb$^{-1}$ 
and Run Ib: $90.35 \pm 3.70$ pb$^{-1}$).

\subsubsection{Inclusive $W$ Muon Selection \label{ds2}}

After passing the three levels of trigger requirements, the following event 
selection cuts are applied to the inclusive muon data sample:

$\bullet$  The event must belong to a good run.

$\bullet$ $p_T^{\mu}\geq$ 20 GeV, \\
where $p_T^{\mu}$ is the beam-constrained transverse momentum of the muon
(determined by a fit to the CTC hits, constrained by the beam line).

$\bullet$ The muon must be fiducial and central (pseudorapidity
$|\eta_{\rm lab}^{\mu}| \leq 1$).

$\bullet$ ISO(0.4)$\equiv E_{\Delta R=0.4}^{\rm Excess}/p_T^{\mu} <0.1$,
where $E_{\Delta R=0.4}^{\rm Excess}$ is the excess transverse energy
in a cone of size $\Delta R=\sqrt{(\Delta \phi_{\rm lab})^2+(\Delta \eta_{\rm lab})^2}$ 
centered on the direction of the muon.

$\bullet$ $E_{\rm HAD} \leq 6$ GeV,\\
where $E_{\rm HAD}$ is the energy deposited in the hadron calorimeter tower traversed by the muon.

$\bullet$ $E_{\rm EM} \leq 2$ GeV,\\
where $E_{\rm EM}$ is the energy deposited in the electromagnetic calorimeter tower traversed by the muon.

$\bullet$ $|\Delta X_{\rm CMU}| < 2$ cm, $|\Delta X_{\rm CMP}| < 5$ cm, $|\Delta X_{\rm CMX}| < 5$ cm,\\
where $\Delta X_{\rm CMU}$, $\Delta X_{\rm CMP}$ and $\Delta X_{\rm CMX}$ are the differences
between the $x$ position of the stub in the muon chambers and the extrapolation of the CTC track to 
these muon chambers.

$\bullet$ $|Z_{\rm VTX}| \leq 60$ cm.

$\bullet$ The event must pass the cosmic ray filter.

$\bullet$ The impact parameter must be $|d_0|\leq 0.2$ cm.

$\bullet$ $|Z_0-Z_{\rm VTX}|\leq 5$ cm,\\
where $Z_0$ is the $z$-position of the muon track.
This cut, combined with the previous two, significantly 
reduces the cosmic muon background.

We next apply the following cuts, as in the electron case:

$\bullet$ $\met > 20$ GeV.

$\bullet$ $M_T^W > 40$ GeV.

$\bullet$ The event must not be consistent with a $Z$ decaying into
two observed leptons, or a $Z$ in which one of the decay tracks has not 
been identified.

The 38601 events passing these cuts constitute our inclusive $W$ muon
data sample [Run Ia (CMUP): 4441 events, Run Ia (CMNP): 955 events,
Run Ib (CMUP): 20527 events, Run Ib (CMNP): 3273 events, and Run Ib (CMX): 9405 events],
corresponding to an integrated luminosity of 107 pb$^{-1}$ 
[Run Ia (CMUP): $18.33 \pm 0.66$ pb$^{-1}$, Run Ia (CMNP): $19.22 \pm 0.69$ pb$^{-1}$,
Run Ib (CMUP): $88.35 \pm 3.62$ pb$^{-1}$, Run Ib (CMNP): $89.20 \pm 3.66$ pb$^{-1}$ and
Run Ib (CMX): $88.98 \pm 3.65$ pb$^{-1}$].

\subsubsection{Inclusive $W$+jet Event Selection \label{ds3}}

Our final analysis dataset consists of those $W$ events which include
at least one jet with $E_T^{\rm jet}>15$ GeV, 
$|\eta_{\rm lab}^{\rm jet}|<2.4$, and $\Delta R_{l-j}^{\rm lab}>0.7$, where 
$\Delta R_{l-j}^{\rm lab}\equiv\sqrt{(\Delta\eta_{l-j}^{\rm lab})^2+(\Delta\phi_{l-j}^{\rm lab})^2}$,
and $\Delta\eta_{l-j}^{\rm lab}$ and $\Delta\phi_{l-j}^{\rm lab}$ are the differences
in pseudorapidity and polar angle between the charged lepton and the 
jet in the laboratory frame.  The results of the analysis 
pertain to the $W^{-}$ boson.
All $W^{+}$ bosons in the sample are $CP$ transformed to be treated as $W^{-}$ bosons
\cite{footnote}.

These requirements leave
12676 electron $W$+jet events and 6941 muon $W$+jet events,
with $15 < p_T^W < 105$ GeV, where $p_T^W$ is the transverse
momentum of the $W$ boson, defined as the vector sum of the
$\met$ and charged lepton transverse momentum.
The data event yields for the four $p_T^W$ bins 
($15 < p_T^W < 25$ GeV, $25 < p_T^W < 35$ GeV, $35 < p_T^W < 65$ GeV, and $65 < p_T^W < 105$ GeV)
are presented in Table \ref{t1}.
\begin{table}
\center
\caption{\label{t1}The electron ($N_e$) and muon ($N_{\mu}$) CDF data event yields for inclusive $W$+jet production.  
The muon event yields are lower relative to the electron event yields, due to lower muon efficiencies and acceptances.}
\begin{tabular}{|c||c|c|}\hline \hline
\multicolumn{3}{|c|}{\bf Data event yields for inclusive \boldmath $W$+jet production}\\  \hline
\multicolumn{1}{|c||}{$p_T^W$ (GeV)} & 
\multicolumn{1}{c|}{\hspace{1.435cm}\mbox{ }$N_e$ \hspace{1.435cm}\mbox{ }} &
\multicolumn{1}{c|}{$N_{\mu}$}\\ \hline 
15--25 & 5166  & 2821 \\
25--35 & 3601  & 1869 \\
35--65 & 3285  & 1880 \\
65--105 & 624  & 371  \\ \hline \hline

\end{tabular}
\end{table}

The actual number of events is not of critical importance for us, because  
we are interested in the shape of the distributions and not the absolute event yields.
We thus analyze the distributions normalized to unity.  We will come back
to the actual event yields after the inclusion of the background (Section \ref{sec:5}) and 
systematic uncertainties (Section \ref{sec:sys}). 

\section{\label{sec:3}The Monte Carlo simulation}

\subsection{The DYRAD Monte Carlo event generator}

DYRAD \cite{giele} is the next-to-leading order $W$+jet 
event generator used to
establish the SM prediction.  We include the ``1-loop'' processes,
since these affect the next-to-leading order 
angular coefficients and more completely
simulate the events we study.  This generator is
of order $\alpha_s^2$ in QCD, generating up to two jets passing the
minimal requirement of $E_T^{\rm jet}>10$ GeV if the Feynman diagram does not
contain any gluon loops, and generates up to one jet with the same requirement 
if a gluon loop is present in the Feynman diagram.  
As a result, DYRAD does not appropriately model events with more than two jets.  
These extra jets in the data occupy low and
high values of the azimuthal angle $\phi$ in our CS frame.  
We are careful not to bias our measurement
due to this effect (see Section \ref{mac}).
The jet transverse energy cut of 10 GeV is required because the theoretical calculations
are unreliable for small jet transverse energies due to 
infrared and collinear divergencies.  
A jet-jet angular separation cut of greater than 0.7 in 
$\eta_{\rm lab}$-$\phi_{\rm lab}$ space is imposed, which is important for 
the definition of a jet.  
No additional kinematic cuts for the jet,
charged lepton, and neutrino are required in order to obtain a reliable 
theoretical prediction of the angular distribution of the $W$.  The cross section
for inclusive $W$+jet production calculated up to order $\alpha_s^2$ is 722.51 $\pm$
3.89 pb for $W^-$ and $W^+$ bosons combined.  
This simulation uses $Q^2=(M_W^{\rm pole})^2$, 
where $M_W^{\rm pole}=80.3$ GeV is the pole mass of the $W$,
CTEQ4M($\Lambda=0.3$ GeV) parton distribution 
functions \cite{cteq4m}, and 0.7-cone jets in $\eta_{\rm lab}$-$\phi_{\rm lab}$ space.

\begin{figure}
\includegraphics[scale=.46]{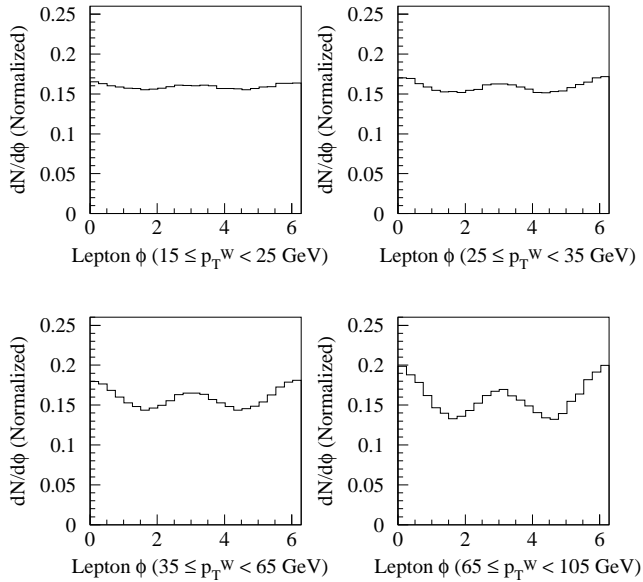}
\caption{The theoretical charged lepton $\phi$ distribution in the Collins-Soper $W$ 
rest-frame for the four $p_T^W$ regions, as generated by DYRAD.  The distributions are normalized to unity.}
\label{fextra1}
\end{figure}

In order to obtain smooth SM kinematic distributions up to $p_T^W=100$ GeV, 
and especially smooth $\cos\theta$ vs. $\phi$ distributions for different $p_T^W$ regions, 
we generated a large sample of DYRAD events ($\sim$250 M).  This Monte Carlo event sample size
was required since events with negative weights, corresponding to the gluon loop matrix elements, 
produce significant fluctuations in the kinematic distributions with limited statistics.
The DYRAD simulation allows us to establish the SM prediction
for the $\phi$ distribution of the charged lepton and the
predictions for the angular coefficients and helicity cross
sections of the $W$ up to order $\alpha_s^2$ \cite{js}.  
The expected $\phi$ distributions for four $p_T^W$
bins are shown in Figure \ref{fextra1}.
For zero $p_T^W$ we expect a flat 
distribution, whereas the QCD effects at higher $p_T^W$ result in
two minima.
In order to simulate the detector response,
we pass the generator events through the fast Monte Carlo
detector simulator, described in the next section.

\subsection{The fast Monte Carlo detector simulation}

The fast Monte Carlo (FMC) CDF detector simulation
includes the detailed geometry of the detector, geometrical and kinematic 
acceptances of all subdetectors, detector resolution effects parameterized
using gaussians obtained explicitly from data, detailed
magnetic field map, and multiple Coulomb scattering effects.
The integrated luminosities, lepton identification and trigger efficiencies, 
and all experimental cuts imposed on the $W$, leptons, $\met$, and jets are
incorporated.  The effect of the underlying event, caused by
interacting spectator quarks, is also included.
The FMC program receives
the particle four-momenta for each generated DYRAD event
along with the next-to-leading order cross section prediction
from DYRAD (which includes gluon loop effects)
and produces kinematic distributions smeared
by detector resolution, and sculpted by geometrical 
and kinematic acceptances and efficiencies.  The FMC 
also reports event yield predictions.  The FMC successfully
reproduces the kinematic features of inclusive $W$ and $Z$ boson production,
as well as the features of vector boson production in association 
with a jet \cite{mythesis,len}.

For the $W$+jet data, we additionally require at least one ``good'' jet
($E_T^{\rm jet}>15$ GeV and $|\eta_{\rm lab}^{\rm jet}|<2.4$)
that also passes the $\Delta R_{l-j}^{\rm lab}>0.7$ cut, where $\Delta R_{l-j}^{\rm lab}$ is the
opening angle in $\eta_{\rm lab}$-$\phi_{\rm lab}$ space between the lepton 
and the leading ``good'' jet.
The FMC event yields for inclusive $W$+jet production up to order $\alpha_s^2$
are presented in Table \ref{t2}.
The Parton Distribution Functions (PDF) systematics
and the renormalization and factorization scale ($Q^2$) systematics will be included in Section \ref{twra}.
\begin{figure}
\includegraphics[scale=.46]{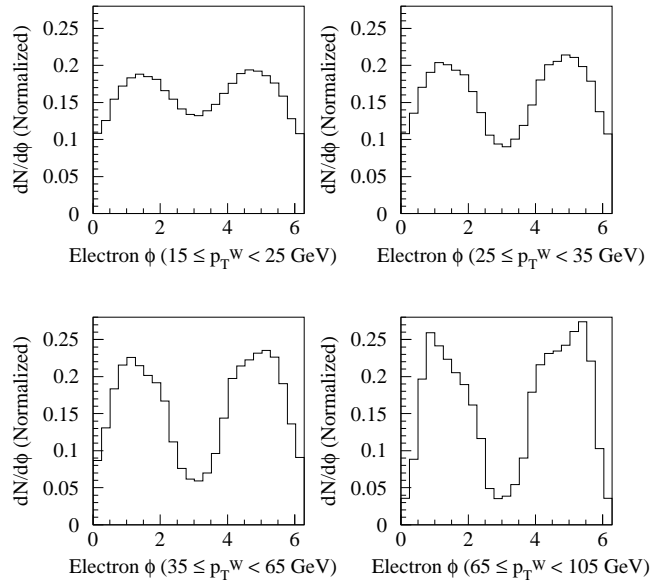}
\caption{The expected CDF electron $\phi$ distribution in the Collins-Soper $W$ rest-frame for the four $p_T^W$ regions, 
after experimental cuts and detector smearing, as generated by the FMC.  The distributions are normalized to unity. The muon distributions are almost identical.}
\label{fextra2}
\end{figure}
\begin{table}
\center
\caption{\label{t2}The electron ($N_e$) and muon ($N_{\mu}$) FMC event yields for inclusive $W$+jet production 
up to order $\alpha_s^2$.  Contributions from backgrounds are not included.  
The uncertainties in the event yield predictions are dominated by the uncertainties associated with
integrated luminosities, lepton identification and trigger efficiencies and the DYRAD prediction of 
the $W$+jet production cross section. 
Systematic uncertainties associated with PDF choice and $Q^2$ scale variation are not included.}
\begin{tabular}{|c||c|c|}\hline \hline
\multicolumn{3}{|c|}{\bf FMC event yields for inclusive \boldmath $W$+jet production}\\  \hline
\multicolumn{1}{|c||}{$p_T^W$ (GeV)} & 
\multicolumn{1}{c|}{\hspace{1.5cm}\mbox{ }$N_e$\hspace{1.5cm}\mbox{ }} &
\multicolumn{1}{c|}{$N_{\mu}$}\\ \hline 
15--25 & 3867 {$\pm$ 137} & 2027 {$\pm$ 102} \\
25--35 & 2632 {$\pm$ 93} &  1384 {$\pm$ 66} \\
35--65 & 2474 {$\pm$ 87} &  1314 {$\pm$ 67} \\
65--105 & 518 {$\pm$ 18} &  279 {$\pm$ 14} \\ \hline \hline
\end{tabular}
\end{table}

The FMC detector simulation, along with DYRAD, shows
how the acceptances and efficiencies of the detector 
and the analysis cuts affect
the $\phi$ distributions that are experimentally observed.  
Figure \ref{fextra2} shows the expected measurement of the $\phi$ distributions for the
electron dataset (the muon distributions are almost identical) for
the four $p_T^W$ bins.
The effects of the 
acceptances and efficiencies are significant; instead of two 
minima we observe two maxima.  The main reason for this is
the charged lepton and neutrino 
$p_T$ cuts, which limit the allowed $(\cos\theta,\phi)$ phase 
space considerably.
The FMC plots are normalized to the FMC signal event yields,
and all experimental cuts have been applied.

\section{\label{sec:4}Acceptances and Efficiencies}

The lepton identification and trigger efficiencies are measured by
using the leptons from CDF Run Ia and Ib $Z$ data and by studying
random cone distributions of leptonic $W$ and $Z$ decay
Run Ia and Ib data samples.  
The kinematic and geometrical acceptances
are calculated using the DYRAD event generator, 
which produces the SM prediction,
and the FMC detector simulation, which 
produces the CDF experimental expectation.  

\begin{figure}[!]
\includegraphics[scale=.46]{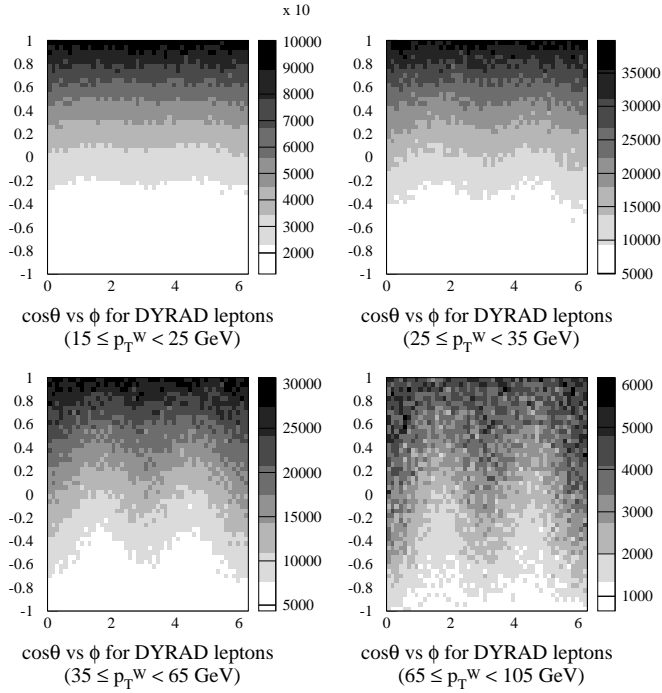}
\caption{The $\cos \theta$ vs. $\phi$ phase space for the
four $p_T^W$ bins, for the DYRAD generator simulation (arbitrary units).}
\label{f1}
\end{figure}
\begin{figure}[!]
\includegraphics[scale=.46]{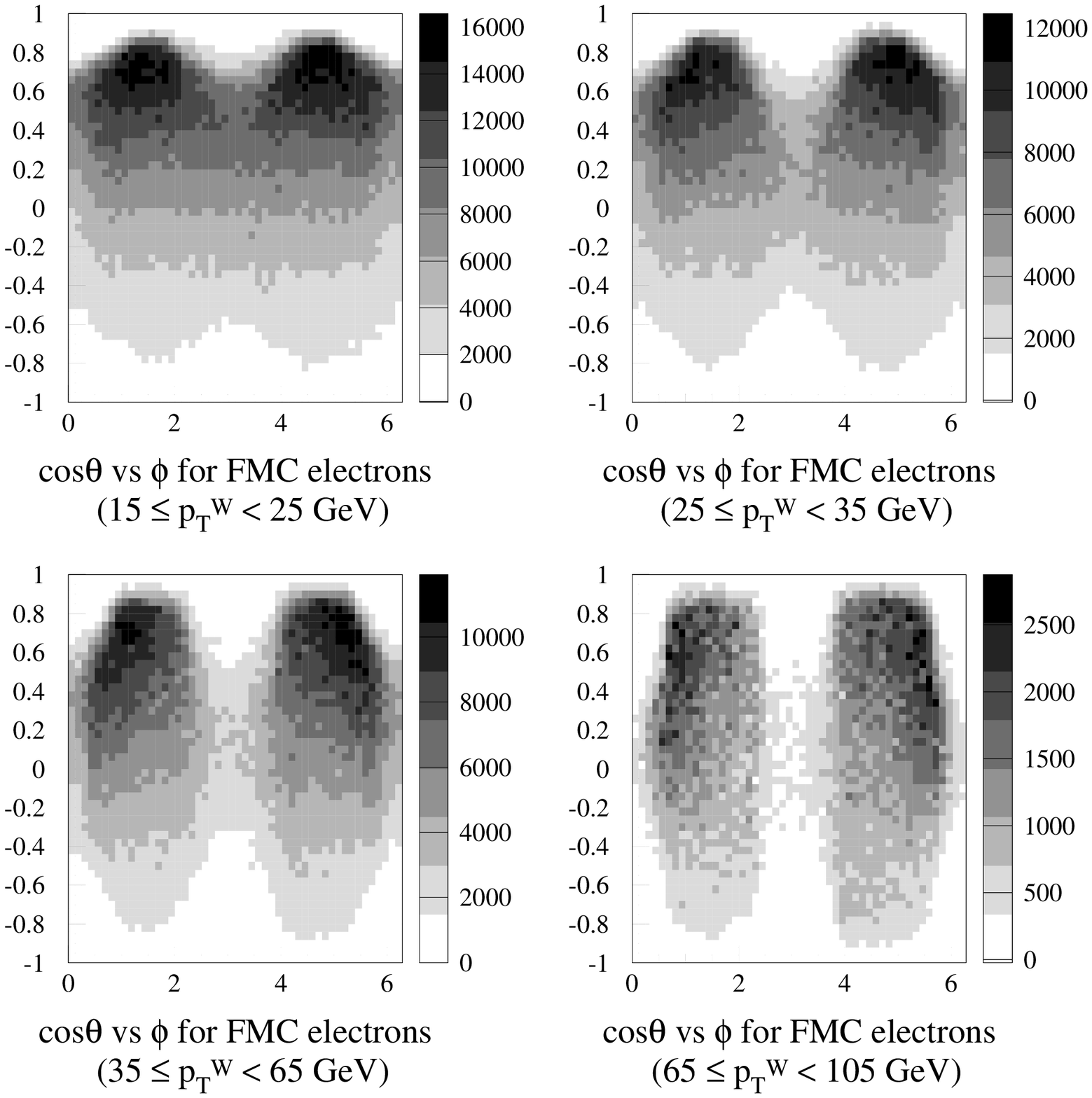}
\caption{The $\cos \theta$ vs. $\phi$ phase space for the
four $p_T^W$ bins, for the electron FMC signal simulation (arbitrary units).}
\label{f2}
\end{figure}	
\begin{figure}[!]
\includegraphics[scale=.46]{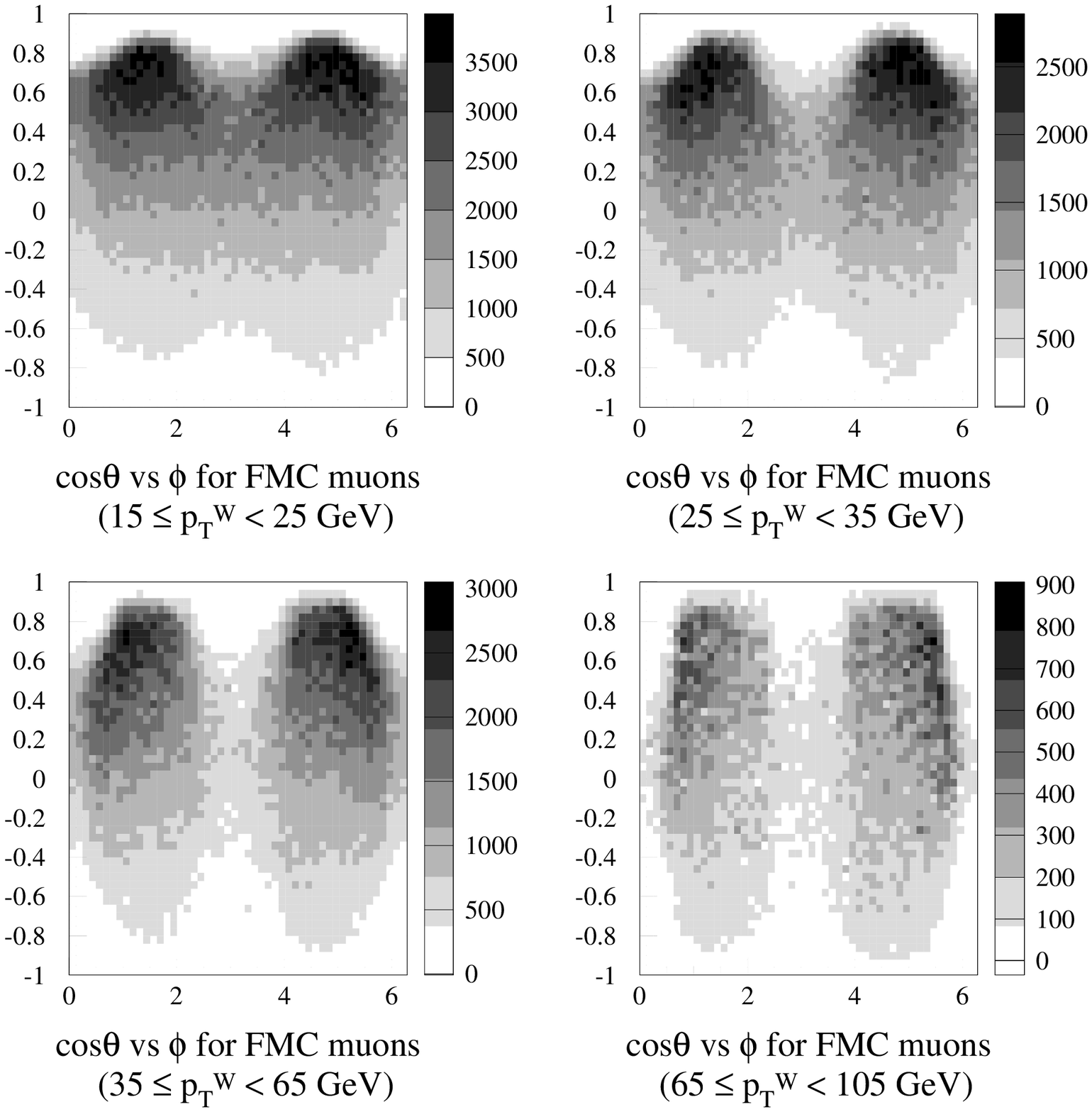}
\caption{The $\cos \theta$ vs. $\phi$ phase space for the
four $p_T^W$ bins, for the muon FMC signal simulation (arbitrary units).}
\label{f3}
\end{figure}	
\begin{figure}[!]
\includegraphics[scale=.46]{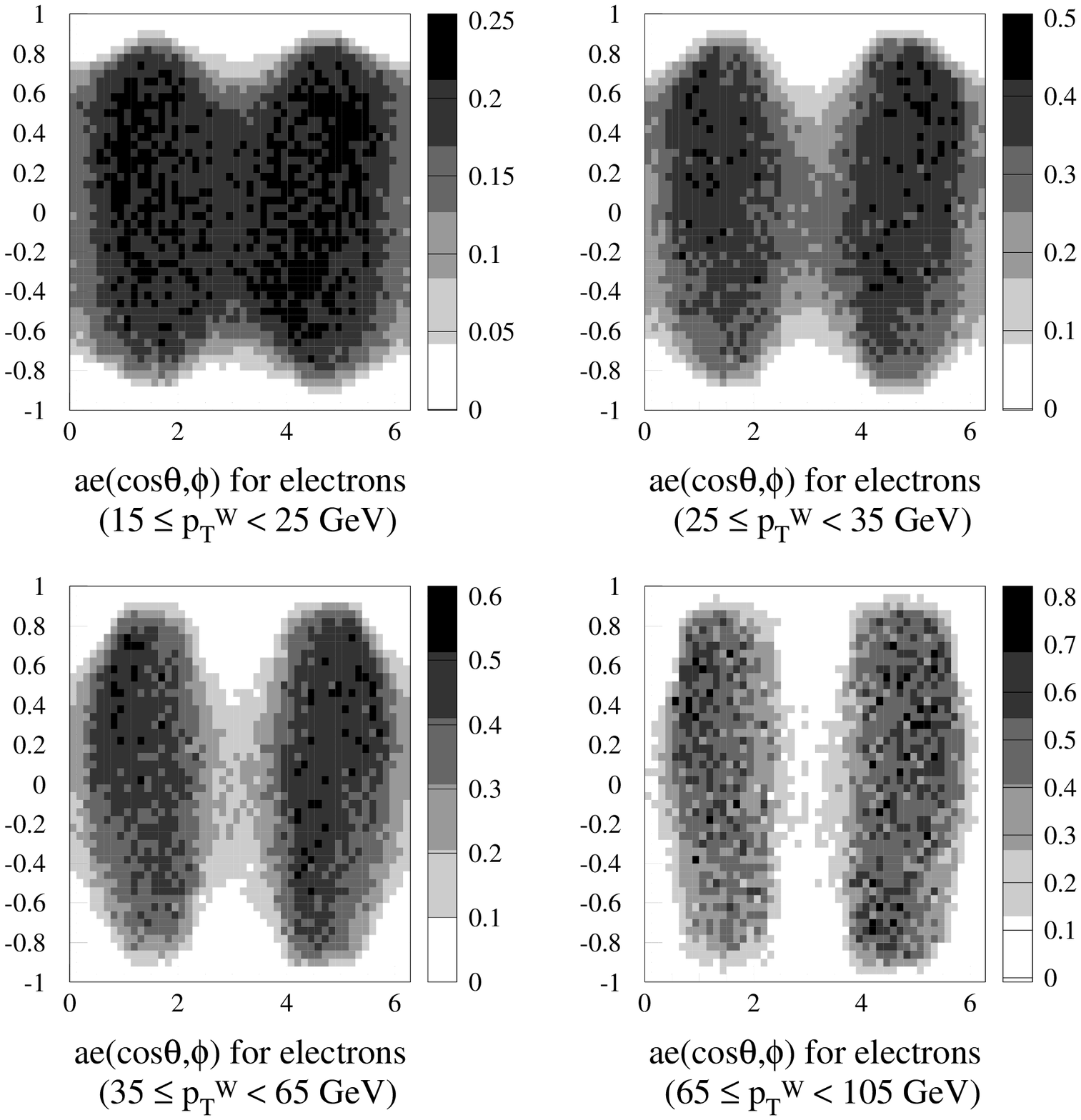}
\caption{Acceptance times efficiency for the electrons
as a function of $\cos \theta$ and $\phi$ in the Collins-Soper frame.}
\label{f4}
\end{figure}
\begin{figure}[]
\includegraphics[scale=.46]{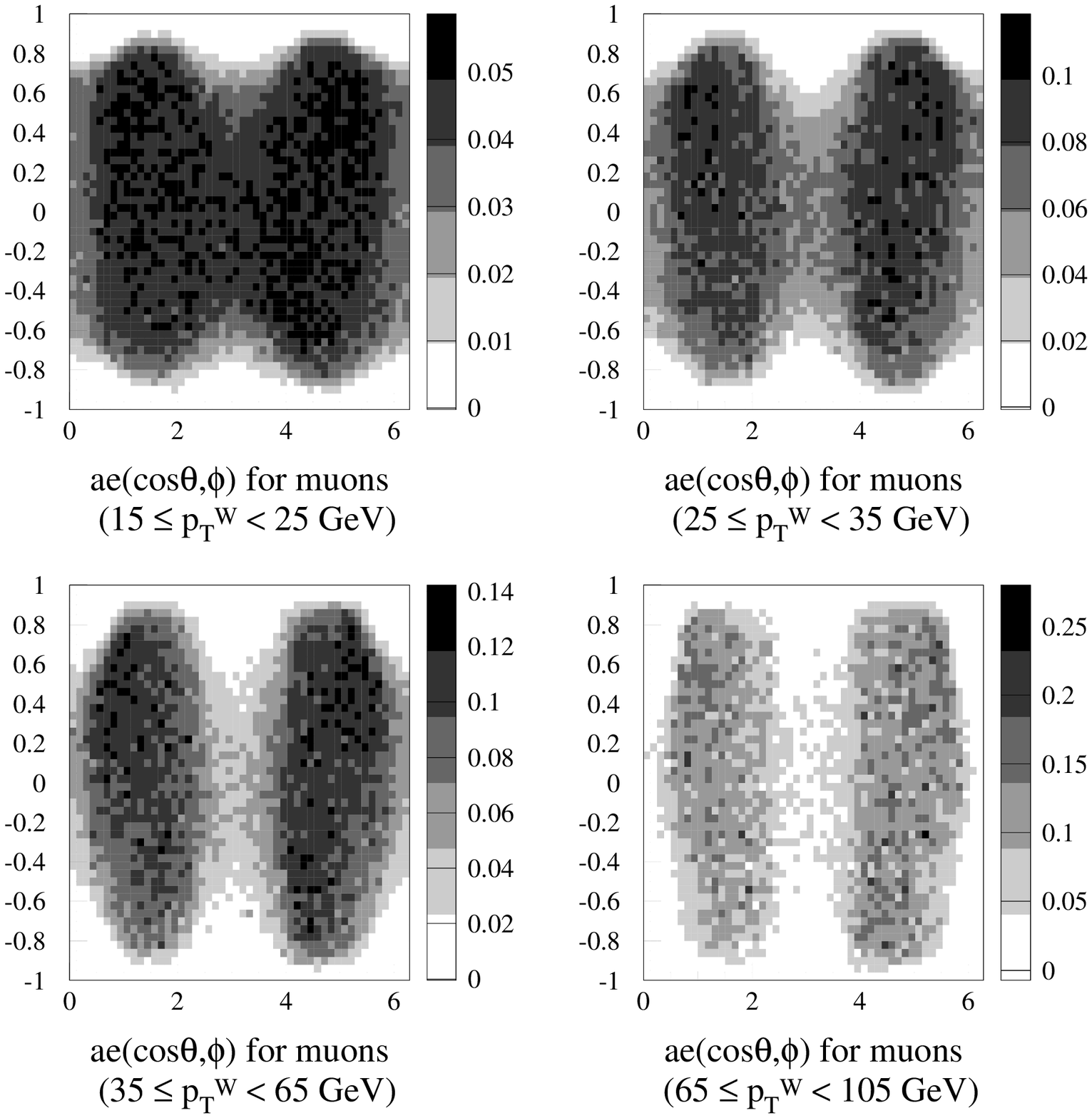}
\caption{Acceptance times efficiency for the muons
as a function of $\cos \theta$ and $\phi$ in the Collins-Soper frame.}
\label{f5}
\end{figure}
We are especially interested in the product of overall acceptance times efficiency ($ae$)
as a function of ($\cos\theta$,$\phi$) associated with each of the four $p_T^W$ bins
We create 2-dimensional histograms of $\cos\theta$ vs. $\phi$
for each of the four $p_T^W$ bins, using the DYRAD simulation.  This procedure is repeated 
after the events pass the FMC simulation, where the appropriate mixture of Run Ia and Run Ib $W$ leptons 
is used, based on FMC event yield predictions for all subdetectors.

The resulting plots
are shown in Figures \ref{f1}, \ref{f2}, and \ref{f3},
for DYRAD, FMC-electrons, and FMC-muons respectively.
We subsequently divide the FMC 
2-dimensional histograms by the corresponding DYRAD ones, producing
the 2-dimensional differential acceptance times efficiency 
$ae(\cos\theta,\phi)$ of Figures \ref{f4} and 
\ref{f5}, for electrons and muons respectively.
The overall acceptance times efficiency is
higher for the electrons.
These $ae(\cos\theta,\phi)$ values are used for the $\cos\theta$-integration
of the cross section, as described in Section \ref{direct}.

\section{\label{sec:5}Background estimation}

The main sources of background in the ($W\rightarrow e\nu_e$)+jet and
($W\rightarrow \mu \nu_{\mu}$)+jet processes are $Z$+jets events where
the $Z$ is misidentified as a $W$ (``one-legged'' $Z$), ($W\rightarrow \tau \nu_{\tau}$)+jet events,
and QCD background resulting from jet misidentification.
The $W+\gamma$ background is only a few events with run and event numbers the same
as the $W+\gamma$ events of \cite{len}; it is treated as a systematic uncertainty, 
which also gives an indication of the radiative effects in the measurement.  
This uncertainty is very small (see Section \ref{sec:sys}).

A small background contribution arises from $t\bar{t}$ production,
where one of the produced $W$ bosons decays leptonically and the 
other $W$ boson decays hadronically to jets.  This background
is estimated to be $30 \pm 7$ events for the electron sample
\cite{dittmann} and $16 \pm 3$ events for the muon sample, a 0.3\% effect.  
An equally small background is the $Z\rightarrow \tau^+ \tau^-$ production, where one of the
tau leptons decays hadronically and the other one leptonically.
This background is estimated to be $47 \pm 1$ events in the electron sample \cite{dittmann},
and $25 \pm 1$ events in the muon sample, a 0.5\% effect.  
To demonstrate the insignificance of the $t\bar{t}$ and $Z\rightarrow \tau^+ \tau^-$
backgrounds, we perform our analysis including the charged lepton $\phi$ distribution
for these background events, in several possible shapes, for the four $p_T^W$ bins.
The resulting change in the extracted values of the angular coefficients is
negligible compared to our systematic and statistical uncertainties.
Thus, we ignore the backgrounds associated with $t\bar{t}$ production 
and $Z\rightarrow \tau^+ \tau^-$ decays.

Finally, the cosmic ray background in the muon $W$+jet datasets is estimated
to be significantly less than 0.1\%, and is therefore neglected.

\subsection{\boldmath One-legged $Z$ background}

To study this background we generate a DYRAD sample of $Z$+jet events and
pass it through the FMC Monte Carlo simulation and the subsequent analysis
program.  This predicts how many $Z$ bosons are misidentified as $W$ bosons.
In these cases, the $Z$ bosons satisfy all kinematic and lepton identification
cuts for $W$ bosons, but one of their decay leptons, or {\it legs}, is undetected.
The DYRAD cross section for $Z$+jet up to order $\alpha_s^2$ is 68.21 $\pm$
0.37 pb.  For this DYRAD simulation we used $Q^2=(M_Z^{\rm pole})^2=91.2$ GeV,
the CTEQ4M($\Lambda=0.3$ GeV) parton distribution functions, 
0.7-cone jets, jet-jet angular separation greater than 0.7 in $\eta_{\rm lab}$-$\phi_{\rm lab}$ space, 
and $E_T^{j}>10$ GeV.  
At the FMC level, we impose our usual $W$ boson event selection cuts and additionally require
at least one ``good'' jet ($E_T^{\rm jet}>15$ GeV and $|\eta_{\rm lab}^{\rm jet}|<2.4$)
that also passes the $\Delta R_{l-j}^{\rm lab}>0.7$ cut.  These results are summarized in 
Table \ref{t3}.  
Overall we expect $123 \pm 5$ electron one-legged-Z+jet events
and $337 \pm 18$ muon one-legged-Z+jet events passing the $W$+jet cuts,
without applying any cut on the W transverse momentum.
Comparing these numbers to the FMC event yields for $W$+jet,
the one-legged-Z+jet background is $(1.14 \pm 0.06)\%$ for the electron $W$+jet
and $(5.90 \pm 0.43)\%$ for the muon $W$+jet sample.  This background is higher
for the muon sample, because of the limited coverage of the muon chambers,
which is responsible for higher yields of one-legged muon $Z$ bosons.  

\begin{figure}
\includegraphics[scale=.46]{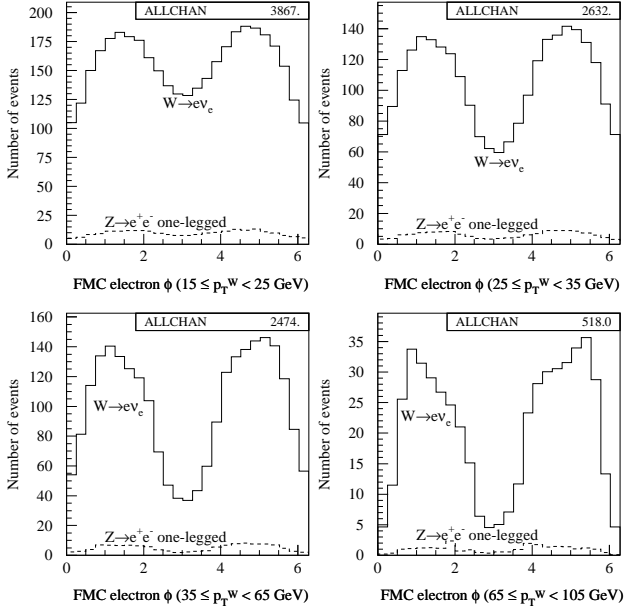}
\caption{Electron $\phi$ distributions for the four $p_T^W$ bins for
electron $W$+jet FMC events (solid histogram) and for
$Z$+jet FMC background (multiplied by 5), 
where one of the electrons from the $Z$ decay
is undetected and the other one passes the detection and 
analysis requirements (dashed histogram).  The histograms are normalized to the electron FMC signal event yields.}
\label{f6}
\end{figure}

\begin{figure}
\includegraphics[scale=.46]{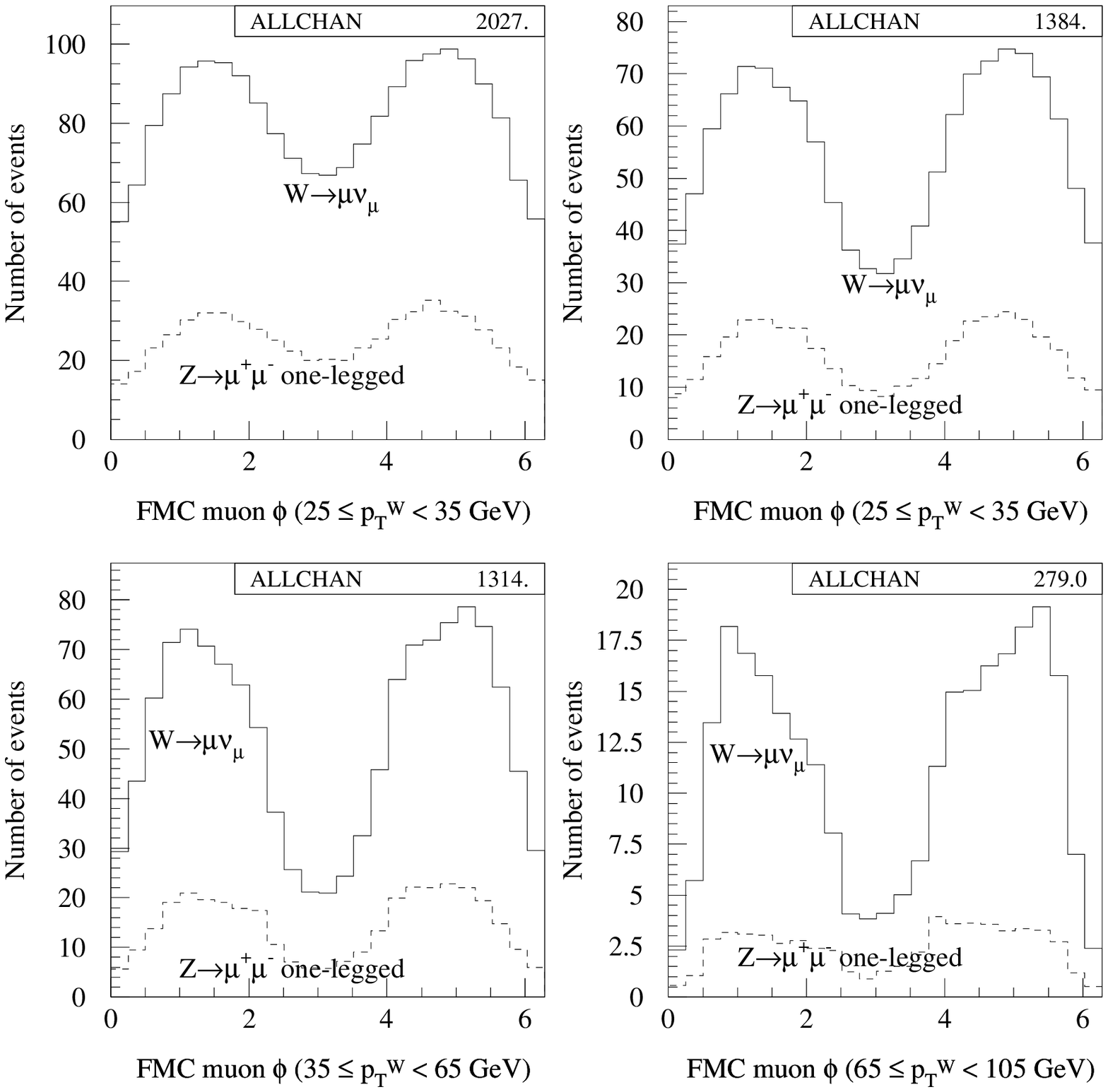}
\caption{Muon $\phi$ distributions for the four $p_T^W$ bins for
muon $W$+jet FMC events (solid histogram) and for
$Z$+jet FMC background (multiplied by 5), 
where one of the muons from the $Z$ decay
is undetected and the other one passes the detection and 
analysis requirements (dashed histogram).  The histograms are normalized to the muon FMC signal event yields.}
\label{f7}
\end{figure}

To examine how this background affects the $W$+jet lepton 
$\phi$ distribution, we
plot the $\phi$ distribution for the leptons from these processes
for the four $p_T^W$ bins (Figures \ref{f6} and \ref{f7}).  
We see that the same pattern of
two maxima at $\frac{\pi}{2}$ and $\frac{3\pi}{2}$ is present.
The background plots are normalized to the expected event yields from the FMC,
multiplied by a factor of five (to make them visible), 
and superimposed on the signal FMC distributions, normalized to the signal
FMC event yields.
We include the one-legged $Z$ FMC $\phi$ distribution 
in the complete theoretical prediction of the $\phi$ distributions,
in order to correctly extract the angular coefficients.
\begin{table}
\center
\caption{\label{t3}Monte Carlo background estimation of the number of electron and muon one-legged $Z$+jet
events.  The background fractions are calculated with respect to the FMC $W$+jet event yields.}
\begin{tabular}{|c||c|c||c|c|}\hline \hline
\multicolumn{5}{|c|}{\bf One-legged $Z$+jet background}\\  \hline
\multicolumn{1}{|c||}{$p_T^W$ (GeV)} & 
\multicolumn{1}{c|}{$N_e$} & 
\multicolumn{1}{c||}{Fraction} &
\multicolumn{1}{c|}{$N_{\mu}$} & 
\multicolumn{1}{c|}{Fraction} \\ \hline 
15--25 & 47 {$\pm$ 2} & 1.22 {$\pm$ 0.07} \% & 127 {$\pm$ 7} & 6.26 {$\pm$ 0.47} \% \\
25--35 & 30 {$\pm$ 1} & 1.14 {$\pm$ 0.05} \% &  82 {$\pm$ 4} & 5.92 {$\pm$ 0.40} \% \\
35--65 & 25 {$\pm$ 1} & 1.01 {$\pm$ 0.05} \% &  72 {$\pm$ 4} & 5.48 {$\pm$ 0.41} \% \\
65--105 & 5 {$\pm$ 0} & 0.96 {$\pm$ 0.03} \% &  12 {$\pm$ 1} & 4.30 {$\pm$ 0.42} \% \\ \hline \hline
\end{tabular}
\end{table}

\subsection{\boldmath ($W\rightarrow \tau \nu$)+jet background}

If the $W$ boson decays to a $\tau$ 
that subsequently decays leptonically,
the three final neutrinos
contribute to the $\met$, which is incorrectly associated
with a single neutrino.  The signal of one 
charged lepton along with the $\met$ mimics that of a 
$W$ directly decaying to the charged lepton. 
Most of the tau background is removed when we
utilize the fact that the charged lepton
and $\met$ coming from the $\tau$ decay
are soft.  As a result, the $W$ transverse mass in the $\tau$ events
is significantly smaller than that in the electron or muon events.
By applying the $p_T$ cuts for the leptons 
and the $W$ transverse mass cut,
we remove $92\%$ of the tau $W$+jet events 
at the DYRAD generator level.

To study the remaining tau background we start with a tau $W$+jet
DYRAD sample ($Q^2=(M_W^{\rm pole})^2$, 
CTEQ4M($\Lambda=0.3$ GeV) parton distribution function
and 0.7-cone jets in $\eta_{\rm lab}$-$\phi_{\rm lab}$ space), and we let the tau decay 
to an electron or a muon.  We then vector-sum the three neutrinos resulting
from the $W$ and tau decays to form a single $\met$.
Subsequently, we pass the events through the FMC detector
simulator to see how many events pass the $W$+jet cuts
after they are weighted by the detector acceptances
and efficiencies.  The branching ratios for the tau
decays we use are 17.83 \% for electrons and 17.37 \% for muons \cite{rpp}.
At the FMC level, we require at least one ``good'' jet
($E_T^{\rm jet}>15$ GeV and $|\eta_{\rm lab}^{\rm jet}|<2.4$)
that also passes the $\Delta R_{l-j}^{\rm lab}>0.7$ cut.
The tau background results are presented in Table \ref{t4}.   
Overall we expect $247 \pm 9$ tau electrons  
and $130 \pm 7$ tau muons to infiltrate the $W$+jet samples,
without applying any cut on the W transverse momentum.
Comparing these numbers to the FMC event yields for the electron
and muon $W$+jet samples,
the tau background is $(2.28 \pm 0.12) \%$ for the electron $W$+jet sample,
and $(2.28 \pm 0.17) \%$ for the muon $W$+jet sample. 

\begin{figure}[t]
\includegraphics[scale=.46]{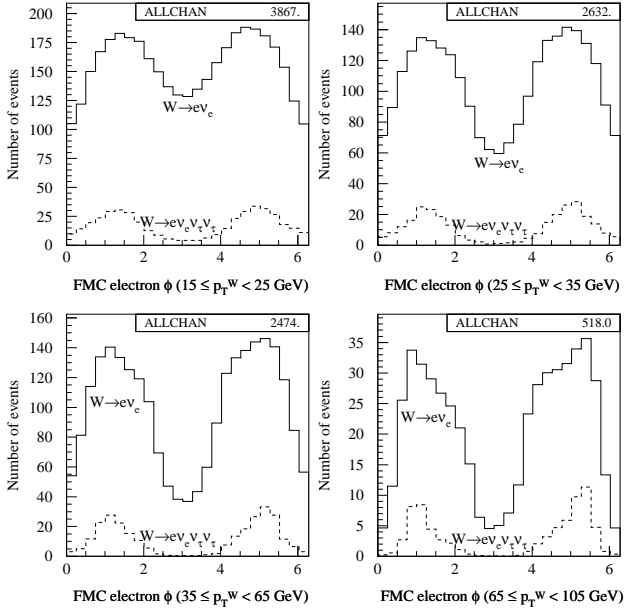}
\caption{Electron $\phi$ distributions for the four $p_T^W$ bins for
electron $W$+jet FMC events (solid histogram) and for
tau $W$+jet FMC background (multiplied by 5), 
where the tau decays to an electron
(dashed histogram).  The histograms are normalized to the FMC signal event yields.}
\label{f8}
\end{figure}

\begin{figure}[t]
\includegraphics[scale=.46]{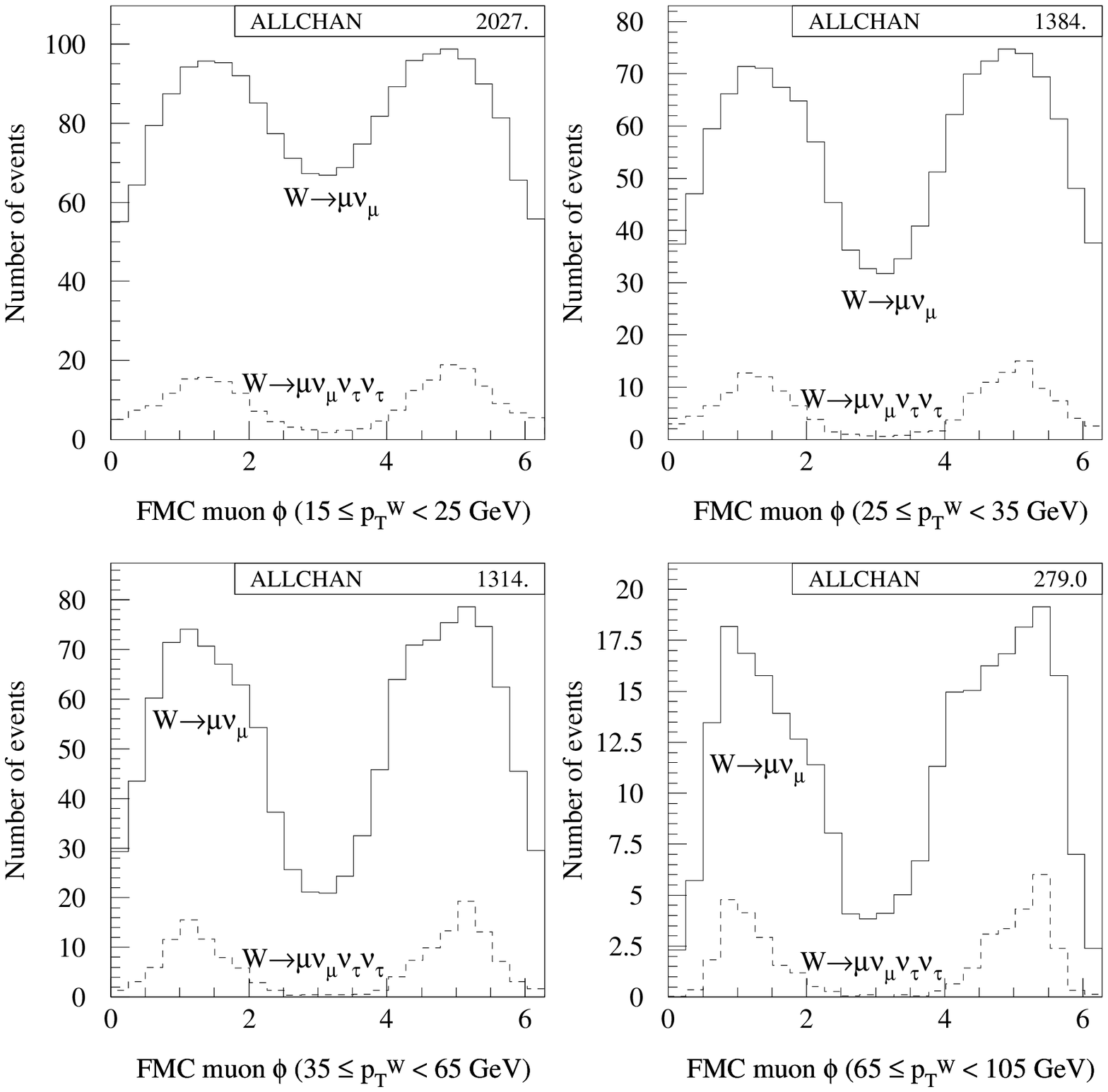}
\caption{Muon $\phi$ distributions for the four $p_T^W$ bins for
muon $W$+jet FMC events (solid histogram) and for
tau $W$+jet FMC background (multiplied by 5), 
where the tau decays to a muon
(dashed histogram).  The histograms are normalized to the FMC signal event yields.}
\label{f9}
\end{figure}

To see how this background affects the $W$+jet lepton $\phi$ distribution, we
plot the $\phi$ distribution for the leptons resulting from
leptonic tau decays in $W$+jets events for the four $p_T^W$ bins 
(Figures \ref{f8} and \ref{f9}).  
We see that the pattern of two maxima at 
$\frac{\pi}{2}$ and $\frac{3\pi}{2}$ is again present.
The background plots are normalized to the expected event yields from the FMC,
multiplied by a factor of five, 
and superimposed on the signal FMC distributions, normalized to the signal FMC
event yields.
We include the $\tau$-background FMC $\phi$ distribution 
in the complete theoretical prediction of the $\phi$ distributions,
in order to correctly extract the angular coefficients.
\begin{table}
\center
\caption{\label{t4}Monte Carlo background estimation of the number of electron and muon $W$+jet
events, where the $W$ decays to a tau and the electron or muon is the 
decay product of the tau.  The fractions of the 
backgrounds are calculated with respect to the FMC $W$+jet event yields.}
\begin{tabular}{|c||c|c||c|c|} \hline \hline
\multicolumn{5}{|c|}{\bf \mbox(\boldmath $W\rightarrow \tau \; \bar{\nu}_{\tau}\rightarrow  \bar{\nu}_{\tau} \; \nu_{\tau}\; \bar{\nu}_{e/\mu} \; e/\mu $)+jet background}\\  \hline
\multicolumn{1}{|c||}{$p_T^W$ (GeV)} & 
\multicolumn{1}{c|}{$N_e$} & 
\multicolumn{1}{c||}{Fraction} &
\multicolumn{1}{c|}{$N_{\mu}$} & 
\multicolumn{1}{c|}{Fraction} \\ \hline 
15--25 & 86 {$\pm$ 3} & 2.22 {$\pm$ 0.11} \% & 45 {$\pm$ 2} & 2.22 {$\pm$ 0.15} \% \\
25--35 & 57 {$\pm$ 2} & 2.16 {$\pm$ 0.10} \% &  30 {$\pm$ 2} & 2.17 {$\pm$ 0.18} \% \\
35--65 & 56 {$\pm$ 2} & 2.26 {$\pm$ 0.11} \% &  30 {$\pm$ 2} & 2.28 {$\pm$ 0.19} \% \\
65--105 & 15 {$\pm$ 1} & 2.89 {$\pm$ 0.22} \% &  8 {$\pm$ 0} & 2.87 {$\pm$ 0.14} \% \\ \hline \hline
\end{tabular}
\end{table}
\subsection{QCD background}

The QCD background in the case of inclusive $W$ production and decay consists
predominantly of dijet events, where one of the jets is misidentified as a lepton
and the other one is not detected, resulting in the creation of 
$\met$.  In the $W$+jet case, the QCD background is 
multijet events, where one of the jets is detected,
one is lost or mismeasured (resulting in $\met$) and 
one is misidentified as a charged lepton to erroneously reconstruct
a $W$.
The number and distribution of QCD background events in the 
four $p_T^W$ bins are determined from the Run Ia and Run Ib
CDF data.

\begin{figure}
\includegraphics[scale=.46]{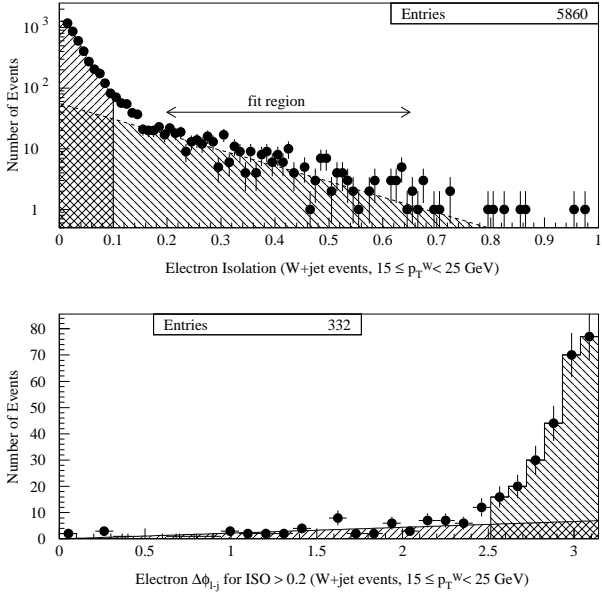}
\caption{Isolation of the electron in $W$+jet events (upper histogram) and the difference in the 
azimuthal angles of the electron and the jet for high isolation events (lower histogram).
These histograms are used to estimate the QCD background in the electron $W$+jet data,  for the
$15 \leq p_T^W \leq 25$ GeV bin (see text).  Corresponding histograms are used for the estimation 
of the QCD background in the three other electron $W$+jet $p_T^W$ bins.  
}
\label{f10}
\end{figure}

To measure the expected number of QCD background events in our data samples
we look at leptons with isolation (ISO), defined in Section \ref{sec:2},
greater than 0.2.
Our signal is in the ISO $<$ 0.1 region and most of the events
with lepton ISO $>$ 0.2, but not all of them, are QCD background events.
The upper histogram of Figure \ref{f10} shows the isolation distribution of 
the electrons from $W$+jet events, for the first $p_T^W$ bin.  
When plotted on a semi-log scale, the ISO $<$ 0.1 and 
the ISO $>$ 0.2 regions can be approximated with two straight lines.
The technique we use extrapolates the ISO $>$ 0.2 line into the
ISO $<$ 0.1 signal region to calculate its integral and obtain the 
number of events in the signal region, using the assumption
that the QCD background shape is not altered in that region. 
This method would give us the true number of QCD background events, 
if the ISO $>$ 0.2 region was filled exclusively with QCD events.  
In reality, only a fraction of these events are true QCD background, 
the rest being $W$+jet events.
Since we expect to have some $W$+jet events in 
the region of lepton isolation from 0.1 to 0.2, we fit the 
area above 0.2 with a straight line (in the semi-log histogram), 
which describes the QCD background.  We also fit five continuous regions
of lepton isolation, around the central region of ISO=0.20 to ISO=0.65 (namely 0.15-0.65,
0.25-0.65, 0.15-0.60, 0.20-0.65, and 0.25-0.70) to obtain a systematic 
uncertainty for this procedure.
\begin{table*}
\hspace*{-13.mm}
\scriptsize
\caption{\label{t5}The linear least-squares fit parameterization of the $\Delta\phi_{l-j}^{\rm lab}$ 
distribution for ISO $>0.2$ electron $W$+jet events (second column) allows us to estimate the number of ISO $>0.2$
$W$+jet events in the $\Delta\phi_{l-j}^{\rm lab} > 2.5$ region.  
The integral of this line, divided by the bin width ($\pi$/30) of the $\Delta\phi_{l-j}^{\rm lab}$ histogram, 
is the number of $W$+jet events with ISO $>0.2$ and $\Delta\phi_{l-j}^{\rm lab} >  2.5$ shown
in the third column.  These events are subtracted from the total number of ($W$+jet + QCD) background 
events in the ISO $>0.2$ and $\Delta\phi_{l-j}^{\rm lab} > 2.5$ region (fourth column).
The result is divided by the total number of events in the ISO $>0.2$ region,
to obtain an estimate of the fraction of true QCD background events (fifth column).}
\begin{tabular}{|c||c|c|c|c|c|c|} \hline \hline
\multicolumn{1}{|c||}{ } & 
\multicolumn{1}{c|}{\bf \boldmath Fit parameterization of electron} & 
\multicolumn{1}{c|}{\bf\boldmath Electron $W$+jet events with} & 
\multicolumn{1}{c|}{\bf\boldmath ($W$+jet)+QCD events with} & 
\multicolumn{1}{c|}{\bf\boldmath Fraction of} \\
\multicolumn{1}{|c||}{\bf\boldmath \rr{$p_T^W$ (GeV)}{2}} & 
\multicolumn{1}{c|}{\bf\boldmath $W$+jet events with ISO $>0.2$} & 
\multicolumn{1}{c|}{\bf\boldmath ISO $>0.2$ and $\Delta\phi_{l-j}^{\rm lab}>2.5$} & 
\multicolumn{1}{c|}{\bf\boldmath ISO $>0.2$ and $\Delta\phi_{l-j}^{\rm lab}>2.5$} & 
\multicolumn{1}{c|}{\bf\boldmath true QCD events} \\ \hline 
15--25  & $2.21\times\Delta\phi_{l-j}^{\rm lab}-0.05$ & 37.9 & 257 & 0.66=(257-37.9)/332 \\
25--35  & $0.70\times\Delta\phi_{l-j}^{\rm lab}+0.78$ & 16.9 & 98 & 0.58=(98-16.9)/141 \\
35--65  & $1.04\times\Delta\phi_{l-j}^{\rm lab}+0.45$ & 20.6 & 49 & 0.31=(49-20.6)/91\\
65--105 & $0.17\times\Delta\phi_{l-j}^{\rm lab}+0.73$ & 7.5 & 10 & 0.13=(10-7.5)/20 \\ \hline \hline
\end{tabular}
\end{table*}
\begin{table*}
\hspace*{-10.mm}
\footnotesize
\caption{\label{t6}The number of electron W+jet events, the number of QCD 
background events before correction 
and their percentage in the signal region, and the fraction of true QCD 
background events and their percentage 
in the signal region, for the four $p_T^W$ bins (see text for details). 
}
\begin{tabular}{|c||c|c|c|c|c|c|} \hline \hline
\multicolumn{1}{|c||}{ } & 
\multicolumn{1}{c|}{\bf\boldmath Number of } & 
\multicolumn{1}{c|}{\bf\boldmath QCD events } & 
\multicolumn{1}{c|}{\bf\boldmath Percentage of QCD} & 
\multicolumn{1}{c|}{\bf\boldmath Fraction of } &
\multicolumn{1}{c|}{\bf\boldmath Percentage of } \\
\multicolumn{1}{|c||}{\bf \boldmath \rr{$p_T^W$ (GeV)}{2}} & 
\multicolumn{1}{c|}{\bf\boldmath electron $W$+jet} & 
\multicolumn{1}{c|}{\bf\boldmath before correction} & 
\multicolumn{1}{c|}{\bf\boldmath before correction} & 
\multicolumn{1}{c|}{\bf\boldmath true QCD events} &
\multicolumn{1}{c|}{\bf\boldmath QCD background} \\ \hline 
15--25  & 5166 & 423$^{+77}_{-42}$ & 8.18$^{+1.11}_{-0.81}\%$ & 0.66$^{+0.04}_{-0.04}$ & 5.40$^{+0.80}_{-0.63}\%$ \\
25--35  & 3601 & 353$^{+26}_{-148}$ & 9.80$^{+0.74}_{-4.1}\%$ & 0.58$^{+0.07}_{-0.08}$ & 5.68$^{+0.81}_{-2.50}\%$ \\
35--65  & 3285 & 54$^{+151}_{-25}$ &  1.64$^{+4.60}_{-0.75}\%$& 0.31$^{+0.13}_{-0.13}$ & 0.51$^{+1.44}_{-0.32}\%$ \\
65--105 &  624 & 14$^{+8}_{-3}$ & 2.24$^{+1.24}_{-0.27}\%$ & 0.13$^{+0.87}_{-0.13}$ & 0.29$^{+1.96}_{-0.29}\%$\\ \hline \hline
\end{tabular}
\end{table*}

Since not all of the extrapolated region is QCD background, 
we obtain a measurement of the percentage of the true QCD background in the electron $W$+jet sample
above electron isolation of 0.1, 
by making a histogram of $\Delta\phi_{l-j}^{\rm lab}$ for the events with ISO $>$ 0.2,
where $\Delta\phi_{l-j}^{\rm lab}$ is the difference in the $\phi$ angle between the electron and 
the highest-$E_T$ jet, with no other requirements for that jet.
We expect the $\Delta\phi_{l-j}^{\rm lab}$ distribution to be almost flat for the $W$+jet events, because
no correlation exists between the jet and the lepton $\phi$ 
directions.  In reality, this distribution decreases at low $\Delta\phi_{l-j}^{\rm lab}$,
due to the application of the lepton isolation cut in our data.
For QCD background, we expect the $\Delta\phi_{l-j}^{\rm lab}$ between the
highest $E_T$ jet and the jet resembling the lepton to peak at $\pi$.
\begin{figure}
\includegraphics[scale=.46]{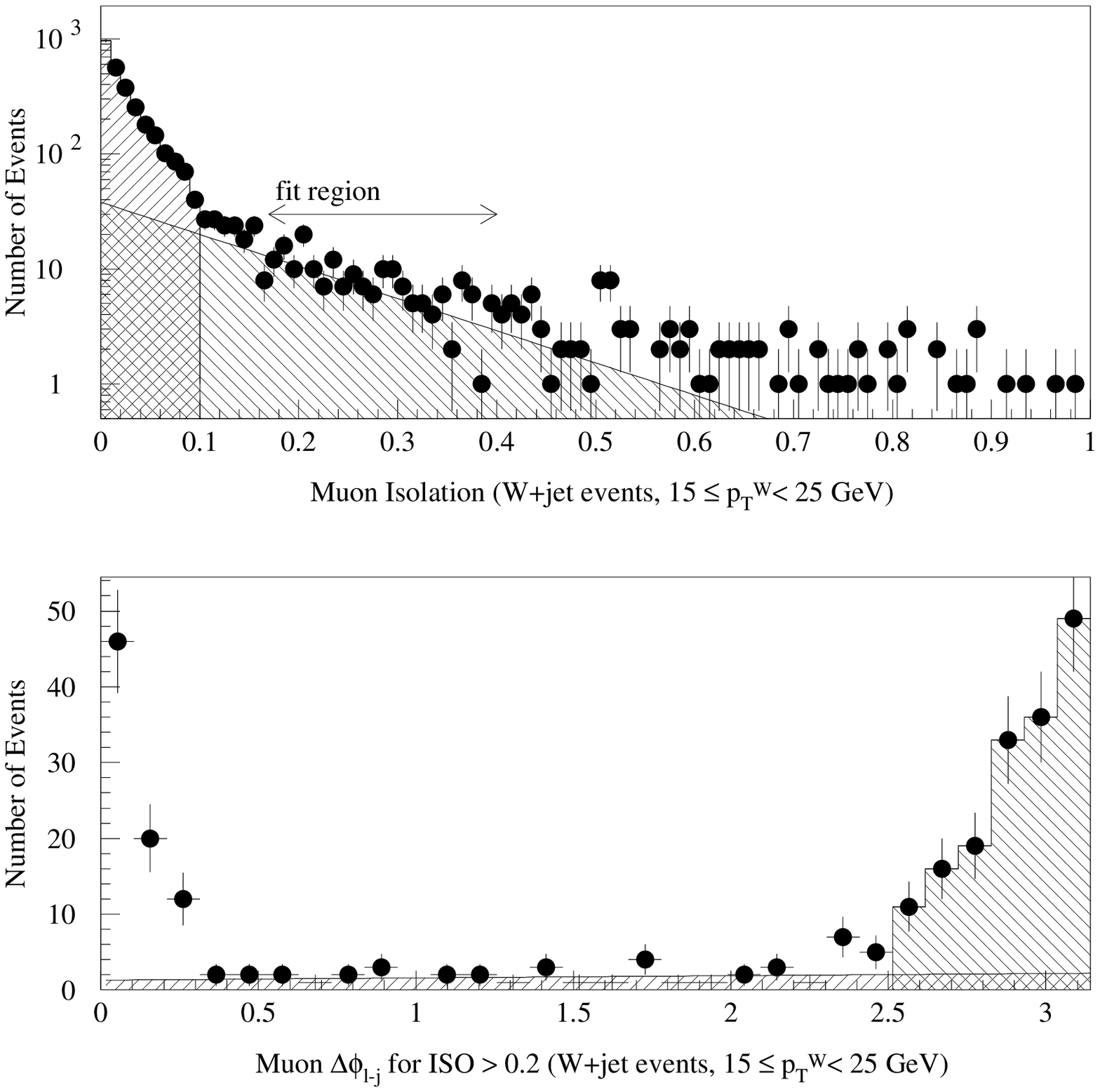}
\caption{Isolation of the muon in $W$+jet events (upper histogram) and the difference in the 
azimuthal angles of the muon and the jet for high isolation events (lower histogram). 
These histograms are used to estimate the QCD background in the muon $W$+jet data, for the 
$15 \leq p_T^W \leq 25$ GeV bin (see text). Corresponding histograms are used for the estimation 
of the QCD background in the three other muon $W$+jet $p_T^W$ bins.
At low $\Delta\phi_{l-j}^{\rm lab}$ the distribution increases, due to bremsstrahlung processes associated
with residual muon $Z$+jet background in the muon $W$+jet sample.  
These
events disappear if we apply the ${\tt zmuo\_veto}$ cut.
(see text).
}
\label{f11}
\end{figure}
The lower histogram of Figure \ref{f10} shows 
the $\Delta\phi_{l-j}^{\rm lab}$ for the events with
lepton isolation greater than 0.2 for electron $W$+jet events and for the first $p_T^W$ bin.
We fit the region $\Delta\phi_{l-j}^{\rm lab} \leq 2.5$ ($W$+jet contribution) with a straight 
line.  The region of the histogram $\Delta\phi_{l-j}^{\rm lab} > 2.5$ above that line corresponds
to true QCD background.  
By dividing this part of the histogram by the total number of events
with ISO $>$ 0.2, 
we determine the true fraction of QCD background in the ISO $>$ 0.2 region.
We expect the same fraction to be valid in the signal region (ISO $<0.1$).
Therefore, the number of true QCD background events is obtained
by multiplying the number of ISO $<$ 0.1 events (as obtained by extrapolating 
the ISO $>$ 0.2 line into the signal region of the lepton isolation plot)
by the QCD background fraction obtained from
the $\Delta\phi_{l-j}^{\rm lab}$ plot.  The procedure is repeated for the four $p_T^W$ bins.
Table \ref{t5} shows the extracted fraction of QCD background in the ISO $>$ 0.2 region for 
the four $p_T^W$ bins.
The electron $W$+jet QCD background results are presented in Table \ref{t6}.

In the study of the QCD background in the muon sample we face a new problem.
We originally apply a cut to muon $W$+jet data (${\tt zmuo\_veto}$) in order to remove 
events that are consistent with the production of a $Z$ boson, where one
of the muons is non-isolated because it fails one (and only one) of the following cuts:
\begin{itemize}
\item The muon isolation cut ISO $<0.1$
\item The electromagnetic calorimeter cut $E^{\rm EM} < 2$ GeV
\item The hadron calorimeter cut $E^{\rm HAD}< 6$ GeV
\end{itemize}
These dimuon events are true $Z$ bosons that look like $W$ bosons because
one muon does not pass one of the above cuts due to inner 
bremsstrahlung or bremmsstrahlung in the electromagnetic or hadronic calorimeters.
The ${\tt zmuo\_veto}$ cut mainly affects the tail of the muon isolation 
distribution (ISO $>0.2$) and causes us to underestimate the QCD background,
since we use that tail to estimate it.
Therefore, for the muon $W$+jet samples, for the purposes of determination of QCD background,
we neglect this cut, in order to remove this bias at high muon isolation 
(ISO $>$ 0.2) and make the transition from the low to high isolation smooth.
Some of the muon $Z$ background is thus counted as QCD background; however we do not
expect it to radically affect our QCD background estimation.
In the isolation method we fit the background starting from ISO=0.17 to ISO=0.40,
to increase the statistical significance of our estimation. 
We also fit five continuous regions
of lepton isolation, around the central region of ISO=0.17 to ISO=0.40 (namely 0.16-0.40,
0.18-0.40, 0.16-0.35, 0.17-0.40, and 0.18-0.45) to obtain a systematic 
uncertainty for this procedure.
The upper histogram of Figure \ref{f11} shows the isolation distribution and fits 
for the muon $W$+jet events and for the first $p_T^W$ bin.  

We obtain a measurement of the percentage of the true QCD background in the muon $W$+jet sample
above muon isolation of 0.2, by making a histogram of $\Delta\phi_{l-j}^{\rm lab}$ for the events 
with ISO $>$ 0.2, where $\Delta\phi_{l-j}^{\rm lab}$ is the difference in the $\phi$ 
angle between the muon and the highest-$E_T$ jet, with no other requirements for that jet.
The lower histogram of Figure \ref{f11} shows 
$\Delta\phi_{l-j}^{\rm lab}$ for the events with
isolation greater than 0.2 for muon  $W$+jet events, for the first $p_T^W$ bin.
The peak in the $\Delta\phi_{l-j}^{\rm lab}=0$ region is due to the muon bremsstrahlung processes
that are not suppressed after we relax the ${\tt zmuo\_veto}$ cut.
We ignore these events when we fit to the straight line describing
the $W$+jet events with high isolation muons.
\begin{table*}
\hspace*{-13.mm}
\scriptsize
\caption{\label{t7}The linear least-squares fit parameterization of the $\Delta\phi_{l-j}^{\rm lab}$ 
distribution for ISO $>0.2$ muon $W$+jet events (second column) allows us to estimate the number of ISO $>0.2$
$W$+jet events in the $\Delta\phi_{l-j}^{\rm lab} > 2.5$ region.  
The integral of this line, divided by the bin width ($\pi$/30) of the $\Delta\phi_{l-j}^{\rm lab}$ histogram, 
is the number of $W$+jet events with ISO $>0.2$ and $\Delta\phi_{l-j}^{\rm lab} > 2.5$ shown
in the third column.  These events are subtracted from the total number of ($W$+jet + QCD) background 
events in the ISO $>0.2$ and $\Delta\phi_{l-j}^{\rm lab} > 2.5$ region (fourth column).
The result is divided by the total number of events in the ISO $>0.2$ region,
to obtain an estimate of the fraction of true QCD background events (fifth column).}
\begin{tabular}{|c||c|c|c|c|c|c|} \hline \hline
\multicolumn{1}{|c||}{ } & 
\multicolumn{1}{c|}{\bf \boldmath Fit parameterization of muon} & 
\multicolumn{1}{c|}{\bf\boldmath Muon $W$+jet events with} & 
\multicolumn{1}{c|}{\bf\boldmath ($W$+jet)+QCD events with} & 
\multicolumn{1}{c|}{\bf\boldmath Fraction of} \\
\multicolumn{1}{|c||}{\bf\boldmath \rr{$p_T^W$ (GeV)}{2}} & 
\multicolumn{1}{c|}{\bf\boldmath $W$+jet events with ISO $>0.2$} & 
\multicolumn{1}{c|}{\bf\boldmath ISO $>0.2$ and $\Delta\phi_{l-j}^{\rm lab}>2.5$} & 
\multicolumn{1}{c|}{\bf\boldmath ISO $>0.2$ and $\Delta\phi_{l-j}^{\rm lab}>2.5$} & 
\multicolumn{1}{c|}{\bf\boldmath true QCD events} \\ \hline 
15--25  & $0.29\times\Delta\phi_{l-j}^{\rm lab}+1.31$ & 13 & 164 & 0.52=(164-13)/288 \\
25--35  & $0.46\times\Delta\phi_{l-j}^{\rm lab}+0.79$ & 12.8 & 69 & 0.40=(69-12.8)/140 \\
35--65  & $0.25\times\Delta\phi_{l-j}^{\rm lab}+1.02$ & 10.5 & 19 & 0.14=(19-10.5)/61\\
65--105 & $0\times\Delta\phi_{l-j}^{\rm lab}+1$ & 6.1 & 1 & {$0=0/6$} \\ \hline \hline
\end{tabular}
\end{table*}
\begin{table*}
\hspace*{-10.mm}
\footnotesize
\caption{\label{t8}
The number of muon W+jet events without the application of the ${\tt zmuo\_veto}$ cut, 
the number of QCD background events before correction and their percentage in the signal region, and the fraction of true QCD background events and their percentage in the signal region, 
for the four $p_T^W$ bins (see text for details).  
}
\begin{tabular}{|c||c|c|c|c|c|c|} \hline \hline
\multicolumn{1}{|c||}{ } & 
\multicolumn{1}{c|}{\bf\boldmath Number of } & 
\multicolumn{1}{c|}{\bf\boldmath QCD events } & 
\multicolumn{1}{c|}{\bf\boldmath Percentage of QCD} & 
\multicolumn{1}{c|}{\bf\boldmath Fraction of } &
\multicolumn{1}{c|}{\bf\boldmath Percentage of } \\
\multicolumn{1}{|c||}{\bf\boldmath \rr{$p_T^W$ (GeV)}{2}} & 
\multicolumn{1}{c|}{\bf\boldmath muon $W$+jet} & 
\multicolumn{1}{c|}{\bf\boldmath before correction} & 
\multicolumn{1}{c|}{\bf\boldmath before correction} & 
\multicolumn{1}{c|}{\bf\boldmath true QCD events} &
\multicolumn{1}{c|}{\bf\boldmath QCD background} \\ \hline 
15--25  & 2779 & 280$^{+102}_{-25}$ & 10.07$^{+3.68}_{-0.90}\%$ & 0.52$^{+0.05}_{-0.04}$ & 5.24$^{+1.98}_{-0.62}\%$ \\
25--35  & 1943 & 103$^{+129}_{-4}$ & 5.30$^{+6.64}_{-0.21}\%$ & 0.40$^{+0.08}_{-0.08}$ & 2.12$^{+2.69}_{-0.43}\%$ \\
35--65  & 2002 & 139$^{+50}_{-21}$ &  6.94$^{+2.50}_{-1.05}\%$& 0.14$^{+0.22}_{-0.14}$ & 0.97$^{+1.57}_{-0.97}\%$ \\
65--105 &  389 & 11$^{+0}_{-0}$ & 2.70$^{+0.00}_{-0.00}\%$ & {$0^{+1}_{-0}$} & 0$^{+2.7}_{-0}\%$\\ \hline \hline
\end{tabular}
\end{table*}
Table \ref{t7} shows the extracted fraction of QCD background in the ISO $>$ 0.2 region for 
the four $p_T^W$ bins.
The muon $W$+jet QCD background results are presented in Table \ref{t8}.
For the highest muon $p_T^W$ bin the predicted number of true $W$+jet events 
is greater than the total number of events with ISO $>$ 0.2 and $\Delta \phi_{l-j}^{\rm lab} > 2.5$,
which results in a fraction of true QCD background events above ISO $>$ 0.2 equal to zero.  

After we calculate the percentage of the QCD background in the signal region,
we multiply it by the CDF $W$+jet event yields to obtain the absolute prediction of
the number of QCD background events in each of the four $p_T^W$ bins, for both and electron and muon
$W$+jet data.  The results are presented in Table \ref{t9}.
\begin{figure}
\includegraphics[scale=.46]{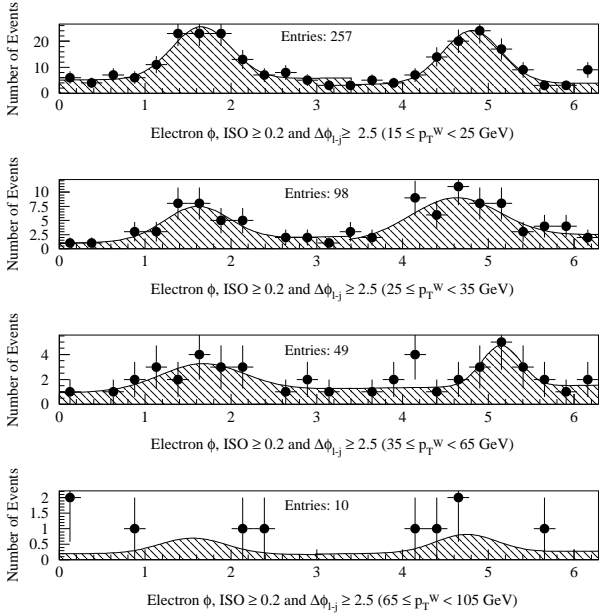}
\caption{The Collins-Soper $\phi$ distribution of electrons from $W$+jet events with ISO $> 0.2$ 
and $\Delta\phi_{l-j}^{\rm lab} > 2.5$
for each of the four $p_T^W$ bins.  These events are predominantly QCD background events.  We fit the distribution of the
first three $p_T^W$ bins with two Gaussians on top of two straight lines.  For the highest $p_T^W$ bin we use the distribution
of the total QCD background, normalized to the number of the QCD events in this bin. (see text for details).
}
\label{f12}
\end{figure}	
\begin{figure}
\includegraphics[scale=.46]{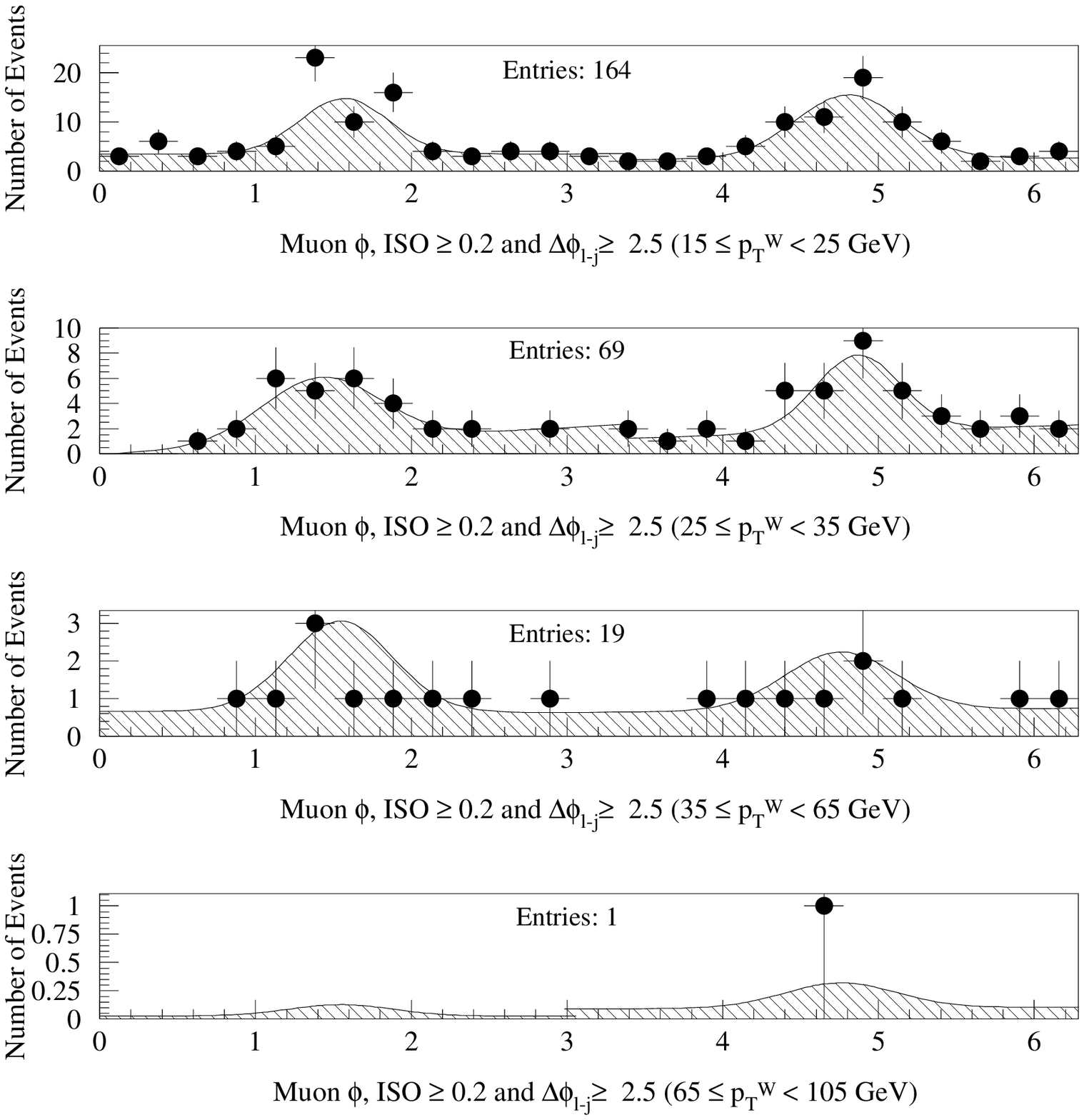}
\caption{The Collins-Soper $\phi$ distribution of muons from $W$+jet events with ISO $> 0.2$ and $\Delta\phi_{l-j}^{\rm lab} > 2.5$
for each of the four $p_T^W$ bins.  These events are predominantly QCD background events.  We fit the distribution of the
first two $p_T^W$ bins with two Gaussians on top of two straight lines.
For the two highest $p_T^W$ bins we use the distribution of the total background, 
normalized to the number of the QCD events in those bins.
}
\label{f13}
\end{figure}	
\begin{table*}
\center
\caption{\label{t9}QCD background estimation for the electron and muon $W$+jet events.
The fractions of the backgrounds are calculated with respect to the CDF Data $W$+jet events.}
\begin{tabular}{|c||c|c||c|c|}\hline \hline
\multicolumn{5}{|c|}{\bf QCD background}\\  \hline
\multicolumn{1}{|c||}{$p_T^W$ (GeV)} & 
\multicolumn{1}{c|}{$N_e$} & 
\multicolumn{1}{c||}{Fraction} &
\multicolumn{1}{c|}{$N_{\mu}$} & 
\multicolumn{1}{c|}{Fraction} \\ \hline 
15--25 & 279 {$^{+41}_{-33}$} & 5.40 {$^{+0.80}_{-0.63}$} \% & 148 {$^{+56}_{-17}$} & 5.24 {$^{+1.98}_{-0.62}$} \% \\
25--35 & 205 {$^{+29}_{-90}$} & 5.68 {$^{+0.81}_{-2.50}$} \% & 40 {$^{+50}_{-8}$} & 2.12 {$^{+2.69}_{-0.43}$} \% \\
35--65 & 17 {$^{+47}_{-11}$} & 0.51 {$^{+1.44}_{-0.32}$} \% & 18 {$^{+30}_{-18}$} & 0.97 {$^{+1.57}_{-0.97}$} \% \\
65--105 & 2 {$^{+12}_{-2}$} & 0.29 {$^{+1.96}_{-0.29}$} \% &  0 {$^{+10}_{-0}$}& 0 {$^{+2.7}_{-0}$}\% \\ \hline \hline
\end{tabular}
\end{table*}
\begin{table*}
\center
\caption{\label{t10}Summary of electron $W$+jet backgrounds.  The background
fractions are calculated with respect to the FMC signal event yields 
for the electroweak backgrounds and with respect to the data for the QCD background.}
\begin{tabular}{|c||c|c|c|c|}\hline \hline
\multicolumn{5}{|c|}{\bf\boldmath Electron $W$+jet Backgrounds} \\ \hline
\multicolumn{1}{|c||}{\bf Background}  &
\multicolumn{1}{c|}{$p_T^W$=15--25 GeV} &
\multicolumn{1}{c|}{$p_T^W$=25--35 GeV} &
\multicolumn{1}{c|}{$p_T^W$=35--65 GeV} &
\multicolumn{1}{c|}{$p_T^W$=65--105 GeV} \\ \hline
{$W \rightarrow \tau \nu_{\tau}$} & 86 $\pm$ 3 (2.22 \%) & 57 $\pm$ 2 (2.16 \%) & 56 $\pm$ 2 (2.26 \%) & 15 $\pm$ 1 (2.89\%) \\
{$Z \rightarrow e^{+} e^{-}$} & 47 $\pm$ 2 (1.22 \%) & 30 $\pm$ 1 (1.14 \%) & 25 $\pm$ 1 (1.01 \%) & 5 $\pm$ 0 (0.96 \%) \\
{QCD} & $279^{+41}_{-33}$ (5.40 \%) & $205^{+29}_{-90}$ (5.68 \%) & $17^{+47}_{-11}$ (0.51 \%) & $2 ^{+12}_{-2}$ (0.29 \%) \\ \hline \hline
\end{tabular}
\end{table*}
\begin{table*}[!]
\center
\caption{\label{t11}Summary of muon $W$+jet backgrounds.  The background fractions are
calculated with respect to the FMC signal event yields for the 
electroweak backgrounds and with respect to the data for the QCD background.}
\begin{tabular}{|c||c|c|c|c|}\hline \hline
\multicolumn{5}{|c|}{\bf\boldmath Muon $W$+jet Backgrounds} \\ \hline
\multicolumn{1}{|c||}{\bf Background}  &
\multicolumn{1}{c|}{$p_T^W$=15--25 GeV} &
\multicolumn{1}{c|}{$p_T^W$=25--35 GeV} &
\multicolumn{1}{c|}{$p_T^W$=35--65 GeV} &
\multicolumn{1}{c|}{$p_T^W$=65--105 GeV} \\ \hline
{$W \rightarrow \tau \nu_{\tau}$} & 45 $\pm$ 2 (2.22 \%) & 30 $\pm$ 2 (2.17 \%) & 30 $\pm$ 2 (2.28 \%) & 8 $\pm$ 0 (2.87\%) \\
{$Z \rightarrow \mu^{+} \mu^{-}$} & 127 $\pm$ 7 (6.26 \%) & 82 $\pm$ 4 (5.92 \%) & 72 $\pm$ 4 (5.48 \%) & 12 $\pm$ 1 (4.30 \%) \\
{QCD} & $148^{+56}_{-17}$ (5.24 \%) & $40^{+50}_{-8}$ (2.12 \%) & $18^{+30}_{-18}$ (0.97 \%) & $0^{+10}_{-0}$ (0 \%) \\\hline \hline
\end{tabular}
\end{table*}
\begin{table*}[!]
\small
\footnotesize
\caption{\label{t12}The expected total event yields for inclusive $W$+jet production.
The signal and electroweak backgrounds are calculated up to order $\alpha_s^2$.  The PDF and
$Q^2$ systematics have also been included (second set of uncertainties).}
\begin{tabular}{|c||c|c|c||c|c|c|}\hline \hline
\multicolumn{7}{|c|}{\bf Expected signal+background event yields for inclusive \boldmath $W$+jet production}\\  \hline
\multicolumn{1}{|c||}{ } &
\multicolumn{3}{c||}{\bf Electrons} &
\multicolumn{3}{c|}{\bf Muons} \\  \cline{2-7}
\multicolumn{1}{|c||}{\rr{$p_T^W$ (GeV)}{2.5}} & 
\multicolumn{1}{c|}{$N_e({\rm Signal})$} &
\multicolumn{1}{c|}{$N_e({\rm Background})$} &
\multicolumn{1}{c||}{$N_e$(Total prediction)} &
\multicolumn{1}{c|}{$N_{\mu}({\rm Signal})$} &
\multicolumn{1}{c|}{$N_{\mu}({\rm Background})$} &
\multicolumn{1}{c|}{$N_{\mu}$(Total prediction)}\\ \hline 
15--25 & 3867 {$\pm$ 137} & {$412^{+41}_{-33}$} & {$4279^{+148}_{-146}$}{$+880$/$-660$} & 2027 {$\pm$ 102} & {$320^{+57}_{-19}$} & {$2347^{+124}_{-112}$}{$+484$/$-330$}\\
25--35 & 2632 {$\pm$ 93} & {$292^{+29}_{-90}$} & {$2924^{+100}_{-132}$}{$+598$/$-408$} & 1384 {$\pm$ 66} & {$152^{+50}_{-10}$} & {$1536^{+88}_{-72}$}{$+329$/$-224$}\\
35--65 & 2474 {$\pm$ 87} & {$98^{+47}_{-11}$} & {$2572^{+102}_{-91}$}{$+562$/$-383$} & 1314 {$\pm$ 67} & {$120^{+31}_{-19}$} & {$1434^{+79}_{-75}$}{$+312$/$-213$}\\
65--105 & 518 {$\pm$ 18} & {$22^{+12}_{-2}$} & {$540^{+22}_{-19}$}{$+118$/$-81$} & 279 {$\pm$ 14} & {$20^{+10}_{-1}$} & {$299^{+18}_{-15}$}{$+66$/$-45$}\\ \hline \hline
\end{tabular}
\end{table*}

To complete the study of the QCD background we need to estimate its
shape to properly include this background in the Standard Model prediction
of the lepton $\phi$ distribution in the CS frame, for each of the four $p_T^W$ bins.
We plot $\phi$ for the events with ISO $>$ 0.2 and
$\Delta\phi_{l-j}^{\rm lab}>2.5$ for the electrons and muon datasets,
as shown in Figures \ref{f12} and \ref{f13}, respectively.  We fit the distributions
to the sum of two Gaussians and two straight lines.  For the last $p_T^W$ bin 
of the electrons and the last two $p_T^W$ bins
of the muons, there are not enough statistics for the fit,
so we use the total distributions (for $15 \leq p_T^W \leq 105$ GeV)
normalized to the number of events for those high $p_T^W$ bins.   
We do not
expect the shape of the QCD background to be significantly altered with
increasing $p_T^W$.  We assume that these distributions are the same as the ones in the 
signal region (ISO $<0.1$) after they are properly normalized.
We use these distributions to add the QCD
background to the Standard Model prediction, after they are normalized to 
the expected number of QCD background events, given by Table \ref{t9}.

\subsection{Summary of backgrounds and Standard Model event yields prediction.}

Backgrounds for electron and muon $W$+jet events for each of the four $p_T^W$ bins are
summarized in Tables \ref{t10} and \ref{t11} respectively.
We obtain the total $W$+jet event yield prediction by adding these backgrounds
to the FMC $W$+jet signal prediction of Table \ref{t2}.
To obtain the final uncertainties, we add linearly the uncertainties associated with the $W$+jet 
signal and electroweak background and add the result to the QCD background uncertainty in quadrature. 
The total $W$+jet event yields after the inclusion of the backgrounds are presented in Table \ref{t12}.
The PDF and $Q^2$ systematic uncertainties are also included (see Section \ref{sec:sys}).
\begin{figure}[!]
\includegraphics[scale=0.44]{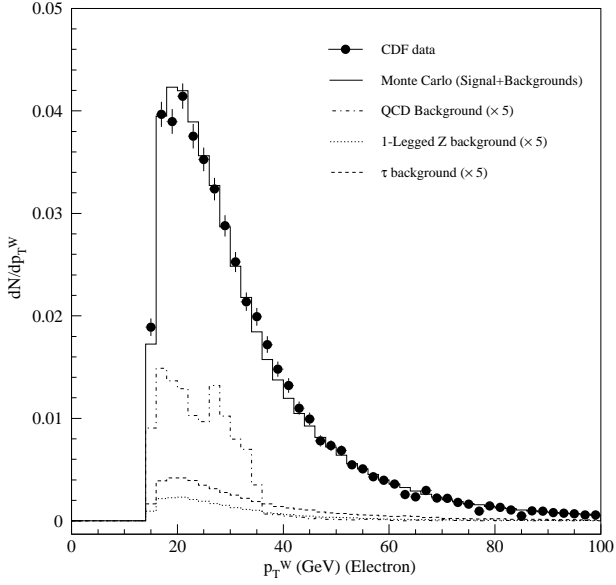}
\caption{The transverse momentum of the $W$ for the electron $W$+jet data sample
(points) along with the FMC signal simulation including backgrounds (solid histogram).  
The backgrounds are multiplied by 5, to be visible.  The data and expected signal+background distributions are normalized to unity.}
\label{wpt1}
\end{figure}
\begin{figure}[!]
\includegraphics[scale=0.44]{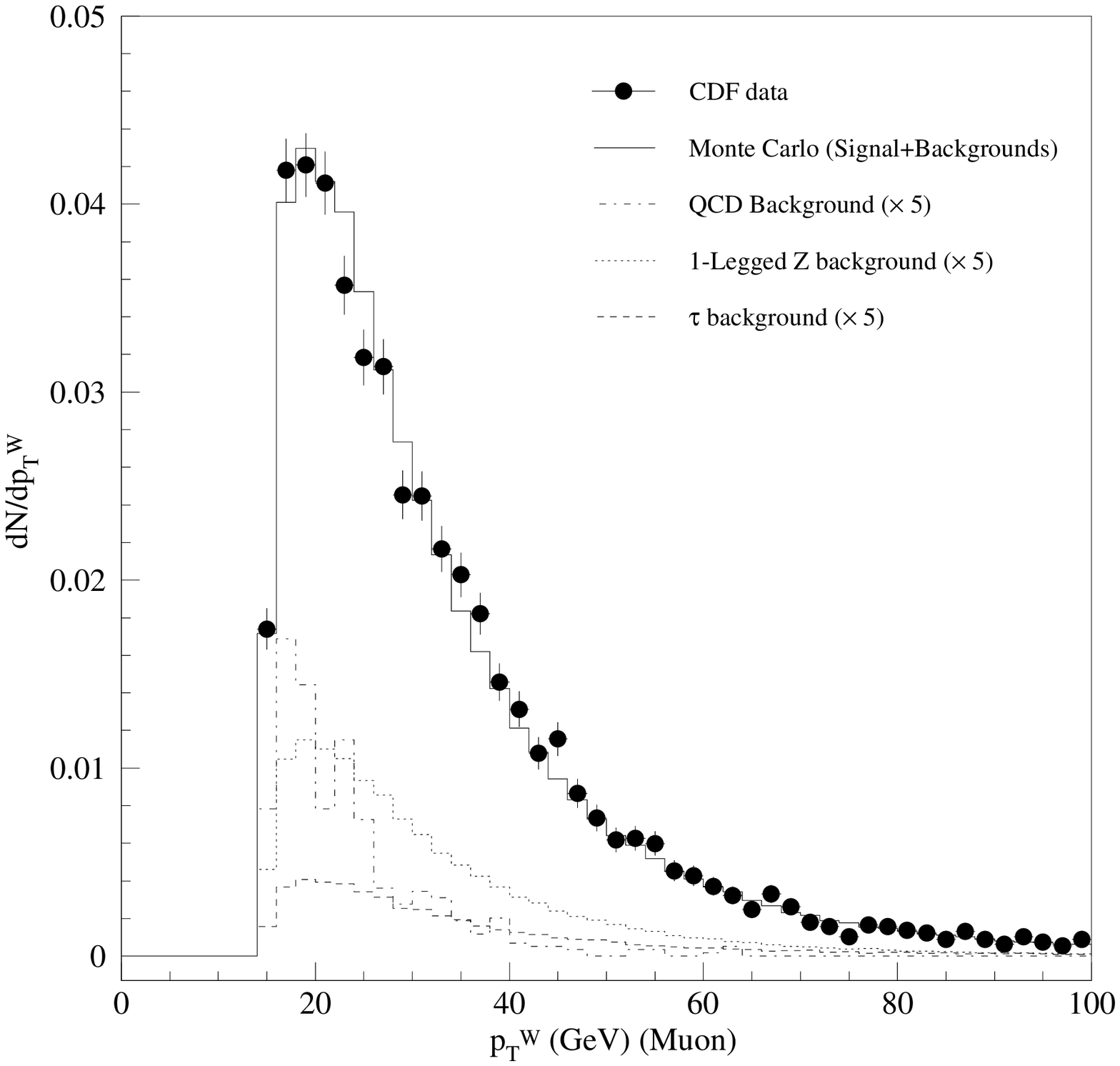}
\caption{The transverse mass of the $W$ for the electron $W$+jet data sample
(points) along with the FMC signal simulation including backgrounds (solid histogram).
The backgrounds are multiplied by 5, to be visible.
The data and expected signal+background distributions are normalized to unity.}
\label{pp1}
\end{figure}
\begin{figure}[!]
\includegraphics[scale=0.44]{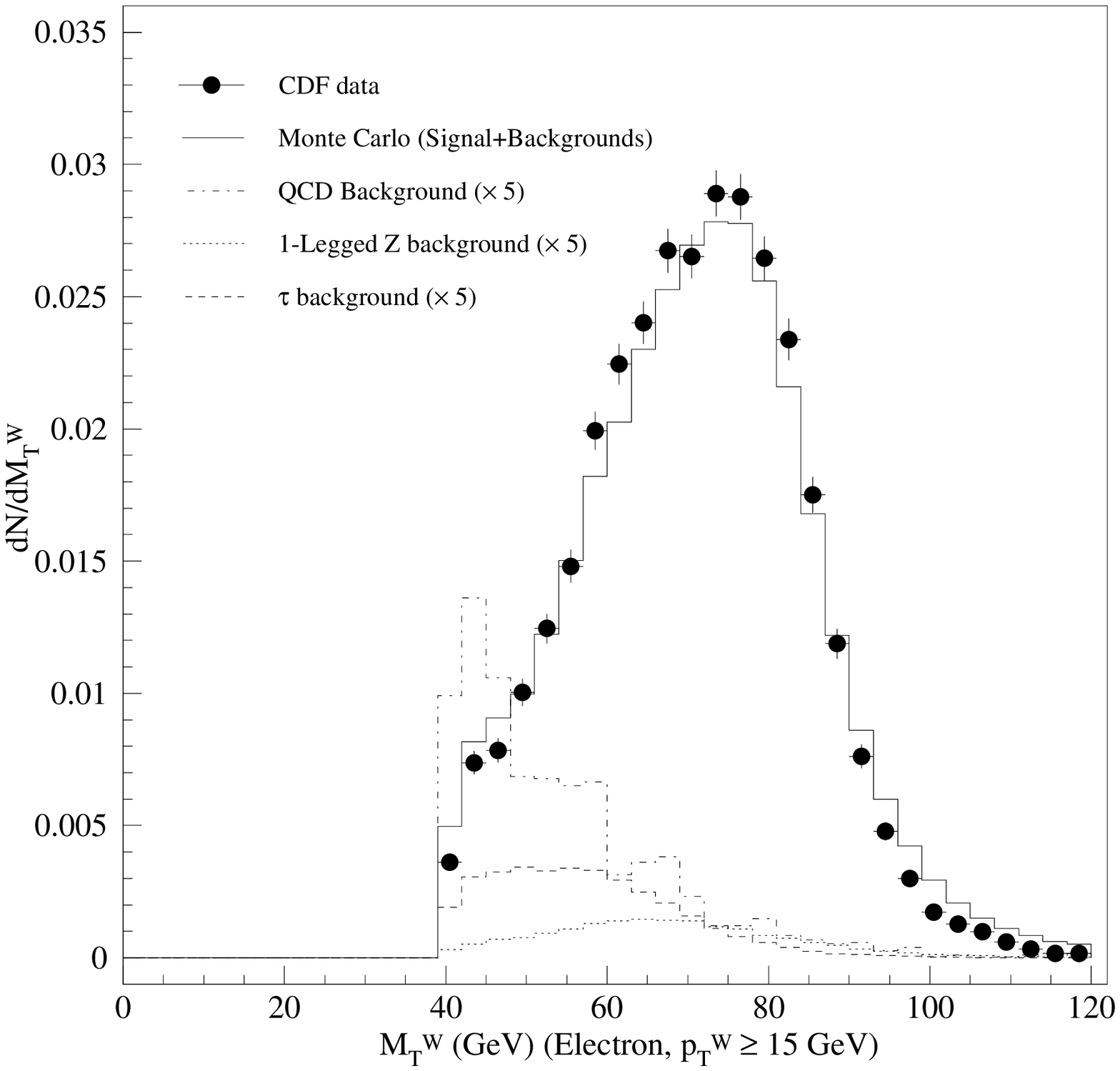}
\caption{The transverse momentum of the $W$ for the muon $W$+jet data sample
(points) along with the FMC signal simulation including backgrounds (solid histogram).  
The backgrounds are multiplied by 5, to be visible.  The data and expected signal+background distributions are normalized to unity.}
\label{wpt2}
\end{figure}
\begin{figure}[!]
\includegraphics[scale=0.44]{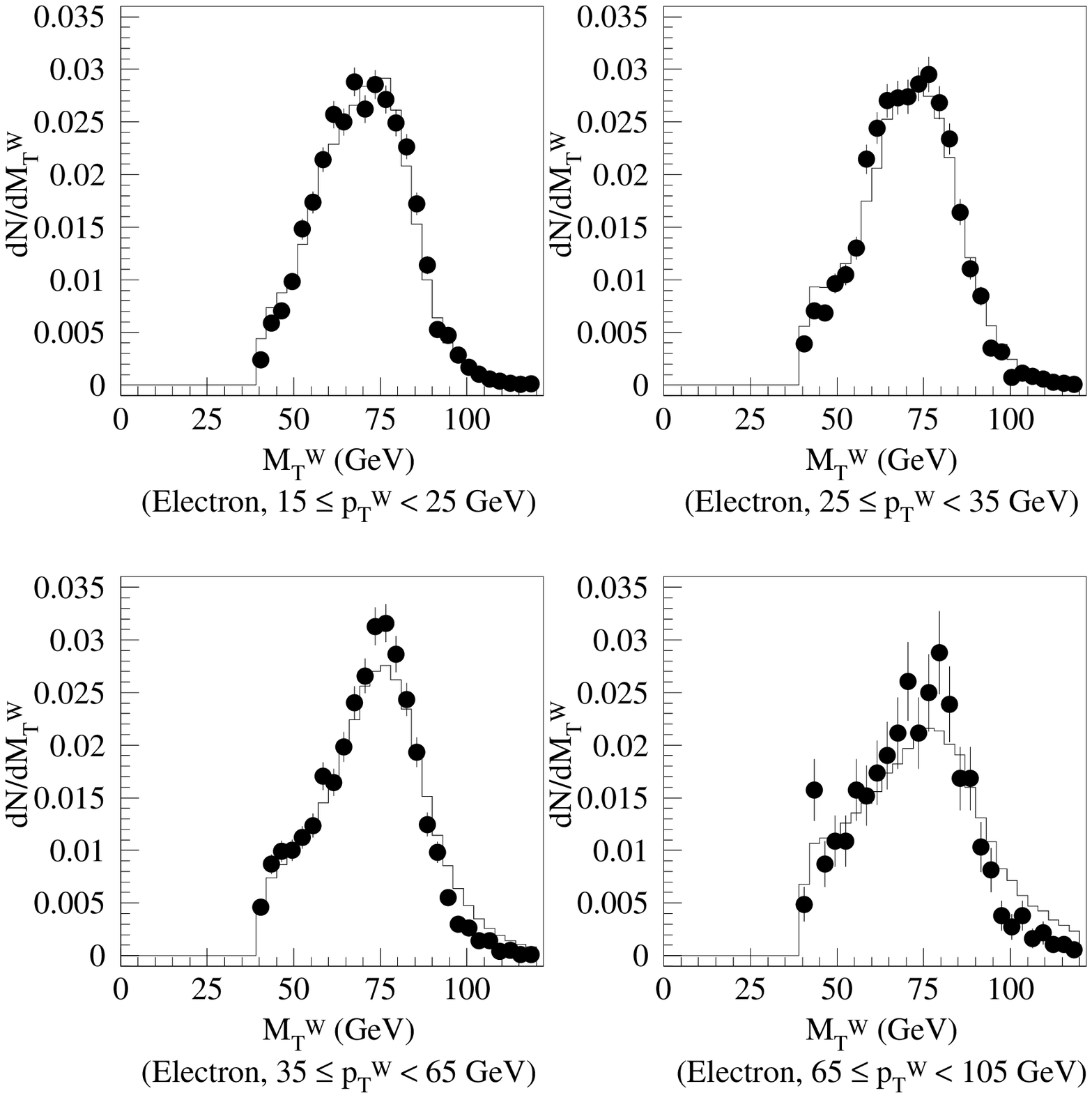}
\caption{The transverse mass of the $W$ for the electron $W$+jet data sample
(points) along with the FMC signal simulation including backgrounds (histogram),
for the four $p_T^W$ bins.  
The data and expected signal+background distributions are normalized to unity.}
\label{pp2}
\end{figure}
\begin{figure}[t]
\includegraphics[scale=0.44]{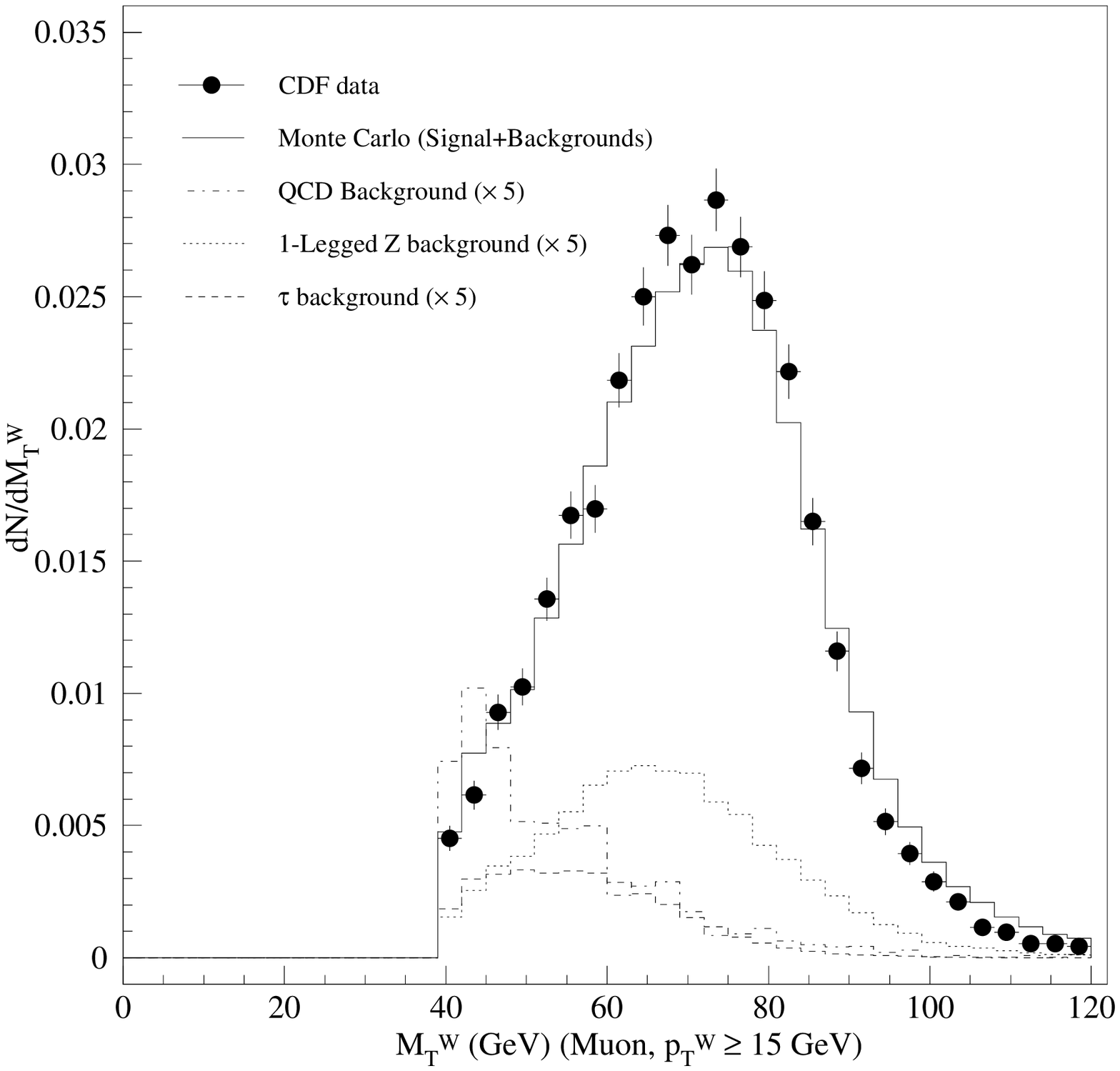}
\caption{The transverse mass of the $W$ for the muon $W$+jet data sample
(points) along with the FMC signal simulation including backgrounds (solid histogram).
The backgrounds are multiplied by 5, to be visible.
The data and expected signal+background distributions are normalized to unity.}
\label{pp3}
\end{figure}
\begin{figure}[t]
\includegraphics[scale=0.44]{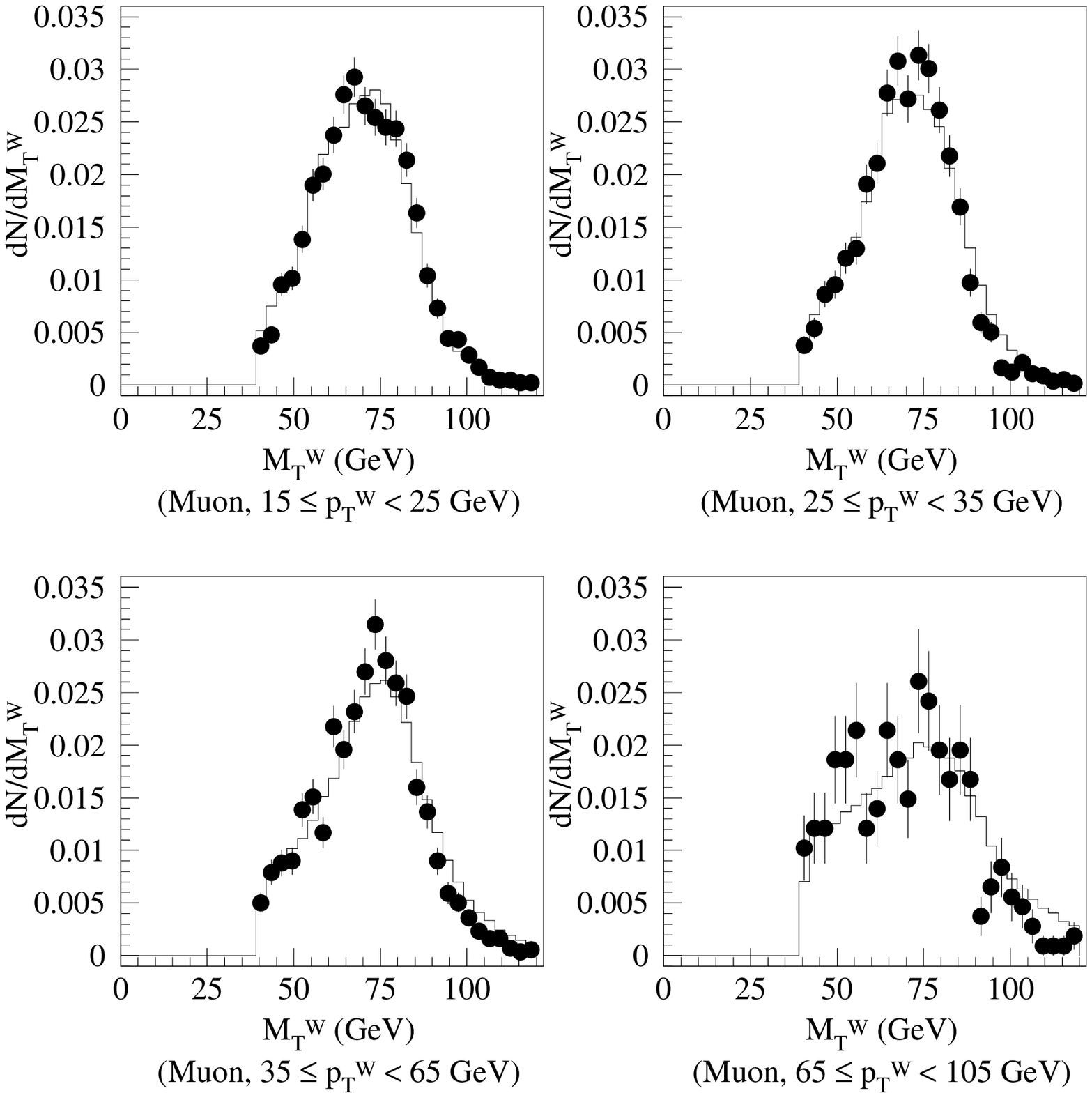}
\caption{The transverse mass of the $W$ for the muon $W$+jet data sample
(points) along with the FMC signal simulation including backgrounds (histogram),
for the four $p_T^W$ bins.  
The data and expected signal+background distributions are normalized to unity.}
\label{pp4}
\end{figure}

\section{\label{twra}\boldmath Comparison Between Expected and Observed $W$ Distributions}

We study the expected (FMC) $W$ kinematical distributions
after the inclusion of backgrounds and compare them to the experimental
distributions.  Figures \ref{wpt1} and \ref{wpt2} show the $W$ transverse
momentum for electrons and muons respectively.  The observed and simulated
distributions have been normalized to unity.  We observe good agreement between
the observed and simulated $p_T^W$ distributions.
Figure \ref{pp1} shows the $W$ transverse mass distribution for 
the electron $W$+jet dataset and for the DYRAD events passed through
the FMC detector simulation.
Figure \ref{pp2} shows the same distributions for the four $p_T^W$ bins.
Figures \ref{pp3} and \ref{pp4} show the same distributions for the muon $W$+jet
datasets.  The observed and simulated distributions are again normalized to unity.
In all of the above plots, the FMC distributions are produced with
properly weighted signal and background contributions, for
electron and muon detector regions.

\section{\label{direct} \boldmath Direct measurement of the azimuthal angle 
of the charged leptons from $W$ decays in the Collins-Soper
frame}

For each $W$ event we boost to the $W$ rest-frame to calculate
the azimuthal angle of the charged lepton.  The longitudinal 
momentum of the $W$ ($p_Z^W$) is
not known, because the longitudinal momentum of the neutrino is not measurable,
so we use the mass of the $W$ to constrain it.
For a particular event, the longitudinal momentum of the neutrino
is constrained by the mass of the $W$, according to the equation:
\begin{equation}
p_z^{\nu} = \frac{1}{(2p_T^l)^2}\left (Ap_z^l \pm E^l\sqrt{A^2-4(p_T^l)^2(p_T^{\nu})^2}\right ),
\end{equation}
where
\begin{equation}
A = M_W^2 +(p_T^W)^2-(p_T^l)^2-(p_T^\nu)^2,
\end{equation}
$E^l$ is the energy of the charged lepton, $p_T^l$ is its transverse momentum,
$p_z^l$ is its longitudinal momentum,
$p_T^{\nu}$ is the neutrino transverse momentum, and $p_T^W$ is
the transverse momentum of the $W$.
This equation is unique for every event, since the kinematics
of the lepton and neutrino, as well as the mass of the $W$, contribute
to the shape of the curve $p_z^{\nu}=f(M_W)$.  If the mass of the $W$ was known on an
event by event basis, there would be a two-fold ambiguity in the
value of ${p}_z^{\nu}$ of the neutrino in the laboratory frame.  Because the $W$ boson
has a finite width given by a PDF-convoluted Breit-Wigner distribution, $BW(M)$, 
we actually have two distributions of possible values of 
$p_z^{\nu}$, $BW(M(p_z^{\nu}))$, where $M(p_z^{\nu})$ is the 
mass of the $W$ as a function of the neutrino longitudinal momentum 
for the particular kinematics of 
the event.  

The choice of one of the two neutrino longitudinal momentum solutions does 
not affect the $\phi$ analysis, since both solutions
result in the same charged lepton $\phi$ in the CS frame.
For this analysis, only the choice of the $W$ mass is of interest.  
The choice is made based on the 2-dimensional $M^W$ vs. $M_T^W$ histograms 
constructed with DYRAD events.  For a specific $M_T^W$ we use a probability
distribution of $W$ masses and randomly select one for each event, 
based on that distribution.  This method
was devised to better reconstruct the $|\cos(\theta)|$ distribution
\cite{mythesis}, since the polar angle is very sensitive to the 
selection of the $W$ mass.  
In our analysis, the azimuthal angle is not affected
by the choice of mass, so the answer is almost the same even
if we choose a mass based on the Breit-Wigner
distribution and the requirement that the mass is greater 
than the measured transverse mass.  

After obtaining a $\phi$ for 
every event, we proceed to analyze our sample.
Theoretically, the $W$ differential cross section, integrated over $\cos\theta$ and
$y$ is given by:
\begin{eqnarray}
\frac{d\sigma}{d(p_T^W)^2d\phi}=C(1+\beta_1\cos{\phi}+\beta_2\cos{2\phi} \nonumber \\
+\beta_3\sin{\phi}+\beta_4\sin{2\phi}), 
\label{eqphi}
\end{eqnarray}
where
\begin{eqnarray}
C=\frac{1}{2\pi}\frac{d\sigma}{d(p_T^W)^2},\: \beta_1=\frac{3\pi}{16}A_3(p_T^W),\:\beta_2=\frac{A_2(p_T^W)}{4},  \nonumber \\
\beta_3=\frac{3\pi}{16}A_7(p_T^W),\:\beta_4=\frac{A_5(p_T^W)}{2}.
 \label{ucd}
\end{eqnarray}
The theoretical $\phi$ distributions for the charged lepton from $W$ boson decay in $W$+jet production 
are shown in Figure \ref{fextra1}.

From Equations (\ref{eqphi}) and (\ref{ucd}), the reader might conclude that only the 
$A_2$, $A_3$, $A_5$, and $A_7$ coefficients are measurable with the $\phi$ analysis, since the 
other angular coefficients are integrated out.  However, in the actual $W$+jet data samples,
what we measure is the number of events:
\begin{eqnarray}
\label{new} & &N(p_T^W,\phi)= \nonumber \\
& &\int \frac{d\sigma}{d (p_T^W)^2 d\phi d\cos\theta} ae(p_T^W,\cos\theta,\phi) 
d\cos\theta \int {\cal L} dt \nonumber \\
&+& N_{bg}(p_T^W,\phi),
\end{eqnarray}
where ${\cal L}$ is the instantaneous luminosity 
and $ae(p_T^W,\cos\theta,\phi)$ is the overall
acceptance times efficiency, determined in Section \ref{sec:4}, for a particular $W$ transverse
momentum and region in the 
$(\cos\theta,\phi)$ phase space.  The quantity $N_{bg}(p_T^W,\phi)$ is the background for
the given $\phi$ bin and $p_T^W$, estimated in Section \ref{sec:5}.
Combining Equations (\ref{new}) and (\ref{eq1}), the measured distribution
is
\begin{eqnarray}
\label{newer} N(p_T^W,\phi)&=& C' (f_{-1}(p_T^W,\phi) + \sum_{i=0}^{7} A_i(p_T^W) f_i(p_T^W,\phi)) \nonumber\\ &+& N_{bg}(p_T^W,\phi),
\end{eqnarray}
where $C' = C\int {\cal L} dt$.  The $f_i$ are fitting functions, 
which are integrals of the product
of the explicit functions $g_i(\cos\theta,\phi)$ and $ae(\cos\theta,\phi)$:
\begin{eqnarray}
\label{fis} f_i(p_T^W,\phi) = \int_{0}^{\pi} g_i(\theta,\phi) ae(p_T^W,\cos\theta,\phi) d\cos\theta ,\\
i=-1,\ldots,7 \nonumber
\end{eqnarray}
where
\begin{eqnarray}
g_{-1}(\theta,\phi)&=&1+\cos^2{\theta} \nonumber \\
g_0(\theta,\phi)&=&\frac{1}{2}(1-3\cos^2{\theta}) \nonumber \\
g_1(\theta,\phi)&=&\sin{2\theta} \cos{\phi} \nonumber \\
g_2(\theta,\phi)&=&\frac{1}{2}\sin^2{\theta} \cos{2\phi} \nonumber \\
g_3(\theta,\phi)&=&\sin{\theta}\cos{\phi} \nonumber \\
g_4(\theta,\phi)&=&\cos{\theta} \nonumber \\
g_5(\theta,\phi)&=&\sin^2{\theta}\sin{2\phi} \nonumber \\
g_6(\theta,\phi)&=&\sin{2\theta} \sin{\phi} \nonumber \\
g_7(\theta,\phi)&=&\sin{\theta}\sin{\phi} \nonumber \\
\end{eqnarray}

Because we multiply the $g_i(\theta,\phi)$ functions by $ae(p_T^W,\cos\theta,\phi)$ before
integrating over $\cos\theta$, no $f_i$ is exactly zero
and all of the angular coefficients $A_i$ are in principle measurable.  
We have verified that the FMC-simulated $\phi$ distributions, 
fitted with a linear combination of $f_i$, result in 
angular coefficient values consistent with the SM predictions \cite{js}.
This result supports the self-consistency of the method.

We use Simpson integration
for the calculation of the $f_i$ fitting functions given by Equation (\ref{fis}).  
The explicit functions $g_i(\theta,\phi)$ are integrated over $\cos \theta$,
after they are weighted with the value of $ae(p_T^W,\cos\theta,\phi)$ extracted 
from the 2-dimensional histograms of Figures \ref{f4} and \ref{f5}.

\begin{figure}
\includegraphics[scale=.46]{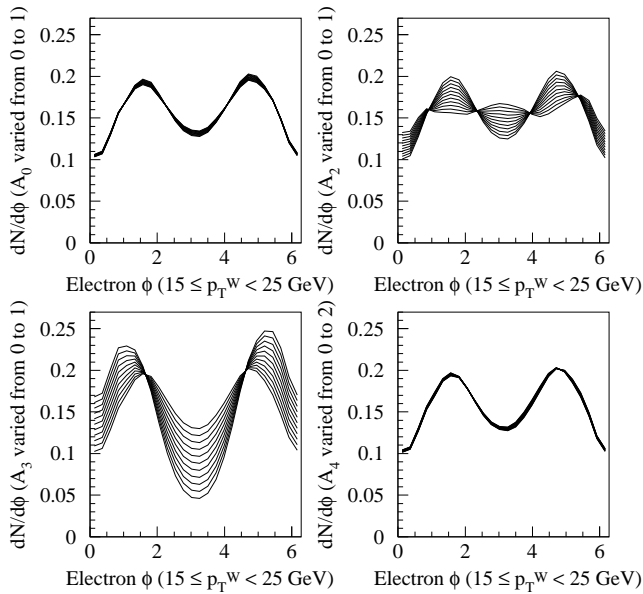}
\caption{The predicted electron $\phi$ distributions in the
Collins-Soper $W$ rest-frame for the first $p_T^W$ bin, varying
only one angular coefficient at a time and keeping the other angular coefficients at their 
Standard Model values.  Only $A_2$ and $A_3$ significantly affect the shape of these distributions.  
The same is true for the other $p_T^W$ bins and also for the muon $\phi$ distributions.}
\label{f14}
\end{figure}

Although the use of 
Equation (\ref{newer}) allows us in principle to measure all of the angular coefficients,
in reality, the current statistics do not allow us to make a
significant measurement of angular coefficients other than $A_2$ and $A_3$.
This is due to the fact that the fitting functions $f_{i \neq 2,3}$ are small, and the
$\phi$ distributions are insensitive to large variations of the corresponding
angular coefficients.  Figure \ref{f14} shows how the expected
electron $\phi$ distributions are modified as the angular coefficients $A_i$ are varied, 
one coefficient at a time (the muon $\phi$ distributions are almost identical).  
Using  Equation (\ref{newer}), we vary $A_0$, $A_2$, and $A_3$ from 0 to 1 with a step size of 0.1,
and $A_4$ from 0 to 2 with a step size of 0.2.
We find that only $A_2$ and $A_3$ strongly affect the azimuthal distributions, thus only
these two angular coefficients are measurable with the our current $\phi$ analysis.  
Large variations of $A_0$ and $A_4$ result in small changes in the $\phi$ distributions, 
hence the uncertainties associated with the measurement of $A_0$ and $A_4$ are large; these two
coefficients cannot be measured in a statistically significant manner with the $\phi$ analysis. 
The same is true for $A_1$, $A_5$, $A_6$, and $A_7$, all of which are consistent with zero for
our current experimental precision.

Figure \ref{f15} shows the observed CS electron $\phi$ distributions for CDF electron 
$W$+jet data for the four $p_T^W$ bins. 
Figure \ref{f16} shows the corresponding $\phi$ distributions of the CDF muon $W$+jet data.
The solid lines are the SM theoretical predictions including backgrounds, whereas the points correspond 
to CDF $W$+jet data (the error bars are statistical only).  The theoretical prediction for the $\phi$ distributions
is constructed using Equations (\ref{newer}) and (\ref{fis}).  The free 
parameters are the angular coefficients $A_i$.
The background $\phi$ shapes are 
given by Figures \ref{f6}, \ref{f8}, and \ref{f12},
for electrons and Figures \ref{f7}, \ref{f9}, and \ref{f13} for muons,
normalized to the event yields of Tables \ref{t10} and \ref{t11} respectively.
The expected signal is normalized to the FMC signal event yields of 
Table \ref{t12}, and subsequently the backgrounds are added to construct 
$N(p_T^W,\phi)$, according to Equation (\ref{newer}).
The total theoretically predicted distributions
along with the experimental ones, are finally normalized to unity.  
The experimental results are in good agreement with the Standard Model prediction,
which includes the effects of $W$ polarization and QCD contributions up to order
$\alpha_s^2$.

\section{Measurement of the Angular Coefficients \label{mac}}

\begin{figure}
\includegraphics[scale=.46]{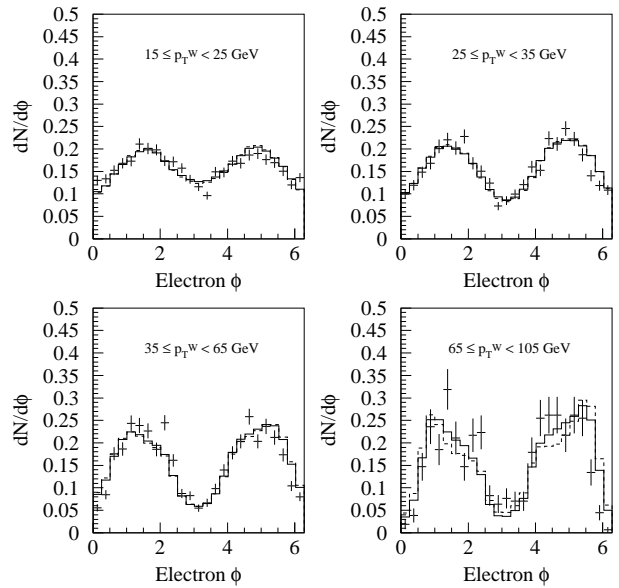}
\caption{The $\phi$ distributions for the electron CDF data (points)
the SM Monte Carlo (solid lines) and the result of the fit (dashed lines) 
for the four bins of $p_T^W$.  The errors are only statistical.
The fit is performed from $\pi/2$ to $3\pi/2$ and resulted in $\chi^2$/dof
equal to 2.32, 1.80, 2.18, and 1.11 for the four bins respectively (11 degrees
of freedom).  
All distributions are normalized to unity.}
\label{f15}
\end{figure}	
\begin{figure}
\includegraphics[scale=.46]{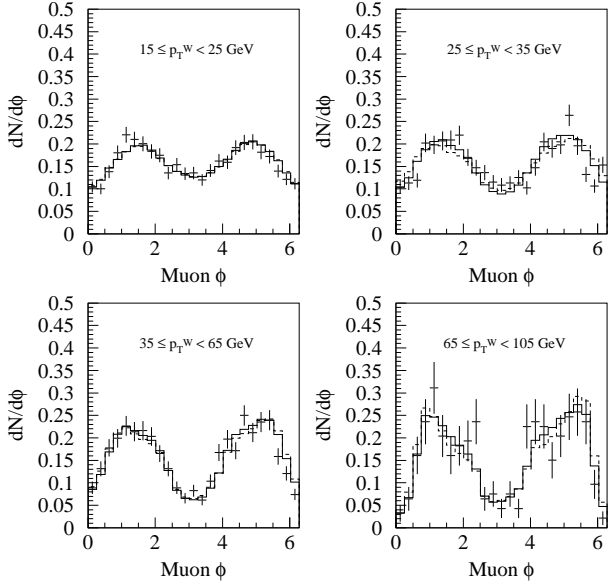}
\caption{The $\phi$ distributions for the muon CDF data (points)
the SM Monte Carlo (solid lines) and the result of the fit (dashed lines) 
for the four bins of $p_T^W$.  The errors are only statistical.
The fit is performed from $\pi/2$ to $3\pi/2$ and resulted in $\chi^2$/dof
equal to 0.41, 1.39, 1.33, and 1.71 for the four bins respectively (11 degrees of
freedom).  All distributions are normalized to unity.}
\label{f16}
\end{figure}
	
The values of the angular coefficients $A_2$ and $A_3$ are extracted
using the least-squares fitting method and the data associated with Figures \ref{f15}
and \ref{f16}.
The least-squares fit is performed over the negative $x$-axis
of the CS frame ($\pi/2<\phi<3\pi/2$) for the following two reasons.

\begin{figure}
\includegraphics[scale=.46]{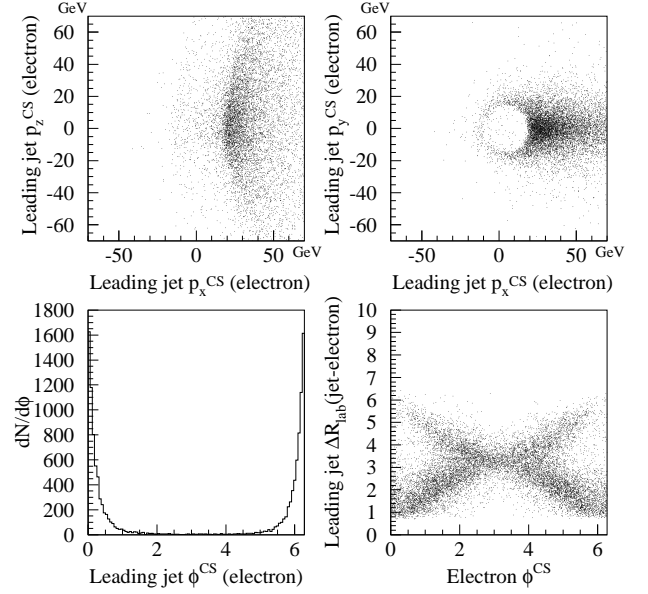}
\caption{The $p_z$ vs. $p_x$, $p_y$ vs. $p_x$, and $\phi$ of the 
leading jet in the CS frame and the $\Delta R$ between the
jet and the lepton in the laboratory frame vs. lepton $\phi$ in the CS frame, 
for electron data.}
\label{f17}
\end{figure}

\begin{figure}
\includegraphics[scale=.46]{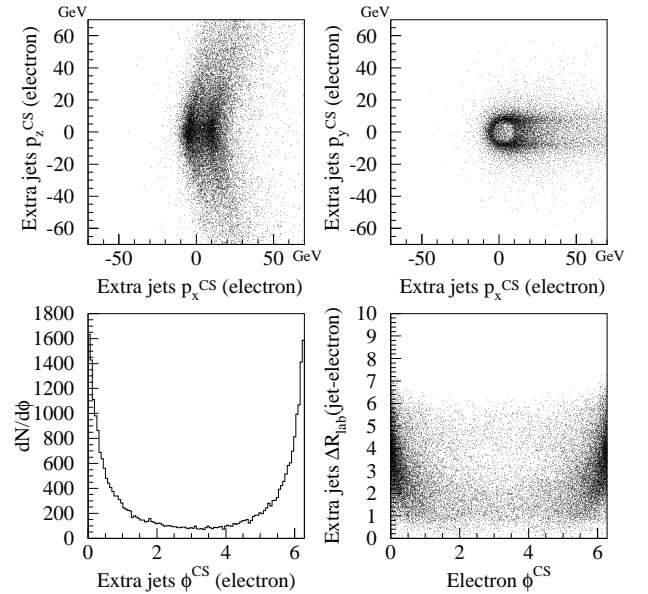}
\caption{The $p_z$ vs. $p_x$, $p_y$ vs. $p_x$, and $\phi$ of the 
extra jets in the CS frame and the $\Delta R$ between the
jets and the lepton in the laboratory frame vs. lepton $\phi$ in the CS frame, 
for electron data.}
\label{f18}
\end{figure}

Firstly, if a single jet perfectly balances the $W$ boson, its momentum will be placed
on the positive $x$-axis in the CS frame.  In reality, the leading jet will be in 
the $x>0$ region of the $z-x$ plane, in proximity to the $x$-axis, as seen in Figure \ref{f17} 
for the electron $W$+jet data.  
The leading jet's $\phi$ in the CS frame will almost always be less than $\pi$/2 or 
greater than 3$\pi$/2, as shown in Figure \ref{f17}.  
A kinematic correlation exists between the angular separation $\Delta R$ between the jet and the 
lepton in the $\phi_{\rm lab}-\eta_{\rm lab}$ space and the CS $\phi$ of the lepton, 
as shown in Figure \ref{f17}.  
The situation is similar for the possible subleading jets in the $W$+jet events (Figure \ref{f18}).
$W$+jet events with more than two jets are not modeled in DYRAD 
simulation; their presence in the data creates extra biases in the low and high regions of the lepton $\phi$ 
distributions.  Because of the lepton-jet angular separation and lepton isolation requirements in our $W$+jet datasets
we obtain a bias-free measurement of the angular coefficients $A_2$ and $A_3$ if we exclude the positive-$x$ half-plane
region of the CS frame.

Secondly, the term $A_3 f_3(\phi)$ in Equation \ref{newer} is the smallest 
measurable term with our data.  Therefore, a more significant
measurement of the angular coefficient $A_3$ is obtained in the CS $\phi$ 
region where the rest of the terms (and mainly the predominant $A_4 f_4(\phi)$ term), 
contribute less.  The ratio $A_3 f_3(\phi)$ / $A_4 f_4(\phi)$ is significantly larger
in the $\pi/2< \phi <3\pi/2$ region, and thus a more sensitive
measurement of $A_3$ is obtained in this region.
We normalize the theory to data from $\pi$/2 to 3$\pi$/2 before we start the fitting
procedure, which is carried out in the $x<0$ region of the $z-x$ plane.

We use the MINUIT $\chi^2$ minimization program \cite{footnote2} to fit 
the electron and muon $\phi$ distributions to the fitting functions $f_i$.
Since these functions are not linearly independent, we cannot fit with all parameters free.
For this reason we keep the angular coefficients $A_0$ and $A_4$ fixed at their SM values and 
allow $A_2$ and $A_3$ to vary.  After we extract values for $A_2$ and $A_3$, we fix these coefficients at
these values, and we repeat the fit procedure varying only the $A_0$ and $A_4$ angular coefficients.  
The angular coefficients $A_1$, $A_5$, $A_6$, and $A_7$ are always kept fixed at their SM values, 
since the theoretical prediction for these coefficients is very close to zero and 
the variation for the first 100 GeV of $p_T^W$ is small in comparison to the experimental precision.  
We expect large statistical uncertainties for the extracted values of 
$A_0$ and $A_4$, since they do not significantly affect the $\phi$ distribution. 
Large variations in their value only slightly alter the leptons' $\phi$ angular distribution.  	

\begin{figure}
\includegraphics[scale=.46]{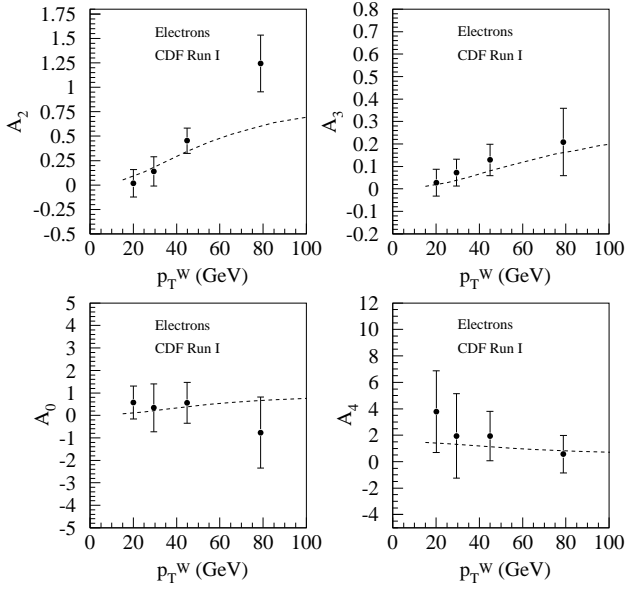}
\caption{The measurement of the angular coefficients 
for the $W$+jet electron data (points) and the SM prediction up to order $\alpha_s^2$
(line).
The errors are only statistical.}
\label{f22}
\end{figure}

\begin{figure}
\includegraphics[scale=.46]{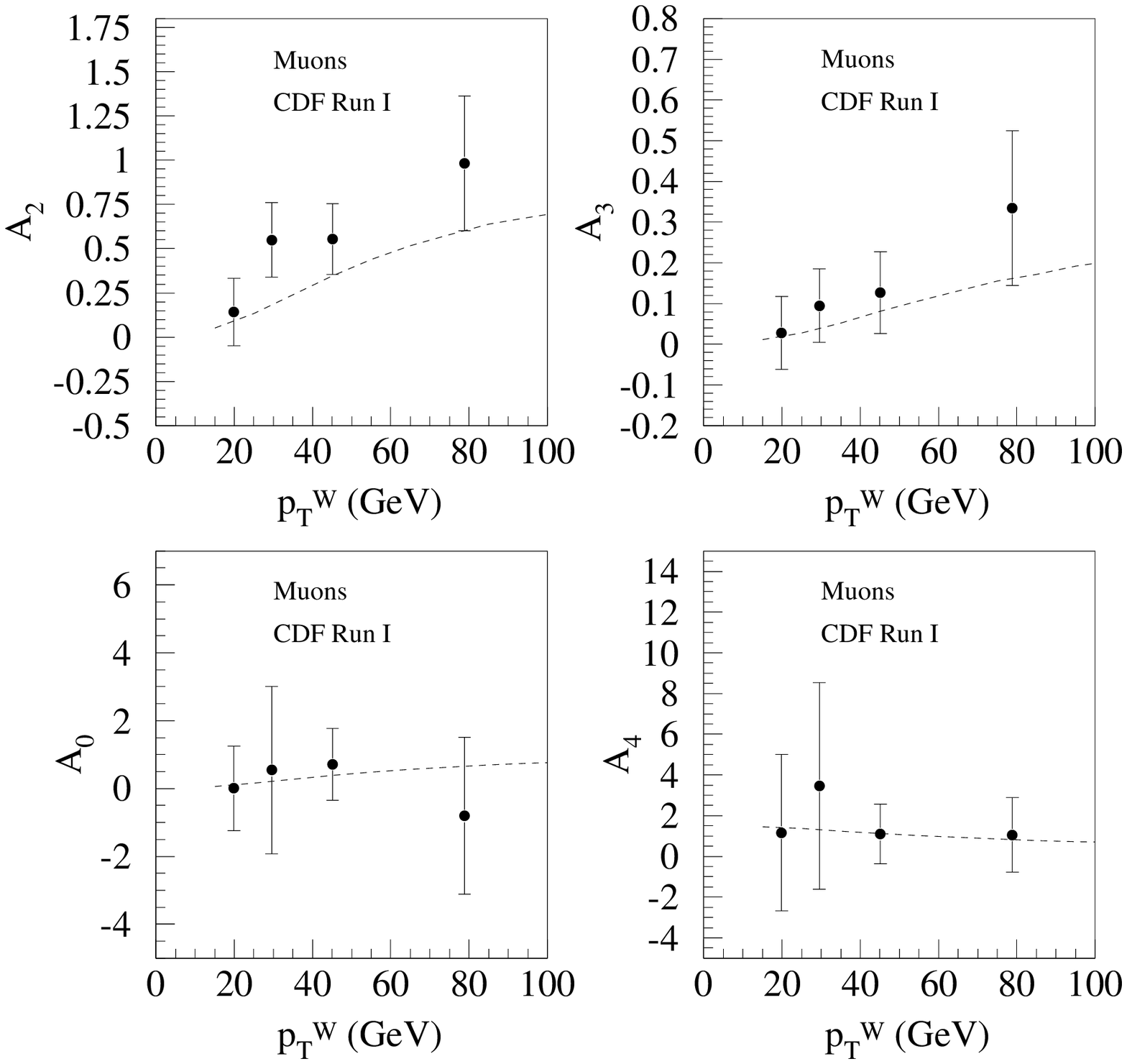}
\caption{The measurement of the angular coefficients 
for the $W$+jet muon data (points) and the SM prediction up to order $\alpha_s^2$
(line).The errors are only statistical.}
\label{f23}
\end{figure}

The results of the MINUIT fits are shown as dashed histograms in Figure \ref{f15} for the 
electron $W$+jet data and Figure \ref{f16} for the muon $W$+jet data.  
Our measurements of the angular
coefficients for the electron and muon $W$+jet data are presented in Figures \ref{f22} and
\ref{f23} respectively.  The bin centers are determined
using the average value of $p_T^W$ for the range of the four $p_T^W$ bins.
The measured angular coefficients associated with the electron and muon $W$+jet data
agree with the SM prediction and with each other.  We emphasize  
that the SM prediction is only up to order $\alpha_s^2$ in QCD.
\begin{figure}
\includegraphics[scale=.46]{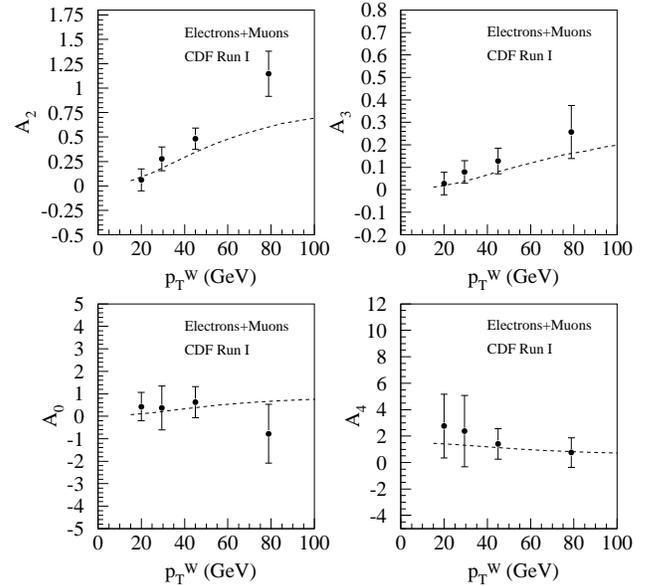}
\caption{The measurement of the angular coefficients 
for the combination of electrons and muons (points) 
and the SM prediction up to order $\alpha_s^2$ (line).  The errors are only statistical.}
\label{f24}
\end{figure}
The statistical uncertainties for $A_0$ and $A_4$ are very large,
as expected, making the measurement of these coefficients unrealistic
using the azimuthal angle analysis.

Assuming weak-interaction lepton-universality,
we combine the measurements of the angular coefficients obtained from the
electron and muon $W$+jet datasets, treating them as the results of two
separate experiments.  If $A_e$ and $A_{\mu}$ are the electron and muon measurements with
statistical uncertainties $\sigma_e$ and $\sigma_{\mu}$ respectively, then the
combined measurement is $A_{\rm comb}=(A_e/\sigma_e^2+A_{\mu}/\sigma_{\mu}^2)/(1/\sigma_e^2+1/\sigma_{\mu}^2)$,
with statistical uncertainty  $\sigma_{\rm comb}=(1/\sigma_e^2+1/\sigma_{\mu}^2)^{-1/2}$.
The result of this statistical combination, along with the SM prediction, 
is presented in Figure \ref{f24}.  
\section{\label{sec:sys} Systematic Uncertainties}

The systematic uncertainties associated with the measurement of the angular
coefficients $A_i$ are related to the jet definition and energy scale, the selection 
of the $W$ mass on an event-by-event basis, the background estimation,
possible presence of $W+\gamma$ events in our datasets, the assumed values of $A_0$ and $A_4$, 
the choice of parton distribution functions, and the renormalization and factorization scale $Q^2$ of the event.
The jet systematic uncertainties, the variation of the $A_0$ and $A_4$
values, and $Q^2$ scale uncertainty are the dominant sources of 
systematics.

\subsection{Jet systematic uncertainties}

The number of data events passing the jet cuts is affected by the
systematic uncertainties associated with the jet $E_T$ scale and the rapidity
requirement.  The same systematic uncertainty has an effect
on the measurement of the angular coefficients.

The uncertainty on jet $E_T$ scale depends on the 
calorimeter stability, relative energy scale corrections, extra interactions, and
underlying event corrections.  The total uncertainty is a quadratic sum of these 
effects.
The systematic uncertainty in the jet energy scale affects the
reconstruction of the $\met$ and the $W$ boson.  For every FMC $W$+jet event,
we shift the energy of the jet by $\sigma_+=85\%\sqrt{E^{\rm jet}}$, where 
$E^{\rm jet}$ is the energy of the jet in GeV,
without changing its direction.  We then
correct the $\met$ value and recalculate all the kinematic variables associated with the $W$ boson, jet, and $\met$.
We subsequently extract the new acceptance times efficiency $ae(\cos\theta,\phi)$
and analyze the data.  We repeat this procedure 
for the energy shifted by $\sigma_-=-85\%\sqrt{E^{\rm jet}}$
and calculate the systematic effect of the jet energy scale
on the measurement of the angular coefficients,
presented in Table \ref{t14} for the electron, muon, and the combination
of the two results.  To obtain the combined results,
we combine the electron and muon measurements for each $p_T^W$ bin and for each 
choice of $E_T^{\rm jet}$ energy shift,
using the statistical uncertainties of the central 
measurements.  The difference between the shifted combined values and the
central combined value determines the systematic uncertainty on the combined measurement.
The same method is used for all the systematic uncertainty estimates.

We vary the jet $E_T$ cut by $(\delta E_T)= \pm 850$ MeV
in both data and MC and repeat the analysis each time, to determine its 
effect on the measurement of the angular coefficients and on the FMC-prediction 
of signal event yields.
Table \ref{t13} shows the systematic uncertainty in the measurement
of the $W$+jet event yields associated with the jet $E_T$ cut variation,
for the four $p_T^W$ bins.  Overall, there is a
$+6.4 \%/-5.8 \%$ effect in the electron event yields and a 
$+6.0 \%/-5.7 \%$ effect in the muon event yields 
due to the jet $E_T$ cut. 

The uncertainty on the rapidity $\eta_{\rm lab}$ of the jet is $\delta\eta_{\rm lab}=\pm 0.2$.  
We vary the jet $\eta_{\rm lab}$ cut from 2.2 to 2.6 to obtain the variation in the data event yields
presented in Table \ref{t13}, for the four $p_T^W$ bins.  
Overall, there is a
$+2.2 \%/-2.4 \%$ effect in the electron event yields and a 
$+0.7 \%/-2.2 \%$  effect in the muon event yields 
due to the jet $\eta_{\rm lab}^{\rm jet}$ cut. 

In order to obtain an estimate of the systematic uncertainty in the
measurement of the angular coefficients associated with the 
jet $E_T$ and $\eta_{\rm lab}$ cuts, we run the analysis
for 11 values of the $E_T^{\rm jet}$ cut, from 14.15 GeV to 15.85 GeV,
and for five values of $|\eta_{\rm lab}^{\rm jet}|$ cut, from 2.2 to 2.6.
We record the variations in the measurement of the angular
coefficients for electrons, muons, and the combination of the
two results.  The results for the four $p_T^W$ bins are presented in 
Table \ref{t14}.  
\begin{table*}
\center
\caption{\label{t13}Systematic uncertainties on $W$+jet event yields due to the $E_T^{\rm jet}$ and
$\eta_{\rm lab}^{\rm jet}$ cuts, and the total systematic uncertainty due to these sources, for the four $p_T^W$ bins.}
\begin{tabular}{|c|c||c|c|c|} \hline \hline
\multicolumn{5}{|c|}{\bf \boldmath $W$+jet event yield systematic uncertainties due to the $E_T^{\rm jet}$ and $\eta_{\rm lab}^{\rm jet}$ cuts}\\  \hline
\multicolumn{1}{|c|}{$p_T^W$ (GeV)} & 
\multicolumn{1}{c||}{Charged Lepton} &
\multicolumn{1}{c|}{$\delta N$ due to $E_T^{\rm jet}$}  &
\multicolumn{1}{c|}{$\delta N$ due to $\eta_{\rm lab}^{\rm jet}$} & 
\multicolumn{1}{c|}{Total systematic uncertainty} \\ \hline 
 & electron & {$^{+593}_{-515}$}  & {$^{+156}_{-154}$}  & {$^{+613}_{-538}$} \\
  \rr{15--25}{2.5}     & muon & {$^{+308}_{-290}$}  & {$^{+34}_{-81}$}  & {$^{+310}_{-301}$} \\ \hline
 & electron & {$^{+179}_{-175}$}  & {$^{+78}_{-104}$}  & {$^{+195}_{-204}$} \\
  \rr{25--35}{2.5}     & muon & {$^{+95}_{-79}$}  & {$^{+12}_{-44}$}  & {$^{+96}_{-90}$} \\ \hline
& electron & {$^{+38}_{-50}$}  & {$^{+50}_{-43}$}  & {$^{+63}_{-66}$} \\
 \rr{35--65}{2.5}      & muon & {$^{+12}_{-29}$}  & {$^{+3}_{-29}$}  & {$^{+12}_{-41}$} \\ \hline
& electron & {$^{+0}_{-1}$}  & {$^{+1}_{-2}$}  & {$^{+1}_{-2}$} \\
  \rr{65--105}{2.5}     & muon & {$^{+0}_{-0}$}  & {$^{+0}_{-0}$}  & {$^{+0}_{-0}$} \\ \hline
 \hline
\end{tabular}
\end{table*}

\subsection{\boldmath Systematic uncertainty due to $W$ mass selection}

As previously discussed, in order to boost to the $W$ rest-frame,
a mass value is selected for the $W$ boson.  We have four different methods
for selecting this mass on an event-by-event basis.  We investigate how each mass selection method
affects our angular coefficients measurement.  The first method selects
a Breit-Wigner mass, which is greater than the measured transverse mass
of the $W$ boson.  The second method selects the greater of the pole mass
and the transverse mass.  In the third method we select
the pole mass or, in case it is less than the transverse mass, we 
select a Breit-Wigner mass, which is greater than the transverse mass.
Finally the fourth method (default) selects a mass based on the
distribution that results from the slice of the theoretical (DYRAD) $M^W$ vs. $M^W_T$ 
2-dimensional histogram (for $W$+jet events) at the measured transverse mass of the $W$ boson.  
This last method is preferred because it removes some biases in the measurement 
of the polar angle $\theta$.  In the $\phi$ analysis, the systematic uncertainty 
on the azimuthal angle $\phi$ due to the selection 
of the mass of the $W$ is minimal.
We run the analysis for the four mass selection methods
and record the variations in the measurement of the angular
coefficients for electrons, muons, and the combination of the
two results.  All methods give almost identical measurements of $\phi$.
The systematic uncertainties for the four $p_T^W$ bins are presented in 
Table \ref{t14}.

\subsection{Backgrounds estimate systematic uncertainty}

There is an uncertainty in the estimation of the backgrounds, 
given by the uncertainties in Tables \ref{t10} and \ref{t11}.  We vary our prediction 
from the highest value to the lowest possible value for 
every background as well as the FMC signal event yields.  These uncertainties do 
not include the PDF and $Q^2$ systematics.  For each 
variation, we rerun the analysis programs 
for the electron and muon case, and we also combine the results.  
The systematic uncertainties are presented for the four $p_T^W$ bins in 
Table \ref{t14}.

\subsection{\boldmath $W+\gamma$ systematic uncertainty}

The $W$+jet angular distribution can be affected by $W$+$\gamma$ production, for a hard
$\gamma$ well-separated from the charged lepton from the $W$ decay. 
Some of the events in our datasets are consistent with $W+\gamma$ production,
according to \cite{len}.  We remove those events and remeasure $A_2$ and $A_3$.
The variation from the original measurement is treated as a systematic uncertainty.
The systematic uncertainties for the four $p_T^W$ bins are presented in 
Table \ref{t14}.

\subsection{\boldmath $A_0$ and $A_4$ variation systematic uncertainty}

In our analysis we keep $A_0$ and $A_2$ fixed at their SM values.
To check how this affects our measurement, we set $A_0$ and $A_2$ at
minimum and maximum values in all possible combinations and
repeat the analysis four times ($A_0$(min)=0, $A_0$(max)=1, $A_4$(min)=0 and
$A_4$(max)=2).  The systematic uncertainties for the four $p_T^W$ bins are presented in 
Table \ref{t14}.

\subsection{PDF systematic uncertainty}

To study the uncertainty associated with the parton distribution functions, 
we use the MRSA$'$ [$\alpha_s(M_Z) = 0.105$ and $\Lambda = 0.150$] PDF
\cite{mrsap} and repeat the analysis.  The systematic uncertainties for the four $p_T^W$ bins 
are presented in Table \ref{t14}.
When we use all PDFs of the MRSA and CTEQ families, we end up with a systematic
uncertainty of $\pm 11\%$ on the DYRAD cross section, which affects both the central
FMC signal event yields and the electroweak backgrounds.  These variations are 
used for the estimation of
the total FMC event yields systematic uncertainty due to choice of PDF.
\subsection{\boldmath $Q^2$ Systematic uncertainty}

Finally we change the renormalization and factorization scale $Q^2$ so that
it is equal to the square of the transverse momentum of the $W$,
instead of the default square of the pole mass of the $W$ boson.  
The systematic uncertainties for the four $p_T^W$ bins are presented in 
Table \ref{t14}.
If we try all $Q^2$ choices provided by DYRAD
(total invariant mass squared, dynamic mass squared, total energy of the $W$ squared,
and transverse energy of the leading jet squared, in addition to
the two mentioned above), we end up with a systematic
uncertainty of $+19\%$~/~$-10\%$ on the DYRAD cross section,
which affects both the central FMC signal event yields and the electroweak
backgrounds.  These variations are used for the estimation of 
the total FMC event yields systematic uncertainty due to $Q^2$ scale variation.
\subsection{Overall analysis systematic uncertainties}

Table \ref{t15}  summarizes the total systematic uncertainties for the $A_{2}$ and $A_{3}$ measurement,
for the four $p_T^W$ bins and for the electron, muon, and combined
results.
To populate this table, we combine the systematics described above and
presented in Table \ref{t14}.

\begin{table*}
 \small
 \center
 \caption{\label{t14}Systematic uncertainties in the measurement of $A_2$ and $A_3$ along with their sources, for electron and muon $W$+jet events and the combination of the electron and muon results.}
 \begin{tabular}{|c||c||c|c||c|c||c|c|} \hline \hline
 \multicolumn{1}{|c||}{ } &
 \multicolumn{1}{c||}{ } &
 \multicolumn{2}{c||}{Electrons} &
 \multicolumn{2}{c||}{Muons} &
 \multicolumn{2}{c|}{Combination} \\ \cline{3-8}
 \multicolumn{1}{|c||}{\rr{Source of systematic uncertainty}{1.5}} &
 \multicolumn{1}{c||}{\rr{$p_T^W$ (GeV)}{1.5}} & 
 \multicolumn{1}{c|}{$\delta A_2$} &
 \multicolumn{1}{c||}{$\delta A_3$} &
 \multicolumn{1}{c|}{$\delta A_2$} &
 \multicolumn{1}{c||}{$\delta A_3$} &
 \multicolumn{1}{c|}{$\delta A_2$} &
 \multicolumn{1}{c|}{$\delta A_3$} \\ \hline

Jet $E_T$ cut             &          15--25 & {$^{+0.0275}_{-0.0533}$} & {$^{+0.0000}_{-0.0064}$} & {$^{+0.0146}_{-0.0502}$} & {$^{+0.0036}_{-0.0032}$} & {$^{+0.0230}_{-0.0523}$} & {$^{+0.0000}_{-0.0043}$} \\
    			  &          25--35 & {$^{+0.0395}_{-0.0109}$} & {$^{+0.0047}_{-0.0039}$} & {$^{+0.0676}_{-0.0559}$} & {$^{+0.0073}_{-0.0017}$} & {$^{+0.0490}_{-0.0261}$} & {$^{+0.0050}_{-0.0027}$} \\
    			  &          35--65 & {$^{+0.0433}_{-0.0000}$} & {$^{+0.0030}_{-0.0033}$} & {$^{+0.0177}_{-0.0131}$} & {$^{+0.0042}_{-0.0011}$} & {$^{+0.0357}_{-0.0000}$} & {$^{+0.0031}_{-0.0022}$} \\
    			  &         65--105 & {$^{+0.0000}_{-0.0617}$} & {$^{+0.0167}_{-0.0153}$} & {$^{+0.1519}_{-0.0000}$} & {$^{+0.0055}_{-0.0808}$} & {$^{+0.0556}_{-0.0299}$} & {$^{+0.0024}_{-0.0374}$} \\ \hline \hline
Jet $\eta_{\rm lab}$ cut            &15--25 & {$^{+0.0043}_{-0.0324}$} & {$^{+0.0000}_{-0.0036}$} & {$^{+0.0098}_{-0.0374}$} & {$^{+0.0044}_{-0.0000}$} & {$^{+0.0002}_{-0.0175}$} & {$^{+0.0000}_{-0.0018}$}\\
     			  &          25--35 & {$^{+0.0007}_{-0.0346}$} & {$^{+0.0012}_{-0.0022}$} & {$^{+0.0299}_{-0.0221}$} & {$^{+0.0000}_{-0.0085}$} & {$^{+0.0106}_{-0.0304}$} & {$^{+0.0000}_{-0.0032}$}\\
    			  &          35--65 & {$^{+0.0093}_{-0.0049}$} & {$^{+0.0023}_{-0.0005}$} & {$^{+0.0070}_{-0.0075}$} & {$^{+0.0038}_{-0.0028}$} & {$^{+0.0070}_{-0.0032}$} & {$^{+0.0013}_{-0.0000}$}\\
    			  &         65--105 & {$^{+0.0000}_{-0.0617}$} & {$^{+0.0022}_{-0.0052}$} & {$^{+0.1021}_{-0.0000}$} & {$^{+0.0000}_{-0.0557}$} & {$^{+0.0112}_{-0.0258}$} & {$^{+0.0000}_{-0.0203}$}\\ \hline \hline
Jet energy scale       	  &	     15--25 & {$^{+0.0594}_{-0.0588}$} & {$^{+0.0042}_{-0.0014}$} & {$^{+0.0499}_{-0.0694}$} & {$^{+0.0075}_{-0.0044}$} & {$^{+0.0560}_{-0.0626}$} & {$^{+0.0052}_{-0.0023}$} \\
       			  &          25--35 & {$^{+0.0851}_{-0.1105}$} & {$^{+0.0096}_{-0.0020}$} & {$^{+0.0860}_{-0.0994}$} & {$^{+0.0261}_{-0.0141}$} & {$^{+0.0854}_{-0.1068}$} & {$^{+0.0147}_{-0.0057}$} \\
    			  &          35--65 & {$^{+0.0017}_{-0.0514}$} & {$^{+0.0204}_{-0.0069}$} & {$^{+0.0142}_{-0.0850}$} & {$^{+0.0183}_{-0.0102}$} & {$^{+0.0054}_{-0.0614}$} & {$^{+0.0197}_{-0.0080}$} \\
    			  &         65--105 & {$^{+0.2468}_{-0.3821}$} & {$^{+0.0836}_{-0.0489}$} & {$^{+0.2989}_{-0.3186}$} & {$^{+0.0382}_{-0.0176}$} & {$^{+0.2660}_{-0.3587}$} & {$^{+0.0662}_{-0.0369}$} \\ \hline \hline
$M_W$ selection           &          15--25 & {$^{+0.0030}_{-0.0000}$} & {$^{+0.0000}_{-0.0011}$} & {$^{+0.0000}_{-0.0016}$} & {$^{+0.0001}_{-0.0000}$} & {$^{+0.0019}_{-0.0000}$} & {$^{+0.0000}_{-0.0007}$} \\
    			  &          25--35 & {$^{+0.0018}_{-0.0000}$} & {$^{+0.0010}_{-0.0000}$} & {$^{+0.0034}_{-0.0000}$} & {$^{+0.0007}_{-0.0000}$} & {$^{+0.0024}_{-0.0000}$} & {$^{+0.0007}_{-0.0000}$} \\
    			  &          35--65 & {$^{+0.0068}_{-0.0046}$} & {$^{+0.0034}_{-0.0002}$} & {$^{+0.0000}_{-0.0120}$} & {$^{+0.0038}_{-0.0000}$} & {$^{+0.0012}_{-0.0040}$} & {$^{+0.0035}_{-0.0000}$} \\
    			  &         65--105 & {$^{+0.0000}_{-0.0561}$} & {$^{+0.0032}_{-0.0000}$} & {$^{+0.0560}_{-0.0000}$} & {$^{+0.0000}_{-0.0598}$} & {$^{+0.0000}_{-0.0149}$} & {$^{+0.0013}_{-0.0210}$} \\ \hline \hline
$\tau$ background      	  &          15--25 & {$^{+0.0009}_{-0.0009}$} & {$^{+0.0001}_{-0.0000}$} & {$^{+0.0014}_{-0.0013}$} & {$^{+0.0001}_{-0.0001}$} & {$^{+0.0011}_{-0.0011}$} & {$^{+0.0001}_{-0.0001}$} \\
    			  &          25--35 & {$^{+0.0008}_{-0.0010}$} & {$^{+0.0000}_{-0.0001}$} & {$^{+0.0023}_{-0.0018}$} & {$^{+0.0002}_{-0.0002}$} & {$^{+0.0013}_{-0.0013}$} & {$^{+0.0001}_{-0.0002}$} \\
    			  &          35--65 & {$^{+0.0008}_{-0.0012}$} & {$^{+0.0001}_{-0.0002}$} & {$^{+0.0018}_{-0.0018}$} & {$^{+0.0002}_{-0.0002}$} & {$^{+0.0011}_{-0.0014}$} & {$^{+0.0001}_{-0.0002}$} \\
    			  &         65--105 & {$^{+0.0024}_{-0.0025}$} & {$^{+0.0003}_{-0.0005}$} & {$^{+0.0000}_{-0.0000}$} & {$^{+0.0000}_{-0.0000}$} & {$^{+0.0015}_{-0.0016}$} & {$^{+0.0002}_{-0.0003}$} \\ \hline \hline
Z background              &          15--25 & {$^{+0.0000}_{-0.0009}$} & {$^{+0.0001}_{-0.0000}$} & {$^{+0.0000}_{-0.0017}$} & {$^{+0.0001}_{-0.0000}$} & {$^{+0.0000}_{-0.0012}$} & {$^{+0.0001}_{-0.0000}$} \\
    			  &          25--35 & {$^{+0.0000}_{-0.0010}$} & {$^{+0.0001}_{-0.0000}$} & {$^{+0.0000}_{-0.0037}$} & {$^{+0.0002}_{-0.0000}$} & {$^{+0.0000}_{-0.0019}$} & {$^{+0.0001}_{-0.0000}$} \\
    			  &          35--65 & {$^{+0.0000}_{-0.0012}$} & {$^{+0.0001}_{-0.0000}$} & {$^{+0.0000}_{-0.0029}$} & {$^{+0.0002}_{-0.0000}$} & {$^{+0.0000}_{-0.0017}$} & {$^{+0.0001}_{-0.0000}$} \\
      			  &         65--105 & {$^{+0.0000}_{-0.0025}$} & {$^{+0.0003}_{-0.0000}$} & {$^{+0.0001}_{-0.0000}$} & {$^{+0.0006}_{-0.0007}$} & {$^{+0.0000}_{-0.0016}$} & {$^{+0.0004}_{-0.0001}$} \\ \hline \hline
QCD background            &          15--25 & {$^{+0.0140}_{-0.0127}$} & {$^{+0.0009}_{-0.0009}$} & {$^{+0.0329}_{-0.0125}$} & {$^{+0.0010}_{-0.0030}$} & {$^{+0.0206}_{-0.0126}$} & {$^{+0.0009}_{-0.0015}$}\\
    			  &          25--35 & {$^{+0.0092}_{-0.0326}$} & {$^{+0.0013}_{-0.0003}$} & {$^{+0.0242}_{-0.0076}$} & {$^{+0.0009}_{-0.0039}$} & {$^{+0.0143}_{-0.0242}$} & {$^{+0.0012}_{-0.0014}$}\\
    			  &          35--65 & {$^{+0.0018}_{-0.0149}$} & {$^{+0.0010}_{-0.0001}$} & {$^{+0.0062}_{-0.0087}$} & {$^{+0.0013}_{-0.0015}$} & {$^{+0.0000}_{-0.0086}$} & {$^{+0.0004}_{-0.0000}$}\\ 
     			  &         65--105 & {$^{+0.0134}_{-0.0050}$} & {$^{+0.0002}_{-0.0014}$} & {$^{+0.0243}_{-0.0000}$} & {$^{+0.0039}_{-0.0007}$} & {$^{+0.0174}_{-0.0031}$} & {$^{+0.0006}_{-0.0001}$}\\ \hline\hline
FMC signal event yield                 &    15--25 & {$^{+0.0031}_{-0.0065}$} & {$^{+0.0000}_{-0.0023}$} & {$^{+0.0000}_{-0.0004}$} & {$^{+0.0012}_{-0.0032}$} & {$^{+0.0019}_{-0.0043}$} & {$^{+0.0000}_{-0.0016}$}\\
    			  &          25--35 & {$^{+0.0190}_{-0.0000}$} & {$^{+0.0003}_{-0.0016}$} & {$^{+0.0355}_{-0.0000}$} & {$^{+0.0020}_{-0.0017}$} & {$^{+0.0246}_{-0.0000}$} & {$^{+0.0008}_{-0.0016}$}\\
    			  &          35--65 & {$^{+0.0173}_{-0.0000}$} & {$^{+0.0023}_{-0.0033}$} & {$^{+0.0030}_{-0.0022}$} & {$^{+0.0039}_{-0.0000}$} & {$^{+0.0115}_{-0.0000}$} & {$^{+0.0028}_{-0.0022}$}\\
    			  &         65--105 & {$^{+0.0000}_{-0.0540}$} & {$^{+0.0009}_{-0.0052}$} & {$^{+0.0830}_{-0.0000}$} & {$^{+0.0000}_{-0.0557}$} & {$^{+0.0110}_{-0.0267}$} & {$^{+0.0000}_{-0.0246}$}\\ \hline \hline
W+$\gamma$                &          15--25 & {$^{+0.0000}_{-0.0060}$} & {$^{+0.0000}_{-0.0031}$} & {$^{+0.0022}_{-0.0000}$} & {$^{+0.0021}_{-0.0000}$} & {$^{+0.0000}_{-0.0031}$} & {$^{+0.0000}_{-0.0015}$} \\
    			  &          25--35 & {$^{+0.0000}_{-0.0064}$} & {$^{+0.0019}_{-0.0000}$} & {$^{+0.0050}_{-0.0000}$} & {$^{+0.0000}_{-0.0039}$} & {$^{+0.0000}_{-0.0025}$} & {$^{+0.0001}_{-0.0000}$} \\
    			  &          35--65 & {$^{+0.0103}_{-0.0000}$} & {$^{+0.0023}_{-0.0000}$} & {$^{+0.0026}_{-0.0000}$} & {$^{+0.0015}_{-0.0000}$} & {$^{+0.0080}_{-0.0000}$} & {$^{+0.0020}_{-0.0000}$} \\
    			  &         65--105 & {$^{+0.0000}_{-0.0704}$} & {$^{+0.0024}_{-0.0000}$} & {$^{+0.0000}_{-0.0060}$} & {$^{+0.0000}_{-0.0166}$} & {$^{+0.0000}_{-0.0467}$} & {$^{+0.0000}_{-0.0049}$} \\ \hline \hline
$A_0$ and $A_4$ variation     &      15--25 & {$^{+0.0296}_{-0.2024}$} & {$^{+0.0074}_{-0.0015}$} & {$^{+0.0323}_{-0.1965}$} & {$^{+0.0080}_{-0.0016}$} & {$^{+0.0305}_{-0.2003}$} & {$^{+0.0076}_{-0.0015}$}\\
    			  &          25--35 & {$^{+0.0691}_{-0.2326}$} & {$^{+0.0155}_{-0.0051}$} & {$^{+0.0553}_{-0.1745}$} & {$^{+0.0240}_{-0.0081}$} & {$^{+0.0644}_{-0.2129}$} & {$^{+0.0181}_{-0.0060}$}\\
    			  &          35--65 & {$^{+0.1053}_{-0.1610}$} & {$^{+0.0218}_{-0.0158}$} & {$^{+0.0917}_{-0.1390}$} & {$^{+0.0170}_{-0.0122}$} & {$^{+0.1013}_{-0.1545}$} & {$^{+0.0202}_{-0.0146}$}\\
    			  &         65--105 & {$^{+0.1189}_{-0.0708}$} & {$^{+0.0150}_{-0.0277}$} & {$^{+0.0963}_{-0.0553}$} & {$^{+0.0167}_{-0.0312}$} & {$^{+0.1106}_{-0.0651}$} & {$^{+0.0157}_{-0.0290}$}\\ \hline \hline
PDF variation             &          15--25 & {$^{+0.0000}_{-0.0159}$} & {$^{+0.0013}_{-0.0000}$} & {$^{+0.0000}_{-0.0419}$} & {$^{+0.0017}_{-0.0000}$} & {$^{+0.0000}_{-0.0251}$} & {$^{+0.0014}_{-0.0000}$}\\
    			  &          25--35 & {$^{+0.0152}_{-0.0000}$} & {$^{+0.0000}_{-0.0034}$} & {$^{+0.0000}_{-0.0018}$} & {$^{+0.0000}_{-0.0003}$} & {$^{+0.0094}_{-0.0000}$} & {$^{+0.0000}_{-0.0024}$}\\
    			  &          35--65 & {$^{+0.0000}_{-0.0068}$} & {$^{+0.0011}_{-0.0000}$} & {$^{+0.0759}_{-0.0000}$} & {$^{+0.0000}_{-0.0003}$} & {$^{+0.0178}_{-0.0000}$} & {$^{+0.0006}_{-0.0000}$}\\
    			  &         65--105 & {$^{+0.0022}_{-0.0000}$} & {$^{+0.0134}_{-0.0000}$} & {$^{+0.4001}_{-0.0000}$} & {$^{+0.0000}_{-0.0931}$} & {$^{+0.1486}_{-0.0000}$} & {$^{+0.0000}_{-0.0275}$}\\ \hline \hline
$Q^2$ variation           &          15--25 & {$^{+0.1043}_{-0.0000}$} & {$^{+0.0028}_{-0.0000}$} & {$^{+0.1200}_{-0.0000}$} & {$^{+0.0000}_{-0.0121}$} & {$^{+0.1098}_{-0.0000}$} & {$^{+0.0000}_{-0.0018}$}\\
    			  &          25--35 & {$^{+0.1319}_{-0.0000}$} & {$^{+0.0000}_{-0.0104}$} & {$^{+0.0668}_{-0.0000}$} & {$^{+0.0000}_{-0.0076}$} & {$^{+0.1099}_{-0.0000}$} & {$^{+0.0000}_{-0.0096}$}\\
    			  &          35--65 & {$^{+0.1347}_{-0.0000}$} & {$^{+0.0000}_{-0.0267}$} & {$^{+0.0386}_{-0.0000}$} & {$^{+0.0000}_{-0.0180}$} & {$^{+0.1062}_{-0.0000}$} & {$^{+0.0000}_{-0.0239}$}\\
    			  &         65--105 & {$^{+0.0000}_{-0.0959}$} & {$^{+0.0101}_{-0.0000}$} & {$^{+0.0000}_{-0.2251}$} & {$^{+0.0162}_{-0.0000}$} & {$^{+0.0000}_{-0.1434}$} & {$^{+0.0124}_{-0.0000}$}\\ \hline \hline
\end{tabular}
 \end{table*}

\begin{table}
 \small
 \center
 \caption{ \label{t15}Total systematic uncertainties in the measurement of $A_2$ and $A_3$, for electron and muon $W$+jet events and the combination of the electron and muon results.  The systematic uncertainties of Table \ref{t14} are combined in quadrature.}
 \begin{tabular}{|c||c|c||c|c||c|c|} \hline \hline
 \multicolumn{7}{|c|}{\bf Total systematic uncertainties} \\ \hline
 \multicolumn{1}{|c||}{ } &
 \multicolumn{2}{c||}{Electrons} &
 \multicolumn{2}{c||}{Muons} &
 \multicolumn{2}{c|}{Combination} \\ \cline{2-7}
 \multicolumn{1}{|c||}{\rr{$p_T^W$ (GeV)}{1.5}} & 
 \multicolumn{1}{c|}{$\delta A_2$} &
 \multicolumn{1}{c||}{$\delta A_3$} &
 \multicolumn{1}{c|}{$\delta A_2$} &
 \multicolumn{1}{c||}{$\delta A_3$} &
 \multicolumn{1}{c|}{$\delta A_2$} &
 \multicolumn{1}{c|}{$\delta A_3$} \\ \hline
          15--25 &  {$^{+0.1275}_{-0.2209}$} & {$^{+0.0092}_{-0.0087}$} & {$^{+0.1390}_{-0.2220}$} & {$^{+0.0128}_{-0.0140}$} & {$^{+0.1307}_{-0.2189}$} & {$^{+0.0094}_{-0.0064}$}\\
          25--35 &  {$^{+0.1779}_{-0.2622}$} & {$^{+0.0191}_{-0.0131}$} & {$^{+0.1492}_{-0.2098}$} & {$^{+0.0362}_{-0.0208}$} & {$^{+0.1641}_{-0.2428}$} & {$^{+0.0239}_{-0.0137}$}\\
          35--65 &  {$^{+0.1779}_{-0.1699}$} & {$^{+0.0305}_{-0.0321}$} & {$^{+0.1276}_{-0.1643}$} & {$^{+0.0263}_{-0.0243}$} & {$^{+0.1530}_{-0.1665}$} & {$^{+0.0289}_{-0.0293}$}\\
         65--105 &  {$^{+0.2743}_{-0.4229}$} & {$^{+0.0883}_{-0.0587}$} & {$^{+0.5503}_{-0.3941}$} & {$^{+0.0453}_{-0.1629}$} & {$^{+0.3297}_{-0.3977}$} & {$^{+0.0692}_{-0.0764}$} \\ \hline \hline
 \end{tabular}
 \end{table}

\begin{table}
\center
\caption{\label{t16}The electron and muon CDF data event yields for inclusive $W$+jet production,
with statistical and systematic uncertainties.  The systematic uncertainties are due to
$E_T^{\rm jet}$ and $\eta_{\rm lab}^{\rm jet}$ cuts.}
\begin{tabular}{|c||c|c|}\hline \hline
\multicolumn{3}{|c|}{\bf Data event yields for inclusive \boldmath $W$+jet production}\\  \hline
\multicolumn{1}{|c||}{$p_T^W$ (GeV)} & 
\multicolumn{1}{c|}{$N_e$} &
\multicolumn{1}{c|}{$N_{\mu}$}\\ \hline 
15--25 & 5166 {$\pm$ 72} {$^{+613}_{-538}$} & 2821 {$\pm$ 53} {$^{+310}_{-301}$}\\
25--35 & 3601 {$\pm$ 60} {$^{+195}_{-204}$} & 1869 {$\pm$ 43} {$^{+96}_{-90}$}\\
35--65 & 3285 {$\pm$ 57} {$^{+63}_{-66}$} & 1880 {$\pm$ 43} {$^{+12}_{-41}$}\\
65--105 & 624 {$\pm$ 25} {$^{+1}_{-2}$} & 371 {$\pm$ 19} {$^{+0}_{-0}$}\\ \hline \hline
\end{tabular}
\end{table}

\subsection{Overall systematic uncertainties in data and Monte Carlo event yields}

Combining the data event yield systematics due to $E_T^{\rm jet}$ and $\eta_{\rm lab}^{\rm jet}$ cut variations in quadrature,
we get the final data event yields presented in Table \ref{t16}.  
Comparing with the FMC event yields of Table \ref{t12}, 
we see that there is a reasonable agreement with the SM prediction.
In Table \ref{t12} we have also included the PDF and $Q^2$ FMC systematic 
uncertainties described earlier, 
combined in quadrature to give a systematic uncertainty of $+22\%$~/~$-15\%$ 
on the FMC signal event yields and electroweak background.
We do not expect perfect agreement since the DYRAD generator produces
up to two jets with $E_T^{\rm jet}>10$ GeV (order $\alpha_s^2$), while in the 
data we have many events with more than two jets with $E_T^{\rm jet}>10$ GeV.
If we impose a cut on the number of jets in the CDF data, by not accepting more
than two jets in an event, and applying strict cuts on at least one jet, 
the disagreement is reduced by more than 50\%.  Nevertheless, we prefer not to constrain the dataset in such a manner.
Note that the event yield measurements do not affect the 
angular coefficient measurements, since we are interested only in the {\it shapes} of the
distributions, in the latter case.

\section{Final results}

\begin{figure}[!]
\includegraphics[scale=.46]{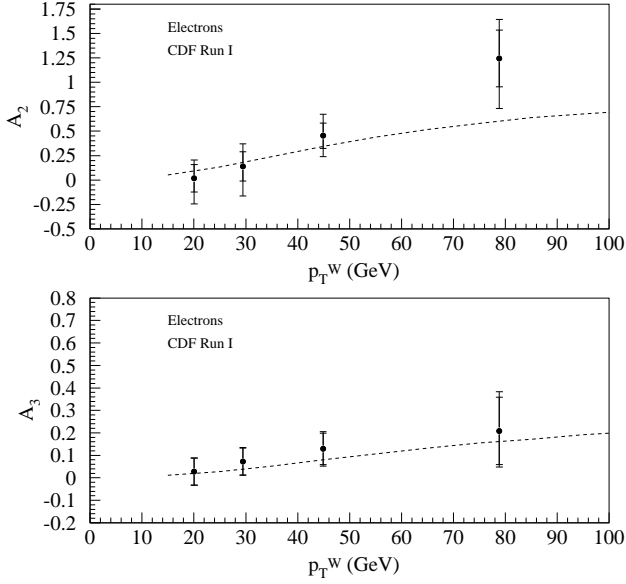}
\caption{Measured $A_2$ and $A_3$ using electron $W$+jet events.  The total (outer) and statistical (inner) uncertainties 
are shown along with the Standard Model 1-loop prediction up to order $\alpha_s^2$ (dashed line).}
\label{f25}
\end{figure}

\begin{figure}[!]
\includegraphics[scale=.46]{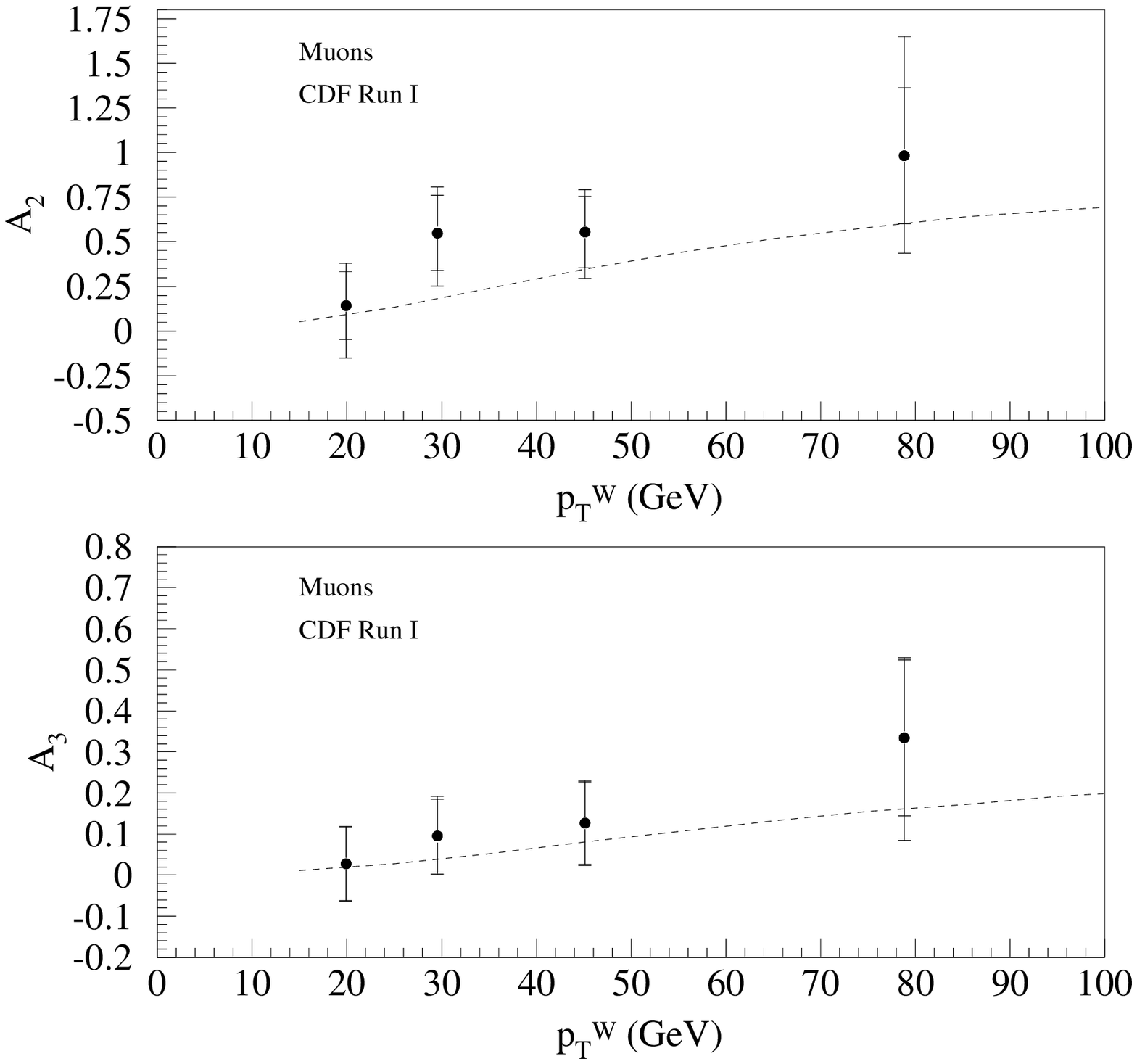}
\caption{Measured $A_2$ and $A_3$ using muon $W$+jet events.  The total (outer) and statistical (inner) uncertainties 
are shown along with the Standard Model 1-loop prediction up to order $\alpha_s^2$ (dashed line).}
\label{f26}
\end{figure}

\begin{figure}[!]
\includegraphics[scale=.46]{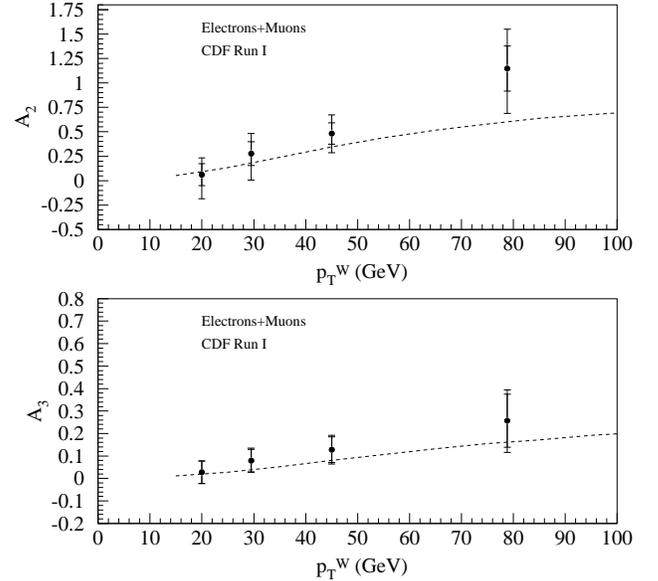}
\caption{Measured $A_2$ and $A_3$ using the combination of electron and muon measurements.  The total (outer) and statistical (inner) uncertainties 
are shown along with the Standard Model 1-loop prediction up to order $\alpha_s^2$ (dashed line).}
\label{f27}
\end{figure}

Combining the statistical and systematic uncertainties associated with the $A_{2}$ and $A_{3}$ measurement,
we obtain our final results, presented in Tables \ref{t17} and \ref{t18}.  Figure \ref{f25} 
shows the measurement of $A_2$ and $A_3$ for the electron $W$+jet data
and Figure \ref{f26} shows the measurement of $A_2$ and $A_3$ for the muon $W$+jet data.
The combination of the electron and muon measurements of the two angular
coefficients is presented in Figure \ref{f27}.
The Standard Model predictions for these angular coefficients, up to order $\alpha_s^2$, are also presented.
\begin{table}[!]
\center
\caption{\label{t17}The measurement of the $A_2$ coefficient  along with the statistical and systematic uncertainties for electron and muon $W$+jet events and the combination of the electron and muon results.
The SM values up to order $\alpha_s^2$ are also included.}
\begin{tabular}{|c||c|c|} \hline \hline
\multicolumn{3}{|c|}{\bf Measurement of {\boldmath $A_2$} angular coefficient}\\  \hline
\multicolumn{1}{|c||}{ } &
\multicolumn{1}{c|}{$p_T^W$ (GeV)} & 
\multicolumn{1}{c|}{$A_2$} \\ \hline
 & 15--25 &                     0.02 {$\pm$ 0.14} {$^{+0.13}_{-0.22}$} \\
 & 25--35 &                     0.14 {$\pm$ 0.15} {$^{+0.18}_{-0.26}$} \\
 & 35--65 &                     0.45 {$\pm$ 0.13} {$^{+0.18}_{-0.17}$} \\
\rr{Electrons}{7.3} & 65--105&  1.24 {$\pm$ 0.29} {$^{+0.27}_{-0.42}$} \\ \hline \hline

 & 15--25 & 		        0.14 {$\pm$ 0.19}{$^{+0.14}_{-0.22}$}    \\
 & 25--35 & 			0.55 {$\pm$ 0.21} {$^{+0.15}_{-0.21}$}  \\
 & 35--65 & 			0.55 {$\pm$ 0.20} {$^{+0.13}_{-0.16}$}  \\
\rr{Muons}{7.3}  & 65--105& 	0.98 {$\pm$ 0.38}{$^{+0.55}_{-0.39}$}    \\ \hline \hline

 & 15--25 & 				0.06 {$\pm$ 0.11}{$^{+0.13}_{-0.22}$}  \\
 & 25--35 & 				0.28 {$\pm$ 0.12}{$^{+0.16}_{-0.24}$}   \\
 & 35--65 & 				0.48 {$\pm$ 0.11} {$^{+0.15}_{-0.17}$} \\
\rr{Combination}{7.3} & 65--105& 	1.15 {$\pm$ 0.23}{$^{+0.33}_{-0.40}$}  \\ \hline \hline

 & 15--25  & 0.09 \\
 & 25--35  & 0.19 \\
 & 35--65  & 0.35 \\
\rr{Standard Model}{7.3} & 65--105 & 0.60\\ \hline \hline
\end{tabular}
\end{table}
\begin{table}[!]
\center
\caption{\label{t18}The measurement of the $A_3$ coefficient along with the statistical and systematic uncertainties for electron and muon $W$+jet events and the combination of the electron and muon results.
The SM values up to order $\alpha_s^2$ are also included.}
\begin{tabular}{|c||c|c|c|} \hline \hline
\multicolumn{3}{|c|}{\bf Measurement of {\boldmath $A_3$} angular coefficient}\\  \hline
\multicolumn{1}{|c||}{ } &
\multicolumn{1}{c|}{$p_T^W$ (GeV)} & 
\multicolumn{1}{c|}{$A_3$} \\ \hline
 & 15--25  & 			 0.03 {$\pm$ 0.06} {$^{+0.01}_{-0.01}$}  \\
 & 25--35  & 			 0.07 {$\pm$ 0.06} {$^{+0.02}_{-0.01}$}  \\
 & 35--65  &  			 0.13 {$\pm$ 0.07} {$^{+0.03}_{-0.03}$}  \\
\rr{Electrons}{7.3} & 65--105 &  0.21 {$\pm$ 0.15} {$^{+0.09}_{-0.06}$}  \\ \hline \hline

 & 15--25   & 				0.03 {$\pm$ 0.09}{$^{+0.01}_{-0.01}$} \\
 & 25--35  & 				0.09 {$\pm$ 0.09}{$^{+0.04}_{-0.02}$}  \\
 & 35--65   &  				0.13 {$\pm$ 0.10}{$^{+0.03}_{-0.02}$} \\
\rr{Muons}{7.3}  & 65--105  & 		0.33 {$\pm$ 0.19} {$^{+0.05}_{-0.16}$} \\ \hline \hline

 & 15--25   & 				0.03 {$\pm$ 0.05}  {$^{+0.01}_{-0.01}$} \\
 & 25--35   & 				0.08 {$\pm$ 0.05}{$^{+0.02}_{-0.01}$}    \\
 & 35--65   &  				0.13 {$\pm$ 0.06} {$^{+0.03}_{-0.03}$}  \\
\rr{Combination}{7.3} & 65--105  & 	0.26 {$\pm$ 0.12} {$^{+0.07}_{-0.08}$}   \\ \hline \hline

 & 15--25   & 0.02\\
 & 25--35   & 0.04\\
 & 35--65  & 0.08\\
\rr{Standard Model}{7.3} & 65--105 & 0.16\\ \hline \hline
\end{tabular}
\end{table}

\section{Conclusions}

We have made the first measurement of the $A_{2}$ and $A_{3}$ 
angular coefficients of $W$ boson production
and decay, 
using the CDF Run Ia and Run Ib electron and muon $W$+jet data.
Our datasets include at least one jet, satisfying the energy and pseudorapidity
requirements.
Due to finite statistical analyzing power of our $W$+jet datasets
and the characteristics of the $W$ decay, only the measurement of $A_2$ and $A_3$ angular
coefficients 
is statistically significant, with the analysis of the azimuthal
angle of the charged lepton in the $W$ rest-frame.  The $A_0$ and $A_4$
coefficients are preferably measurable with a polar angle analysis, while $A_1$ and the next-to-leading
order coefficients -- $A_5$, $A_6$, and $A_7$ -- are not measurable, with any meaningful statistical 
significance, with Run I $W$+jet data.

At leading order, the $A_2$ and $A_3$ angular coefficients fully describe the azimuthal $W$ angular
distribution in the Collins-Soper $W$ rest-frame.  These angular coefficients are also part of the 
total $W$ differential cross section, and can be expressed as ratios of the corresponding helicity cross sections
of the $W$ to its total unpolarized cross section.
This measurement tests the Standard Model prediction for $W$ polarization,
and the associated QCD corrections present in the production of $W$
bosons at high transverse momenta.
We observe good agreement with the Standard Model prediction up to order $\alpha_s^2$ in QCD.
\begin{acknowledgments}
We thank Walter Giele for DYRAD support and useful discussions.
We also thank the Fermilab staff and the technical staffs of the participating institutions for their vital contributions. This work was supported by the U.S. Department of Energy and National Science Foundation; the Italian Istituto Nazionale di Fisica Nucleare; the Ministry of Education, Culture, Sports, Science and Technology of Japan; the Natural Sciences and Engineering Research Council of Canada; the National Science Council of the Republic of China; the Swiss National Science Foundation; the A.P. Sloan Foundation; the Bundesministerium fuer Bildung und Forschung, Germany; the Korean Science and Engineering Foundation and the Korean Research Foundation; the Particle Physics and Astronomy Research Council and the Royal Society, UK; the Russian Foundation for Basic Research; the Comision Interministerial de Ciencia y Tecnologia, Spain; and in part by the European Community's Human Potential Programme under contract HPRN-CT-2002-00292, Probe for New Physics.
\end{acknowledgments}


\begin{thebibliography}{20}
\bibitem[{foo({\natexlab{a}})}]{footnote1}
\bibinfo{note}{If we $CP$ transform the $W^-$ into $W^+$ keeping the momenta of
  the colliding hadrons fixed (in order to have a fixed CS frame), Equation
  (\ref{eq1}) describes the $W^+$ if the $A_1$, $A_4$ and $A_6$ coefficients
  are multiplied by $(-1)$; these coefficients multiply $P$-odd functions of
  $\theta$ and $\phi$, since the Parity transformation takes $\phi \rightarrow
  \phi$ and $\theta \rightarrow \pi-\theta$ in the CS frame.}
\bibitem[{uni()}]{units}
\bibinfo{note}{In this paper, we use natural units, according to which
  $c=\hbar=1$. Therefore, the energy, momentum, and mass are measured in eV.}
\bibitem[{foo({\natexlab{b}})}]{footnote}
\bibinfo{note}{This analysis has been performed on $W^{+}$ and $W^{-}$
  separately with consistent results.}
\end{thebibliography}
\end{document}